\documentclass[11pt,a4paper]{article}
\usepackage[a4paper, margin=0.9in]{geometry}
\usepackage[T1]{fontenc}
\usepackage[latin1]{inputenc}
\usepackage[english]{babel}
\usepackage{amsmath,amsfonts,amssymb,mhsetup,mathtools}
\usepackage{lmodern}
\usepackage[sc]{mathpazo}
\usepackage{euscript}
\usepackage{dsfont}
\usepackage{natbib}
\usepackage{graphicx}
\usepackage{setspace}
\usepackage[activate={true,nocompatibility},kerning=true,final,verbose=errors,babel=true]{microtype}
\usepackage{graphicx}
\usepackage{float}
\usepackage{caption,subcaption}
\usepackage[flushleft]{threeparttable}
\usepackage{booktabs}
\usepackage{makecell}
\usepackage{placeins}
\usepackage{flafter}
\usepackage{chngcntr}
\graphicspath{{figures/}}
\usepackage[titletoc,title,toc]{appendix}

\DeclareMathOperator*{\argmin}{\arg\!\min}

\usepackage{array}
\newcommand{\tabNote}[2]{XX.}

\newtheorem{algo}{Algorithm}
\usepackage{enumitem}

\usepackage{multirow}

\usepackage[affil-it]{authblk}
\usepackage[pdftex,pdfauthor={ Florens Odendahl},pdftitle={Density Forecast Transformations}]{hyperref}
\usepackage[noabbrev,nameinlink,capitalise]{cleveref}
\hypersetup{
	colorlinks   = true, 
	urlcolor     = blue, 
	linkcolor    = blue, 
	citecolor   = blue 
}
\crefname{appsec}{Appendix}{Appendices}
\providecommand{\keywords}[1]{\textbf{\textit{Keywords:}} #1}
\providecommand{\jelcodes}[1]{\textbf{\textit{JEL codes:}} #1}

\author[1]{Matteo Mogliani}
\author[2]{Florens Odendahl}
\affil[1]{Banque de France \thanks{Address: 31 Rue Croix des Petits Champs, 75001 Paris, France. Email: \href{mailto:matteo.mogliani@banque-france.fr}{matteo.mogliani@banque-france.fr}.}}
\affil[2]{Banco de Espa\~{n}a\thanks{Address: Calle de Alcal\'{a} 48, 28014 Madrid, Spain. Email: \href{mailto:florens.odendahl@bde.es}{florens.odendahl@bde.es}. We thank Mohammed Chahad and participants at the 12th ECB Conference on Forecasting Techniques, the 2023 Annual Conference on Real-Time Data Analysis, Methods, and Applications, the 17th International Conference on Computational and Financial Econometrics, the 14th Research Workshop Banco de Espa\~{n}a-CEMFI, the 2024 RCEA ICEEF, the IAAE, the 44th ISF, the EEA 2024, and Banco de Espa\~{n}a and Banca d'Italia seminars for insightful discussions. The paper previously circulated under the title ``Density forecast frequency transformation via copulas''. The views expressed herein are those of the authors and should not be attributed to the Banco de Espa\~{n}a, the Banque de France or the Eurosystem.}}

\title{Density forecast transformations}
\date{\today}  

\begin{document}
\newgeometry{top=0.5in,bottom=0.9in,right=0.9in,left=0.9in}	
\maketitle
\begin{abstract}
		\noindent The popular choice of using a \textit{direct} forecasting scheme implies that the individual predictions do not contain information on cross-horizon dependence. However, this dependence is needed if the forecaster has to construct, based on \textit{direct} density forecasts, predictive objects that are functions of several horizons (\textit{e.g.} when constructing annual-average growth rates from quarter-on-quarter growth rates). To address this issue we propose to use copulas to combine the individual $h$-step-ahead predictive distributions into a joint predictive distribution. Our method is particularly appealing to practitioners for whom changing the \textit{direct} forecasting specification is too costly. In a Monte Carlo study, we demonstrate that our approach leads to a better approximation of the true density than an approach that ignores the potential dependence. We show the superior performance of our method in several empirical examples, where we construct (i) quarterly forecasts using month-on-month \textit{direct} forecasts, (ii) annual-average forecasts using monthly year-on-year \textit{direct} forecasts, and (iii) annual-average forecasts using quarter-on-quarter \textit{direct} forecasts.\\
		\newline \newline
		\newline
		\jelcodes{C53, C32, E37.}
\end{abstract}

\newpage
\restoregeometry

\section{Introduction}\label{sec:Intro}	
Forecasting models are often specified to produce \textit{direct} $h$-step-ahead forecasts, which implies that the predictions do not contain information on their cross-horizon dependence. As a consequence, individual h-step-ahead predictive distributions cannot easily be transformed into predictive objects that depend on several horizons. For example, the literature on macroeconomic risk often uses quantile regression models that, in brief, produce \textit{direct} density forecasts of quarter-on-quarter (qoq) real GDP growth \citep{Adrian2019,Ferrara2022}. However, it is not straightforward in this framework to construct, for instance, density forecasts for annual-average growth rates from the estimated quarter-on-quarter density forecasts because of their potential serial dependence.  

To address this issue, we propose using Gaussian copulas to combine the information of the marginal \textit{direct} $h$-step-ahead predictive densities into a joint distribution, which reflects the serial dependence between the marginals. This enables the practitioner to draw from the multi-horizon distribution and, therefore, to construct predictive distributions that are functions of several horizons, which we label target-frequency predictive densities.\footnote{Note that for models that produce \textit{iterative} $h$-step-ahead predictive densities, for instance Vector Autoregressions, forecasts conditional on specific paths or the annual-average frequency transformation are in general readily available since the iterative approach allows to draw conditionally on all the other forecast horizons.} The resulting multi-horizon predictive objects provide forecasts that are coherent with the underlying marginal predictive densities, in the sense that the moments of the multi-horizon objects are a function of the marginals and the copula parameters. 

While separate forecasting specifications could be used to make predictions for the different predictive objects, there are several reasons why a single specification can be preferable. First, the scenario of having a single forecasting specification is particularly common among professional forecasters and institutions, such as central banks, where changing the forecasting process is costly and yet transformations of the existing forecasts to other target-frequencies are often required. Second, using a single forecasting specification leads to predictive objects with moments that are coherent across the target-frequencies. For instance, using our approach, the mean of the annual-average forecast is consistent with the mean of the quarter-on-quarter predictions. 
Third, some density forecasts are derived from surveys that only report one frequency but not the required target-frequency. In general, our approach helps to broaden the usability of already available individual predictive densities that are based on a \textit{direct} forecasting scheme.

As an alternative to our approach, the researcher could use a simple approach that assumes independence between the different marginal predictive densities, i.e., no correlation between the \text{direct} $h$-step-ahead predictions at different horizons. However, this approach neglects the serial dependence typically present in macroeconomic indicators, which is important for the tails of the multi-horizon distribution. For example, accounting for the positive serial correlation in quarter-on-quarter GDP growth forecasts leads to considerably fatter tails of the annual-average predictive distributions. This is because an approach that constructs the annual-average growth while taking into account the path of quarterly GDP growth over several quarters and a positive correlation would reflect the feature that large positive (negative) growth is typically followed by positive (negative) growth. This makes our approach particularly appealing for macroeconomic and financial risk applications. 

In this paper, we show analytically that the proposed copula approach for density forecast transformations may outperform substantially a benchmark approach that neglects the cross-horizon dependence of the marginal densities. We also show how the resulting gain in the forecasting performance depends on the persistence of the underlying process. Further, we show in several Monte Carlo studies that our approach provides a better approximation to the true underlying annual-average density forecasts under different DGPs. This result holds under misspecified forecasting models, i.e., when the true multivariate distribution is not Gaussian, and for small training samples for the copula parameter estimation.

For the application of our approach, the researcher only needs to compute the correlation between the empirical PITs of the individual $h$-step-ahead predictive distributions for different horizons in a training sample. In particular, the forecaster needs to (i) compute the sequence of realized PITs for the marginal predictive densities at each forecast horizon $h=1,\dots,H$, from a pseudo out-of-sample exercise over a training sample, and to (ii) combine the marginal distributions into a joint distribution via a multivariate Gaussian copula, where the maximum-likelihood estimator of the copula parameters is equal to the rank correlation of the realized PITs. 

We demonstrate the usefulness of our methodology in three empirical applications, where we compare the forecasting performance of the proposed copula approach with a benchmark approach that ignores the serieal dependence between the multi-horizon marginal distributions. The first application is a large-scale forecasting exercise based on monthly data from FRED-MD \citep{McCracken2016} entering bivariate ARDL models regressing a large number of pairs of randomly selected variables. We first compute density forecasts for month-on-month values from these regressions and we then use these predictive densities to compute quarter-on-quarter density forecasts. Results show that the copula approach outperforms the benchmark for the majority of randomly specified bivariate models.
The second empirical application aims at emulating a situation in which a forecaster dispose of predictive densities for year-on-year U.S. monthly inflation, but she needs to summarize the picture of the expected inflation environment by converting the target frequency of the predictive densities from monthly year-on-year rate to annual-average rate. Importantly, the year-on-year and annual-average predictions need to be coherent, i.e., they should be based on the same predictors and model type, and the central tendency of the forecasts across the two frequencies should be very similar. The results show that our copula approach provides significantly better density forecasts of annual-average inflation, in particular at the tails of the distributions, than the benchmark.
Finally, in the third empirical application, we use the predictive densities of quarter-on-quarter U.S. real GDP growth from \cite{Adrian2019}, which are based on \textit{direct} forecasts, and transform them into annual-average forecasts. We find again that the annual-average forecasts based on our copula approach leads to more accurate predictive densities than the benchmark approach.

Our paper contributes to the literature on density forecasts and economic risk prediction. While \cite{Patton2006} introduced conditional copulas to economics with a focus on modeling the cross-sectional dependence of predictive objects (see \citealp{Patton2014}, for review), we show how copulas can be used for density forecast transformations. In  particular, compared to existing contributions, the proposed copula-based approach is designed to combine the marginal forecast densities, unimodal \citep{Adrian2019,Ferrara2022} or multimodal \citep{Mitchell2024}, into new predictive objects of several horizons. Differently from \cite{Smith2016}, who use copulas to model both the time series and cross-sectional correlation of U.S. macroeconomic variables, we propose to use copulas for modeling the time series correlation and transforming marginal density forecasts into multi-horizon objects. Finally, compared with recent works also proposing to model the joint distribution of predictive objects (e.g., \citealp{Clark2020}, for the joint distribution of point forecast errors obtained from surveys via multi-variate stochastic volatility models; \citealp{Grothe2023}, for joint forecasts of hourly electricity prices from point forecasts; \citealp{Ganics2024}, for the construction of fixed-horizon density forecasts out of fixed-event survey density forecasts), we propose to use Gaussian copulas to combine marginal predictive densities of macroeconomic indicators to obtain predictive objects that are transformed to a new target-frequency. 

The paper is structured as follows. \cref{sec:MotivatingExample} provides an analytical example of our forecasting environment. \cref{sec:Method} describes the methodological framework. \cref{sec:MonteCarlo} presents Monte Carlo results. In \cref{sec:Discussion} we provide a robustness analysis and discuss alternative simpler approaches. \cref{sec:Empirical} presents the results from the three empirical exercises. Finally, \cref{sec:Conclusion} concludes.

\FloatBarrier

\section{Motivating example}\label{sec:MotivatingExample}
Consider the following simple mean-zero autoregressive model:
$$Y_{t+1} = \rho Y_{t} + \varepsilon_{t+1}$$
with $\vert\rho\vert<1$ and $\varepsilon_{t}\sim\mathcal{N}(0,\sigma_{\varepsilon}^2)$. The optimal $h$-steap ahead prediction (under both iterated and direct forecasting approach) is given by
$$Y_{t+h\vert t} = \rho^{h} Y_{t}$$
It follows that the forecast error $e_{t+h\vert t}=\sum_{j=0}^{h-1}\rho^{j}\varepsilon_{t+h-j}$ has second moment
\begin{equation*}
\mathbb{V}(e_{t+h\vert t}) = \sigma_{\varepsilon}^2\left(\frac{1-\rho^{2h}}{1-\rho^{2}}\right)
\end{equation*}
and auto-covariance and auto-correlation functions:
\begin{align*}
Cov(e_{t+h\vert t},e_{t+h-k\vert t}) &=\sigma_{\varepsilon}^2\rho^{k}\left(\frac{1-\rho^{2(h-k)}}{1-\rho^{2}}\right) \\
Corr(e_{t+h\vert t},e_{t+h-k\vert t}) &=\rho^{k}\sqrt{\frac{1-\rho^{2(h-k)}}{1-\rho^{2h}}}
\end{align*}
for $h>k>0$. $Corr$ denotes the Pearson correlation. Thus, the process has the following conditional predictive distribution:
$$ Y_{t+h\vert t}| \rho,\sigma_{\varepsilon} \sim \mathcal{N}\left(\rho^{h} Y_{t}, \sigma_{\varepsilon}^2\frac{1-\rho^{2h}}{1-\rho^{2}} \right)$$
with the pdf of $Y_{t+h\vert t}$ denoted by $\phi_{Y_{t+h\vert t}}(y_{t+h}\vert \rho,\sigma_{\varepsilon})$. Now consider a linear transformation of the forecast sequence $\{Y_{t+j\vert t}\}_{j=1}^{h}$, such as 
\begin{equation}\label{eq:Z}
Z_{t+h\vert t}=w_{1}Y_{t+1\vert t} + w_{2}Y_{t+2\vert t} + \dots + w_{h}Y_{t+h\vert t},
\end{equation}
with $\textbf{w}=(w_{1},\dots,w_{h})$ a vector of weights. This transformation is often useful in macroeconomic applications when the original forecasts need to be converted into a different target periodic measure of the same variable. For instance, if $Y_{t}$ is a month-on-month growth rate sampled at monthly frequency and $Y_{t+h\vert t}$ is its $h$-step ahead forecast, then for $h=12$ and $\{w_{j}\}_{j=1}^{h}=1$, the transformed forecast 
\begin{equation}\label{eq:approx1}
Z_{t+12\vert t}=Y_{t+1\vert t} + Y_{t+2\vert t} + \dots + Y_{t+12\vert t}
\end{equation}
is (approximately) the 12-months ahead forecast of the year-on-year growth rate. 

Following the example in \eqref{eq:Z}, the conditional predictive distribution of the ``dependence-attentive'' transformed forecast is:
\begin{equation}\label{eq:ideal}
Z_{t+h\vert t}\vert \rho,\sigma_{\varepsilon},\textbf{w} \sim \mathcal{N}\left(\sum_{j=1}^{h}w_{j}\rho^{j}Y_{t}, \sigma_{\varepsilon}^2\sum_{j=1}^{h}w_{j}^{2}\left(\frac{1-\rho^{2j}}{1-\rho^{2}}\right) + 2\sigma_{\varepsilon}^2\sum_{j=2}^{h}\sum_{k=1}^{j-1}w_{j}w_{j-k}\rho^{k}\left(\frac{1-\rho^{2(j-k)}}{1-\rho^{2}}\right) \right),
\end{equation}
which is the sum of random variables from the joint multivariate Normal forecast distribution
\begin{equation}\label{eq:ideal_joint}
(Y_{t+1\vert t},\dots, Y_{t+h\vert t})'|\rho,\sigma_{\varepsilon},\textbf{w}\sim\mathcal{N}\left(\boldsymbol\mu, \boldsymbol\Sigma \right)\end{equation}
where $\boldsymbol\mu = \left(w_{1}\rho Y_{t},w_{2}\rho^2 Y_{t},w_{3}\rho^3 Y_{t},\cdots,w_{h}\rho^h Y_{t}\right)^{\prime}$ and
\begin{equation*}
\boldsymbol\Sigma = \sigma_{\varepsilon}^2
\left(\begin{array}{ccccc}
w_{1}^{2} & w_{1}w_{2}\rho & w_{1}w_{3}\rho^2 & \cdots & w_{1}w_{h}\rho^{h-1} \\
w_{2}w_{1}\rho & w_{2}^{2}\left(\frac{1-\rho^{4}}{1-\rho^{2}}\right) & w_{2}w_{3}\rho\left(\frac{1-\rho^{4}}{1-\rho^{2}}\right) & \cdots & w_{2}w_{h}\rho^{h-2}\left(\frac{1-\rho^{4}}{1-\rho^{2}}\right) \\
w_{3}w_{1}\rho^2 & w_{3}w_{2}\rho\left(\frac{1-\rho^{4}}{1-\rho^{2}}\right) & w_{3}^{2}\left(\frac{1-\rho^{6}}{1-\rho^{2}}\right) & \cdots & w_{3}w_{h}\rho^{h-3}\left(\frac{1-\rho^{6}}{1-\rho^{2}}\right) \\
\vdots & \vdots & \vdots & \ddots & \vdots\\
w_{h}w_{1}\rho^{h-1} & w_{h}w_{2}\rho^{h-2}\left(\frac{1-\rho^{4}}{1-\rho^{2}}\right) & w_{h}w_{3}\rho^{h-3}\left(\frac{1-\rho^{6}}{1-\rho^{2}}\right) & \cdots & w_{h}^{2}\left(\frac{1-\rho^{2h}}{1-\rho^{2}}\right)\\
\end{array}\right)
\end{equation*}

\begin{figure}[!t]
\graphicspath{{./}{Figures/}} 
	\centering
	\caption{Scores for density forecasts: "dependence-attentive" \textit{vs} "dependence-inattentive"}\label{fig:Scores_AR_pos}
		\includegraphics[scale=0.45]{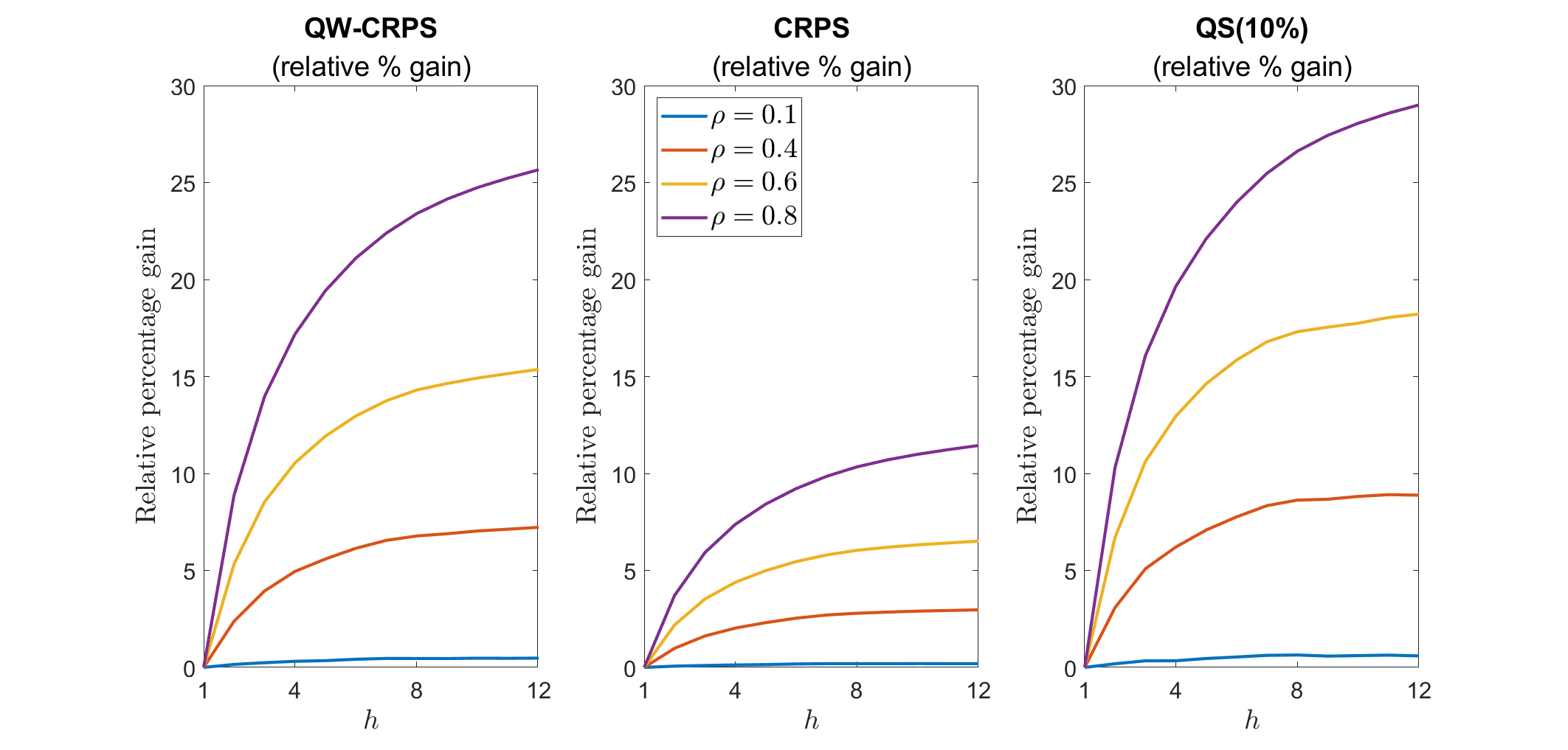}
    \vspace{0.4cm}
	\caption*{\footnotesize \textit{Note}: 
    QW-CRPS denotes the quantile weighted versions of the continuous ranked probability score, with emphasis on the tails. CRPS denotes the continuous ranked probability score. QS($10\%$) denotes the quantile score at the $10\%$ quantile. CRPS, QW-CRPS, and QS($10\%$) are expressed in relative $\%$ gain of the "dependence-attentive" forecaster with respect to the "dependence-inattentive" forecaster.}
\end{figure}

The pdf of $Z_{t+h\vert t}$ is denoted by $\phi_{Z_{t+h\vert t}}(z_{t+h} \vert \rho,\sigma_{\varepsilon},\textbf{w})$. The predictive distribution in \eqref{eq:ideal} can be compared to the predictive distribution of the forecaster who ignores the cross-horizon dependence of the forecasts. We label this forecaster the ``dependence-inattentive'' forecaster because no attention is paid to the potential correlation structure of the forecast errors across horizons. This forecaster draws from the following conditional forecast distribution:
\begin{equation}\label{eq:inattentive}
\tilde{Z}_{t+h\vert t}\vert \rho,\sigma_{\varepsilon},\textbf{w}  \sim \mathcal{N}\left(\sum_{j=1}^{h}w_{j}\rho^{j}Y_{t}, \sigma_{\varepsilon}^2\sum_{j=1}^{h}w_{j}^{2}\left(\frac{1-\rho^{2j}}{1-\rho^{2}}\right)\right),
\end{equation}
where the pdf of $\tilde{Z}_{t+h\vert t}$ is denoted by $\tilde{\phi}_{\tilde{Z}_{t+h\vert t}}( \tilde{z}_{t+h} \vert \rho,\sigma_{\varepsilon},\textbf{w})$. Note that for $\{w_{j}\}_{j=1}^{h}=1$, the conditional predictive distribution of the ``dependence-attentive'' forecaster in \eqref{eq:ideal} simplifies to:
\begin{equation}\label{eq:ideal_simple}
Z_{t+h\vert t} \vert \rho,\sigma_{\varepsilon},\textbf{w} \sim\mathcal{N}\left(\sum_{j=1}^{h}\rho^{j}Y_{t}, \sigma_{\varepsilon}^2\sum_{j=1}^{h}\left(\frac{1-\rho^{j}}{1-\rho}\right)^{2} \right).
\end{equation}

The panels of \cref{fig:Scores_AR_pos} show the average relative accuracy of the "dependence-attentive" density forecast in \eqref{eq:ideal_simple} compared to the "dependence-inattentive" density forecast in \eqref{eq:inattentive} with $\{w_{j}\}_{j=1}^{h}=1$, evaluated through proper scoring rules, such as the quantile-weigthed continuous ranked probability score (QW-CRPS; \citealp{Gneiting2011}) which emphasizes the tails, the continuous ranked probability score (CRPS; \citealp{Gneiting2007}), and the 10\% quantile score (QS(10\%); \citealp{Giacomini2005}). All these metrics show robust gains for the "dependence-attentive" forecast, which increases monotonically with $\rho$ and $h$. For instance, with $\rho=0.6$ and $h=12$ (the year-on-year growth rate transformation for month-on-month predictions), the gain would stand about 7\% according to the CRPS, while the QW-CRPS and the QS(10\%) point to a gain of about 15-18\%. Not surprisingly, the latter suggest that substantial accuracy gains can be obtained when evaluating the tails of the predictive densities, rather than their central region. This is due to the fact that the two densities differ solely in their variance, while the other moments are identical.\footnote{It is nevertheless worth noting that these results depend on the sign of $\rho$. With $\rho<0$, the monotonicity feature is in part lost, in particular for large (negative) autoregressive coefficients and small $h$. However, the "dependence-attentive" forecaster outperforms the "dependence-inattentive" forecaster even under this parameterization, in particular for large forecast horizons. These results are reported in \cref{fig:Scores_AR_neg} in the Appendix.}    

In this stylized example, the researcher knows the dependence structure across forecasting horizons and, therefore, the joint distribution of the $Y_{t+h|t}$. However, in practice when using direct forecasting schemes, this dependence structure is unknown and not estimated alongside the predictions. The next section, therefore, introduces a methodology to estimate the dependence structure for a given set of direct density forecasts.

\FloatBarrier
\section{Constructing multivariate densities with copulas}\label{sec:Method}
This section first introduces the necessary notation and then explains the methodology to transform the marginal density forecasts into the desired target multi-horizon object.
\subsection{Notation} 
$\{Y_{t+j\vert t}\}_{j=1}^h$ denotes the higher frequency prediction that the researcher would like to transform into $Z_{t+h\vert t}$. The variable $Z_{t+h\vert t}$ is a function of $Y_{t+1\vert t},..,Y_{t+h\vert t}$ and potentially of $Y_t$ and its lags. The time subscript $t$ and the forecast horizon subscript $h$ denote the time units of $Y_{t+h\vert t}$. For instance, if $Y_{t+h\vert t}$ denotes monthly predictions then $Z_{t+h\vert t}$ denotes a transformation of the monthly forecasts using at most predictions up to month $t+h$. Recalling the example of the previous section, if $Y_{t}$ is a month-on-month growth rate sampled at monthly frequency, then for $h=12$ and $\{w_{j}\}_{j=1}^{h}=1$, the transformed forecast 
\begin{equation}\label{eq:approx2a}
Z_{t+12\vert t}=Y_{t+1\vert t} + Y_{t+2\vert t} + \dots + Y_{t+12\vert t}
\end{equation}
is approximately the 12-month-ahead forecast of the monthly year-on-year growth rate. In this case, we shall denote $Z_{t+h\vert t}$ a \textit{periodic-transformation} of the forecast sequence $\{Y_{t+j\vert t}\}_{j=1}^{h}$. 

Similarly, a linear transformation can map the forecasts generated at the original sampling frequency of the data into a forecast sequence sampled at a desired lower target frequency. For instance, let $Y_{t}$ be a month-on-month growth rate sampled at monthly frequency. Then, for $h=6$, the transformed forecast
\begin{equation}\label{eq:approx2b}
Z_{t+6\vert t}=\frac{0}{3}Y_{t+1\vert t}+\frac{1}{3}Y_{t+2\vert t}+\frac{2}{3}Y_{t+3\vert t} + Y_{t+4\vert t} + \frac{2}{3}Y_{t+5\vert t} +\frac{1}{3}Y_{t+6\vert t}
\end{equation}
is (approximately) the two-quarter-ahead forecast of the quarterly quarter-on-quarter growth rate (see \citealp{Mariano2003}).\footnote{This formula can be generalised to other frequency transformations, such as from month-on-month or quarter-on-quarter growth rates to annual-average growth rates, by changing the sequence of weights $\textbf{w}$.}
In this case, we shall denote $Z_{t+h\vert t}$ a \textit{frequency-transformation} of the forecast sequence $\{Y_{t+j\vert t}\}_{j=1}^{h}$.

Throughout the paper, we use a boldface notation for vectors, matrices, and functions that take vectors as input.

\subsection{Methodology}
From the previous section, it is clear that constructing the predictive density of transformed forecasts requires drawing from the joint predictive distribution across the forecasting horizon. However, this can be often impractical in empirical applications, such as those relying on direct multi-step forecasting.

To address this issue, in this paper we propose to resort to (Gaussian) copulas \citep{Sklar1959}, which allow to model the marginals and the dependence separately while ensuring that the researcher obtains a valid multivariate distribution; see \cite{Nelsen2006} for an introduction to copulas. Since we consider a forecasting environment, we work with the conditional copulas defined in \cite{Patton2006} but drop the conditioning sets from the notation for simplicity.

Continuing with the example in \cref{sec:MotivatingExample}, the predictive density $\phi_{Z_{t+h\vert t}}( \cdot \vert \rho,\sigma_{\varepsilon},\textbf{w})$, i.e. the density of $Z_{t+h\vert t}$, can be constructed by drawing from the joint distribution of $(Y_{t+1\vert t},\dots, Y_{t+h\vert t})' \vert  \rho,\sigma_{\varepsilon}$ described in eq. \eqref{eq:ideal_joint}. Denoting by $\Phi_{Y_{t+h\vert t}}(y_{t+h})$ the CDF of the Gaussian predictive distribution of $Y_{t+h\vert t}$ evaluated at $y_{t+h}$, we note that:
\begin{equation}\label{eq:corr_PITs}
Corr\left(\Phi_{Y_{t+h\vert t}}(y_{t+h}),\Phi_{Y_{t+h-k\vert t}}(y_{t+h-k})\right) = \frac{6}{\pi}\arcsin\left(\frac{1}{2}\rho^{k}\sqrt{\frac{1-\rho^{2(h-k)}}{1-\rho^{2h}}} \right)
\end{equation}
which is the Pearson correlation coefficients of the probability integral transforms (PITs). The Gaussian copula is given by:
$$\boldsymbol{C}_{\boldsymbol{R}}=\boldsymbol{\Phi}_{\boldsymbol{R}}\left( \Phi^{-1}(U_{1}),\dots,\Phi^{-1}(U_{h})\right)$$
where $\Phi^{-1}$ denotes the inverse CDF of a standard Normal, $\boldsymbol{\Phi_{R}}$ the joint CDF of a standard multivariate Normal with covariance matrix $\boldsymbol{R}$, and $U_{i}$, $i=1,...,h$, are the PITs of the univariate predictive densities over the $h$ forecasting horizons. Note that $\boldsymbol{R}$ is hence the correlation matrix, whose elements are defined in \eqref{eq:corr_PITs}. Further note that the predictive distribution of the ``dependence-inattentive'' forecaster in \eqref{eq:inattentive} is also equivalent to the joint distribution of the forecasts constructed through a Gaussian copula, but with $\boldsymbol{R}=\textbf{I}_{h}$. Since $\boldsymbol{C_{R}}$ is the multivariate distribution of the random variables $\Phi^{-1}(U_{i})$, $i=1,...,h$, given the copula, it is easy to resample the $Y_{t+1\vert t},\dots,Y_{t+h\vert t}$ from their joint distribution:
$$\left(Y_{t+1\vert t},\dots,Y_{t+h\vert t}\right)=\left(\Phi^{-1}(U_{1}),\dots,\Phi^{-1}(U_{h}) \right)$$
and then compute the desired (periodic or frequency) transformed density forecast from the sampled joint forecasts.

In the following we extend this approach to a more general forecasting environment. Assume the forecaster has a set of \textit{direct} $h$-step-ahead predictive densities for $T$ forecast origins, denoted by $\{\{g_{t,h}\}_{h=1}^H\}_{t=1}^T$ and with predictive cumulative distribution functions (CDF) $\{\{G_{t,h}\}_{h=1}^H\}_{t=1}^T$, for outcome variable $Y_{t+h}$; the subscript $h$ denotes the forecast horizon and the subscript $t$ denotes the forecast origin. Further assume that the set of predictive distributions, $\{\{g_{t,h}\}_{h=1}^H\}_{t=1}^T$, is taken as given, for instance, due to institutional restrictions on the forecasting model to be used.

To illustrate the application of our methodology, but without loss of generality, we will assume that the predictive density $g_{t,h}$ is a predictive density for quarter-on-quarter growth rates.\footnote{Predictive distributions for month-on-month growth rates, monthly or quarterly (log-)levels or year-on-year growth rates can be handled analogously with our approach.} In period $T$, the forecaster is asked to provide predictive densities for the annual-average growth rates as well as for the conditional predictive density $\tilde{g}_{T,h}(y_{T+h}|y_{T+h-1},...,y_{T-1})$, henceforth called path-forecast, based on $\{\{g_{t,h}\}_{h=1}^H\}_{t=1}^T$.

We propose to do this by using copula functions, developed by \cite{Sklar1959}. A copula can be described as a function such that for any $\boldsymbol{Q}(y_1,...,y_d)$, where $\boldsymbol{Q}$ is the multivariate distribution function of the random vector $(Y_1,...,Y_d)$, there is a copula function $\boldsymbol{C}(\cdot | \boldsymbol{R})$, such that $\boldsymbol{Q}(y_1,...,y_d)$ = $\boldsymbol{C}( G_{Y_1}(y_1),..., G_{Y_d}(y_d) | \boldsymbol{R} )$, where $G_{Y_1},...,G_{Y_d}$ are the marginal CDFs of $Y_1,...,Y_d$, respectively, and $\boldsymbol{R}$ denotes the parameter(s) that governs the dependence between $G_{Y_1}(y_1),...,G_{Y_d}(y_d)$. Inversely, a copula function $\boldsymbol{C}$, combined with marginal CDFs $G_{Y_1},...,G_{Y_d}$, gives a multivariate distribution. 

A popular copula family is the Gaussian copula, denoted by $\boldsymbol{C_{\text{Ga}}}$, where the dependence between the $d$ variables is governed by the correlation matrix $\boldsymbol{R}$, with ones on the diagonal and the rank correlation of variable $i$ and $j$ as the respective off-diagonal element $(i,j)$. 

Let then $\boldsymbol{Q}_T(y_{T+1},...,y_{T+h} | \boldsymbol{R})$ denote the joint predictive CDF of $Y_{T+1},...,Y_{T+h}$ for forecast origin $T$, conditional on the correlation matrix $\boldsymbol{R}$ and constructed using $\boldsymbol{C_{\text{Ga}}}$. In other words, we define $\boldsymbol{Q}_T(y_{T+1},...,y_{T+H} | \boldsymbol{R}) = \boldsymbol{C_{\text{Ga}}} (G_{T,1}(y_{T+1}),...,G_{T,H}(y_{T+H}) | \boldsymbol{R}) $. Further, let  $\text{PIT}_{t,h} = G_{t,h}(y_{t+h})$ the probability integral transform of the predictive densities, where $y_{t+h}$ is the realized value. The forecaster can obtain an estimate of $\boldsymbol{Q}_T(y_{T+1},...,y_{T+H} | \boldsymbol{R}) $ by implementing the algorithm described below. \\

\begin{algo}\label{algo1}{\textbf{Joint Predictive Distribution}}
\begin{enumerate}
\item Compute the realized PITs, $\{\{\text{PIT}_{t,h}\}_{{h=1}}^{H}\}_{t=1}^{T-H}$, of the predictive CDFs $\{\{G_{t,h}\}_{{h=1}}^{H}\}_{t=1}^{T-H}$. 
\item Compute the rank correlations of $\text{PIT}_{t,h}$ across the different $h$ to get an estimate of $\widehat{\boldsymbol{R}}$.
\item Use $\widehat{\boldsymbol{R}}$ in combination with $\boldsymbol{C_{\text{Ga}}}$ to obtain the joint distribution $\boldsymbol{\widehat{Q}_T}(y_{T+1},..,y_{T+H} | \widehat{\boldsymbol{R}})$. 
\end{enumerate}
\end{algo}

The resulting multivariate distribution allows to sample the \textit{direct} $h$-step-ahead predictions jointly, such that predictive objects that are functions of several horizons can be constructed. Note that the maximum likelihood estimator of $\boldsymbol{R}$ is $\widehat{\boldsymbol{R}}$, i.e. the maximum likelihood estimator of the correlation matrix under a Gaussian copula reduces to the rank correlation of the PITs. 

To illustrate the use of Algorithm 1, consider the following example. The forecaster is asked in $T$, which is the last quarter of the year, to provide a predictive distribution of the annual-average growth for the next year. The forecaster, however, has only a set of \textit{direct} quarter-on-quarter $h$-step-ahead growth rate predictions available, for $h=1,...,4$. To transform the quarter-on-quarter growth rates into annual-average predictions, the forecaster can obtain the set $\{[Y_{T,1}^{(s)}, Y_{T,2}^{(s)}, Y_{T,3}^{(s)}, Y_{T,4}^{(s)}]'\}_{s=1}^S$ of draws, for $s=1,...,S$, from $\boldsymbol{\widehat{Q}^{-1}_T}(y_{T+1},..,y_{T+4} | \widehat{\boldsymbol{R}})$. This can be easily implemented in standard statistics packages. First, do step 1 to 2 of Algorithm 1. Then, step 3 amounts to the following: compute the lower Cholesky decomposition of $\widehat{\boldsymbol{R}}$, denoted by $\boldsymbol{P}$. Next, for each $s=1,...,S$:
\begin{enumerate}[label=(\alph*)]
\item Draw a $4 \times 1 $ vector of independent standard Normals, i.e. draw $\boldsymbol{X} \sim_{iid} \mathcal{N}(\boldsymbol{0},\mathbf{I}_4)$ where $\mathbf{I}_4$ is the $4 \times 4 $ identity matrix.
\item Compute the vector $\boldsymbol{U} = [U_1,U_2,U_3,U_4]= [\Phi(Z_1),\Phi(Z_2),\Phi(Z_3),\Phi(Z_4)]$, where $\Phi(\cdot)$ is the CDF of a standard Normal distribution and $[Z_1,Z_2,Z_3,Z_4]' = \boldsymbol{Z} = \boldsymbol{P}\boldsymbol{X}$.
\item Evaluate $G_{T,1}(U_1),...,G_{T,4}(U_4)$ to get the vector of joint draws $[Y_{T,1}^{(s)}, Y_{T,2}^{(s)}, Y_{T,3}^{(s)}, Y_{T,4}^{(s)}]'$. 
\end{enumerate}

The set of vectors of joint draws $\{[Y_{T,1}^{(s)}, Y_{T,2}^{(s)}, Y_{T,3}^{(s)}, Y_{T,4}^{(s)}]'\}_{s=1}^S$ approximates $\boldsymbol{\widehat{Q}_T}(y_{T+1},..,y_{T+4} | \widehat{\boldsymbol{R}})$ and can be used, alongside with the observations $y_{T}, y_{T-1}, y_{T-2}$, to obtain the predictive distribution of the annual-average growth rate using, for instance, an exact formula or a linear approximation formula as in eq. \eqref{eq:approx2b}, for each $s=1,...,S$. The multivariate distribution also allows to obtain draws of $Y_{t,h}$ conditional on $ Y_{t,h-j}$ for $j = 1,...,h-1$. 

Our approach aims at constructing well performing forecasts but we note that there are several potential sources of misspecification. First, the choice of the Gaussian copula might not reflect the true underlying multivariate distribution of the data. Second, if the marginal distributions are misspecified, the joint distribution will also be misspecified. Third, a potential source of misspecification comes from the conditional copulas. As shown by \cite{Patton2006}, if the conditioning set is not identical for the marginals and the copula, then the constructed conditional joint distribution might not reflect the actual conditional joint distribution. Following \citeauthor{Patton2006}'s (\citeyear{Patton2006}) example, if the researcher conditions $Y_1$ on $W_1 = w_1$, $Y_2$ on $W_2= w_2$ and the copula on $W_1,W_2$, such that $\boldsymbol{\widetilde{Q}_{Y_1,Y_2|W_1,W_2}} (y_1,y_2|w_1,w_2) = \boldsymbol{C}(G_{Y_1|W_1}(y_1|w_1),G_{Y_2|W_2}(y_2|w_2)|w_1,w_2,\boldsymbol{R})$, then $\boldsymbol{\widetilde{Q}}$ will only denote the conditional joint distribution of $(Y_1,Y_2) | W_1,W_2$ if the marginal $Y_1$ is independent of $W_2$ and the marginal of $Y_2$ is independent of $W_1$ (for details see \citealp{Patton2006}). However, given the macroeconomic applications that we consider, this is unlikely to restrict the implementation of our approach for mainly two reasons. First, the marginals come from \textit{direct} forecasting models that use the same predictors for each forecasting horizon. Second, if the predictors differ across forecasting horizons that is typically because different predictors are relevant at that horizon, i.e., the marginals are independent of the non-included predictors. 

Further, our estimation algorithm assumes that (i) dependence parameters do not depend on the predictors used for the marginals and (ii) restricts the copula dependence parameters in $R$ to be constant. Assumption (i) is common in the literature and across model classes. For instance, flexible models such as VARs with stochastic volatility do not make the time-varying variance-covariance matrix an explicit function of dependent variables. However, while we propose a simple algorithm for practitioners, both assumption (i) and (ii) can be relaxed. For instance, \cite{Patton2006} suggests an ARMA-like specification to allow for time-variation in the copula parameters. \cite{Hafner2012} propose a stochastic process for the copula parameters. Both specification could be extended to include exogeneous variables in the copula parameter equations.

Despite the various potential sources of misspecification, the Monte Carlo results in the next section show that our modelling approach outperforms the competitor approach and leads to forecasting performances that are often indistinguishable from the correctly specified predictive distribution.

\FloatBarrier

\section{Monte Carlo study}\label{sec:MonteCarlo}	
We study the performance of our suggested copula approach via Monte Carlo simulations in a scenario where the forecaster has a model that produces \textit{direct} $h$-step-ahead predictive densities for qoq growth rates and then needs to transform the predictive densities into annual-average growth rates and yoy growth rates. The absolute and relative performance of the proposed approach with respect to a simple benchmark ignoring cross-horizon dependence of the forecasts (the  ``dependence-inattentive'' forecaster described in \cref{sec:MotivatingExample}) are evaluated through an out-of-sample exercise. 

\subsection{Monte Carlo design}
The underlying DGP of qoq growth rates, denoted by $Y_t$, takes the form of a VAR(1):
\begin{equation}\label{eq:DGP}
\left[\begin{array}{c}Y_{t}\\X_{t}\end{array}\right]=\left[\begin{array}{c}\tau_{1}\\ \tau_{2}\end{array}\right] + \left[\begin{array}{cc}\theta_{1} & \theta_{2} \\ 0 & \gamma\end{array}\right] \left[\begin{array}{c}Y_{t-1}\\X_{t-1}\end{array}\right]+ \left[\begin{array}{c}\varepsilon_{1,t}\\ \varepsilon_{2,t}\end{array}\right]
\end{equation}
where $\{\varepsilon_{j,t}\}_{t=1}^{T}$ are two uncorrelated sequences of independent and identically distributed (iid) structural shocks. We set $\varepsilon_{2,t}\overset{\text{\tiny iid}}{\sim}\mathcal{N}\left(0, \sigma^{2}_{\epsilon_2} \right)$, with $\sigma_{\epsilon_2}=0.3$, but we consider three different specifications for the error term $\varepsilon_{1,t}$: ($i$) a Normal distribution, ($ii$) a Skew-Normal distribution, or ($iii$) a Skew-$t$ distribution. 
For cases ($ii$) and ($iii$), we adopt a location-scale-shape parameterization \citep{Azzalini2003}. All the distributions are calibrated to have mean zero and standard deviation $\sigma_{\epsilon_1}=0.5$, as well as negative skewness for cases ($ii$) and ($iii$) (with shape parameter $\alpha = -3$). For the Skew-$t$ distribution, the degrees of freedom parameter is set to $\nu = 8$, which implies somewhat heavier tails than the Normal distribution. We consider these different specifications to allow for a varying degree of complexity in the DGP. As for the remaining parameters, we set $\tau_{1}=0.2$, $\tau_{2}=0$, and $\theta_{2}=\gamma=0.5$. To account for different degrees of serial correlation, and hence cross-horizon dependence in the multi-step forecasts, $\theta_1$ takes one of the following values: $\theta_{1} \in \{0.1,0.4,0.7\}$.

We consider two types of forecasting models, both of which are misspecified AR(1) and produce \textit{direct} $h$-step-ahead forecasts. The first forecasting model is used when $\varepsilon_{1,t}$ in the DGP is drawn from a Normal distribution:
\begin{equation}\label{eq:FM1} 
Y_{t+h} = \tau_{h} + \beta_{h} Y_t + u_{t+h},  
\end{equation}
with $u_{t+{h}} \overset{\text{\tiny iid}}{\sim} \mathcal{N}(0,\sigma_{u,{h}}^2)$. 

The second forecasting model is a quantile regression specification used when $\varepsilon_{1,t}$ in the DGP is drawn from the Skew-Normal or the Skew-$t$ distribution:
\begin{equation}\label{eq:FM2} 
Y_{t+{h}}(q) = \tau_{h}(q) + \beta_{h}(q)Y_t + u_{t+{h}}(q),
\end{equation}
where $q$ denotes the quantile with $q \in \{0.05,0.25,0.5,0.75,0.95\}$, and $\tau_h(q)$ and $\gamma_h(q)$ denote respectively the quantile specific intercept and autoregressive parameter. To obtain a full predictive distribution, we smooth the five predicted quantiles using the Skew-$t$ of \cite{Azzalini2003}. Note that the latter introduces a second potential source of model misspecification, in addition to that implied by the specification in \eqref{eq:FM2}.

The parameters $\{ {\tau}_{h}, {\beta}_{h}\, {\sigma^{2}_{u,{h}}}\}$ and $\{ {\tau}_{h}(q), {\beta}_{h}(q)\}_{q \in Q}$ are estimated using a rolling-window estimation scheme with sample size $T_{\text{is}}$, which we set to 200 (quarters). Estimated parameters are used to compute qoq predictive densities up to 12 quarters ahead, from which we get $S$ forecast paths. We then use simple approximating formulas to construct annual-average predictive distributions from those paths. For instance, assuming that the forecast origin is the last quarter of the year, for each forecast path $s=1,\dots,S$ the one year-ahead $(h_{a}=1)$ annual-average forecast is given by:
\begin{equation} \label{eq:TransAA}
Z_{t+4 \vert t}^{(s)}=\frac{1}{4}Y_{t-2}+\frac{2}{4}Y_{t-1}+\frac{3}{4}Y_{t} + Y_{t+1\vert t}^{(s)}+ \frac{3}{4}Y_{t+2\vert t}^{(s)}+\frac{2}{4}Y_{t+3\vert t}^{(s)}+\frac{1}{4}Y_{t+4\vert t}^{(s)}
\end{equation}
and the four quarters-ahead year-on-year forecast:
\begin{equation} \label{eq:TransYoY}
Z^{(s)}_{t+4 \vert t}=Y_{t+1\vert t}^{(s)}+Y_{t+2\vert t}^{(s)}+Y_{t+3\vert t}^{(s)}+Y_{t+4\vert t}^{(s)}
\end{equation}
 
Two competing approaches are here considered: \textit{i)} the benchmark approach, which constructs the annual-average and year-on-year forecasts by directly transforming the quarter-on-quarter forecast paths with \eqref{eq:TransAA} and \eqref{eq:TransYoY}, i.e., without accounting for the cross-horizon dependence; and \textit{ii)} the copula approach, which in turn constructs the predictive distributions for the annual-average and yoy growth rates using first the methodology described in Section \ref{sec:Method} to obtain joint draws of the quarter-on-quarter growth rates, and then expressions \eqref{eq:TransAA} and \eqref{eq:TransYoY} to transform the joint draws.

The out-of-sample size used to compute the correlation of the empirical PITs for the copula approach is denoted by $T_{R}$ and set to 50 (quarters).\footnote{To reduce the computational costs in the Monte Carlo, for each simulation we compute the historical correlation of the empirical PITs once (the first forecast round) and we then use this estimate in all forecasting iterations, instead of updating it at every iteration.}
Similarly, the out-of-sample size used to evaluate the annual-average (resp. year-on-year) density forecasts is denoted by $T_{\text{oos}}$ and also set to 50. In particular, we simulate 200 periods of quarterly data and we then produce annual-average (resp. year-on-year) forecasts every four quarters for horizons one, two, and three years ahead.\footnote{This strategy aims at replicating the calendar year predictions typically used in practice by professional forecasters.} 
Results are based on 500 Monte Carlo iterations. For each iteration we evaluate the performance of the benchmark and copula approach using both relative and absolute forecasting performance measures. For this purpose, in addition to the qoq growth rates, we also simulate the true annual-average and yoy growth rates generated by the DGP described in \eqref{eq:DGP}.  
As in \cref{sec:MotivatingExample}, the relative performance measures include the QW-CRPS (calibrated to evaluate the predictive performance at the tails of the distribution) and the CRPS.\footnote{Results for quantile-score at the 10\% lower tail, QS(10\%), reported in \cref{fig:MC_TickLoss} in the Appendix, are very similar to those for the QW-CRPS, and they are hence not commented for the sake of parsimony.}
The relative performance is evaluated both by directly comparing the two approaches and against the forecasts one could generate with knowledge of the true underlying DGP and the true parameter values. The absolute forecasting performance is measured through the test for the correct specification of the predictive distribution proposed by \cite{Rossi2019}, which is a test for uniformity of the PITs.

\FloatBarrier
\subsection{Monte Carlo results}

\begin{figure}[!t]
\caption{Monte Carlo results for annual-average forecasts: relative scores and EPA tests} \label{fig:MC_AA}
\begin{subfigure}[b]{0.49\textwidth}
	\centering
		  \includegraphics[width=\linewidth]{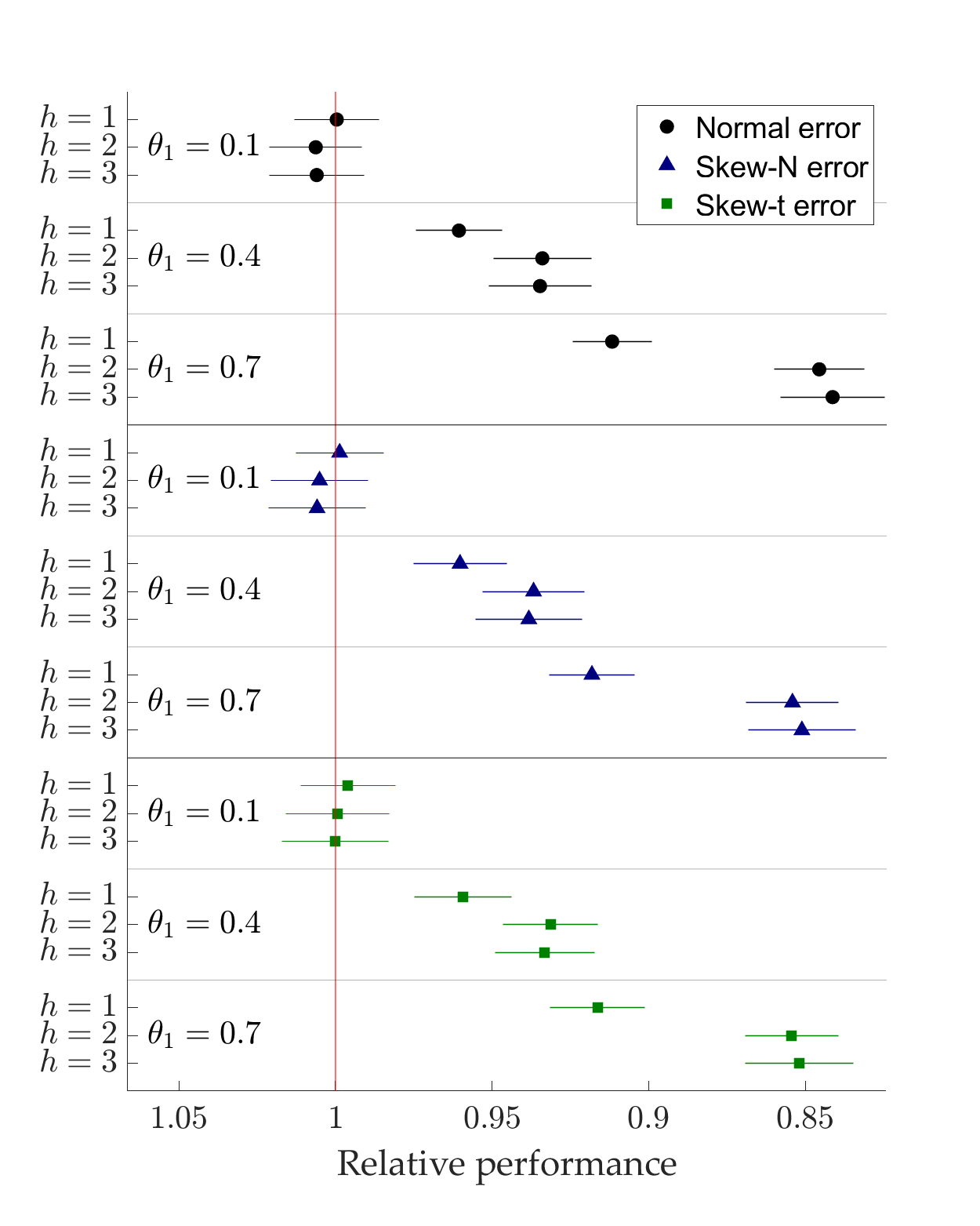}
    \vspace{-1cm}
	\caption{Relative QW-CRPS: copula vs benchmark} 
\end{subfigure} 
\begin{subfigure}[b]{0.49\textwidth}
	\centering
		  \includegraphics[width=\linewidth]{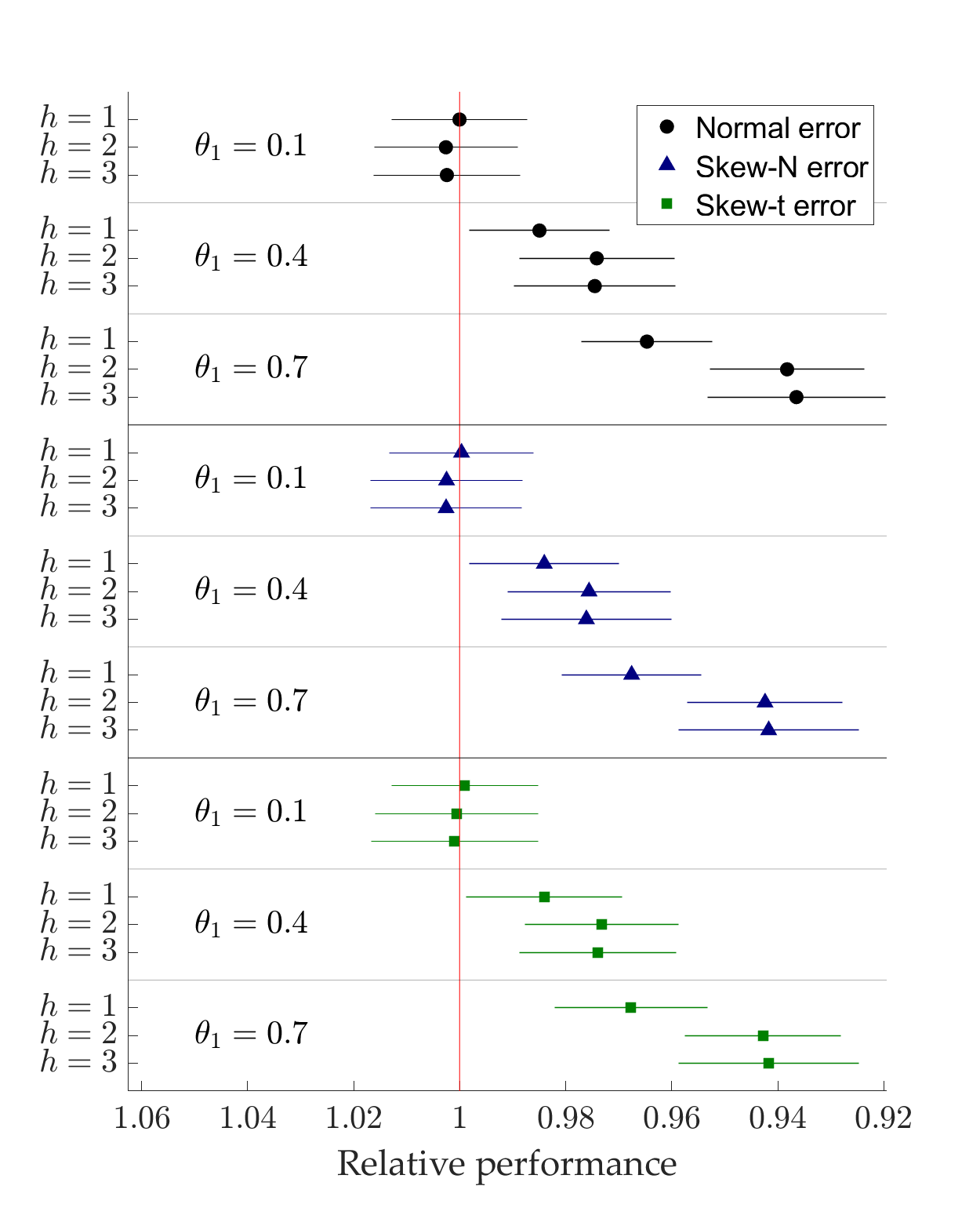}
    \vspace{-1cm}    
	\caption{Relative CRPS: copula vs benchmark}
\end{subfigure}
\begin{subfigure}[b]{0.49\textwidth}
	\centering
		  \includegraphics[width=\linewidth]{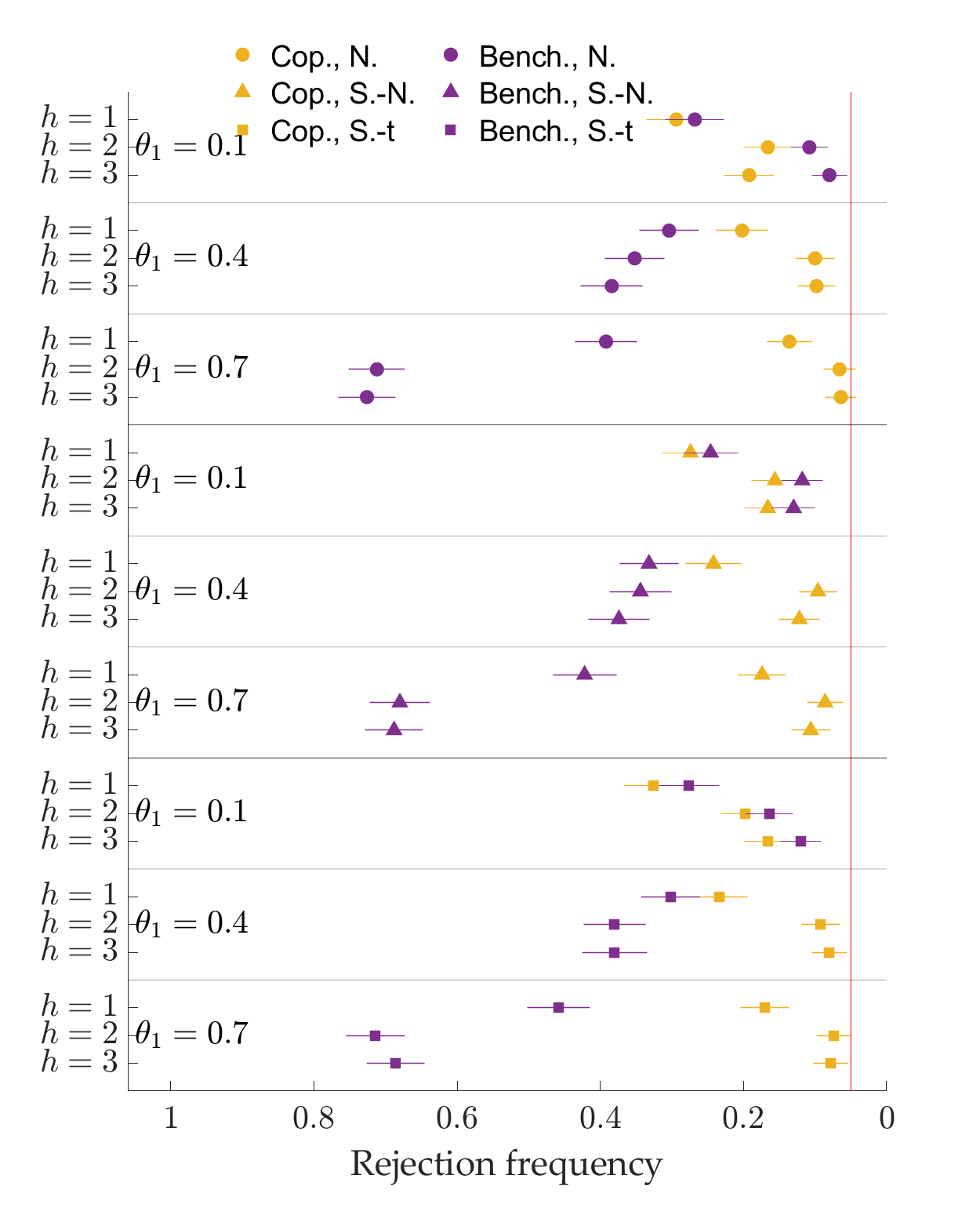}
    \vspace{-1cm}    
	\caption{EPA test: QW-CRPS}
\end{subfigure} 
\begin{subfigure}[b]{0.49\textwidth}
	\centering
		  \includegraphics[width=\linewidth]{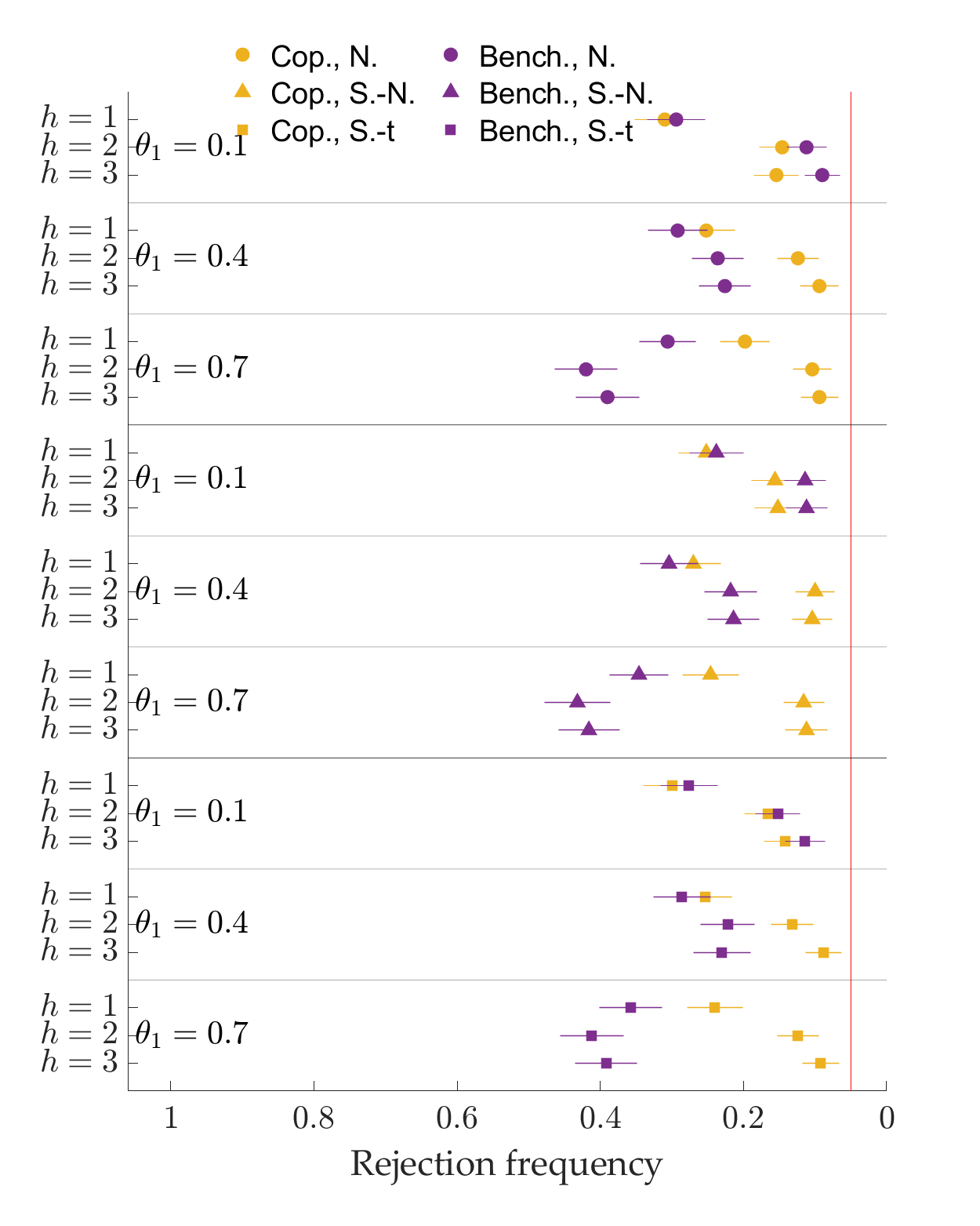}
        \vspace{-1cm}
	\caption{EPA test: CRPS}
\end{subfigure}
	\caption*{\footnotesize \textit{Note}: The $\theta_1$ indicates the autoregressive parameter of $Y_{t}$ in the DGP. The $y$-axis label $h$ denotes the annual-average horizon, i.e., one-year-, two-years-, and three-years-ahead. The $x$-axis in Panel (a) and (b) indicates the QW-CRPS (CRPS) of the copula relative to the benchmark approach, i.e., numbers smaller than one indicate a superior performance of the copula approach. The markers indicate the average score ratio across all Monte Carlo iterations. In Panel (c) and (d), the $x$-axis denotes the rejection frequency of the null hypothesis of a \cite{Giacomini2006} test of unconditional equal predictive ability against the optimal forecasts. The markers indicate the average rejection frequency across all Monte Carlo iterations. The nominal size is 5\%. The horizontal lines around the markers indicate $\pm 2$ bootstrap standard errors of the average score ratio or rejection frequency. N., S.-N., and S.-$t$ indicate the Normal, Skew-Normal, and Skew-t distribution of the error term $\varepsilon_{1,t}$ in the DGP. Standard errors of the tests were computed using a HAC estimator with $\text{bandwidth}=h_{A}-1$.}  
\end{figure}

\begin{figure}[!t]
\caption{Monte Carlo results for year-on-year forecasts: relative scores and EPA tests}
\label{fig:MC_YOY}
\begin{subfigure}[b]{0.49\textwidth}
	\centering
		  \includegraphics[width=\linewidth]{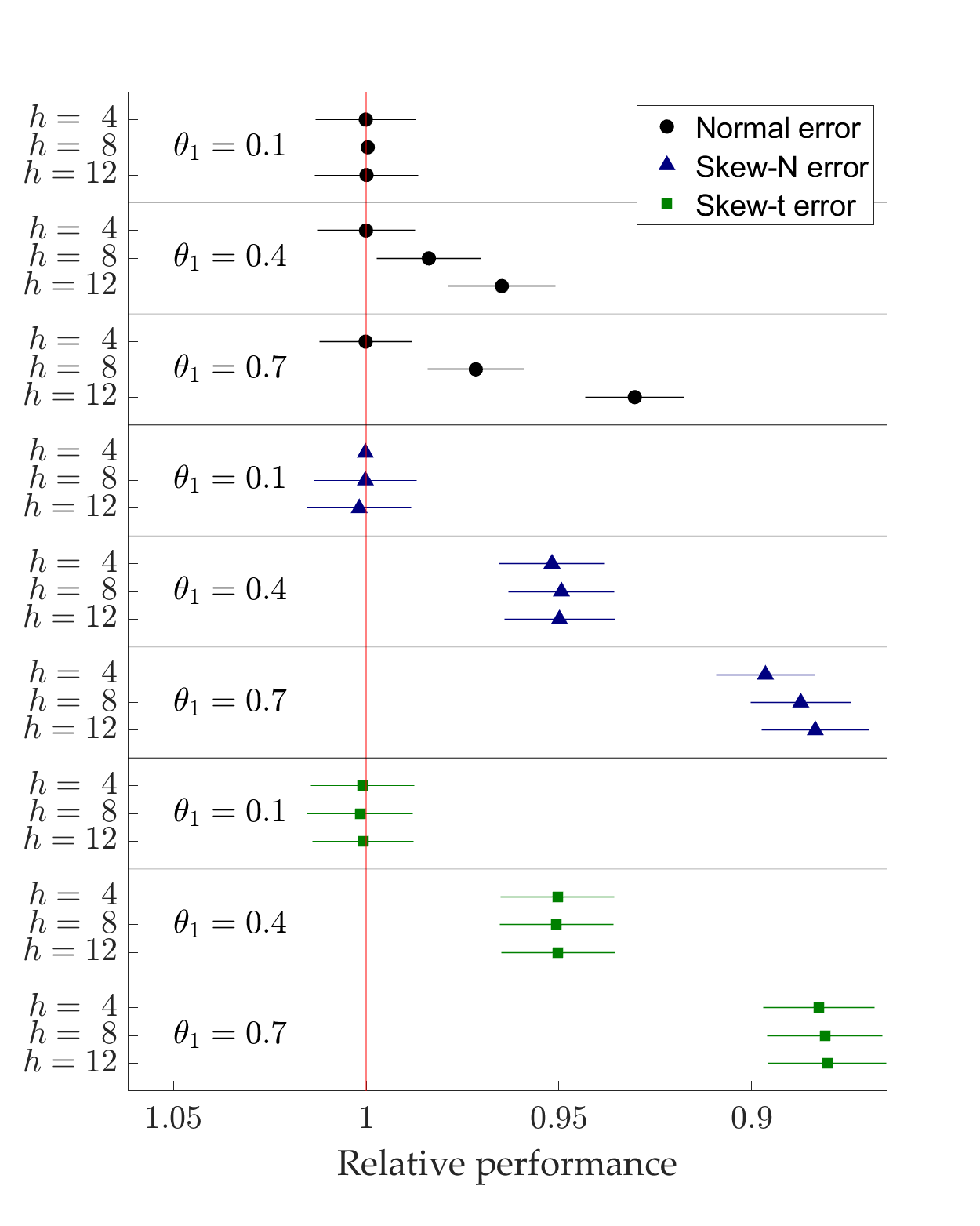}
    \vspace{-1cm}
	\caption{Relative QW-CRPS: copula vs benchmark} 
\end{subfigure} 
\begin{subfigure}[b]{0.49\textwidth}
	\centering
		  \includegraphics[width=\linewidth]{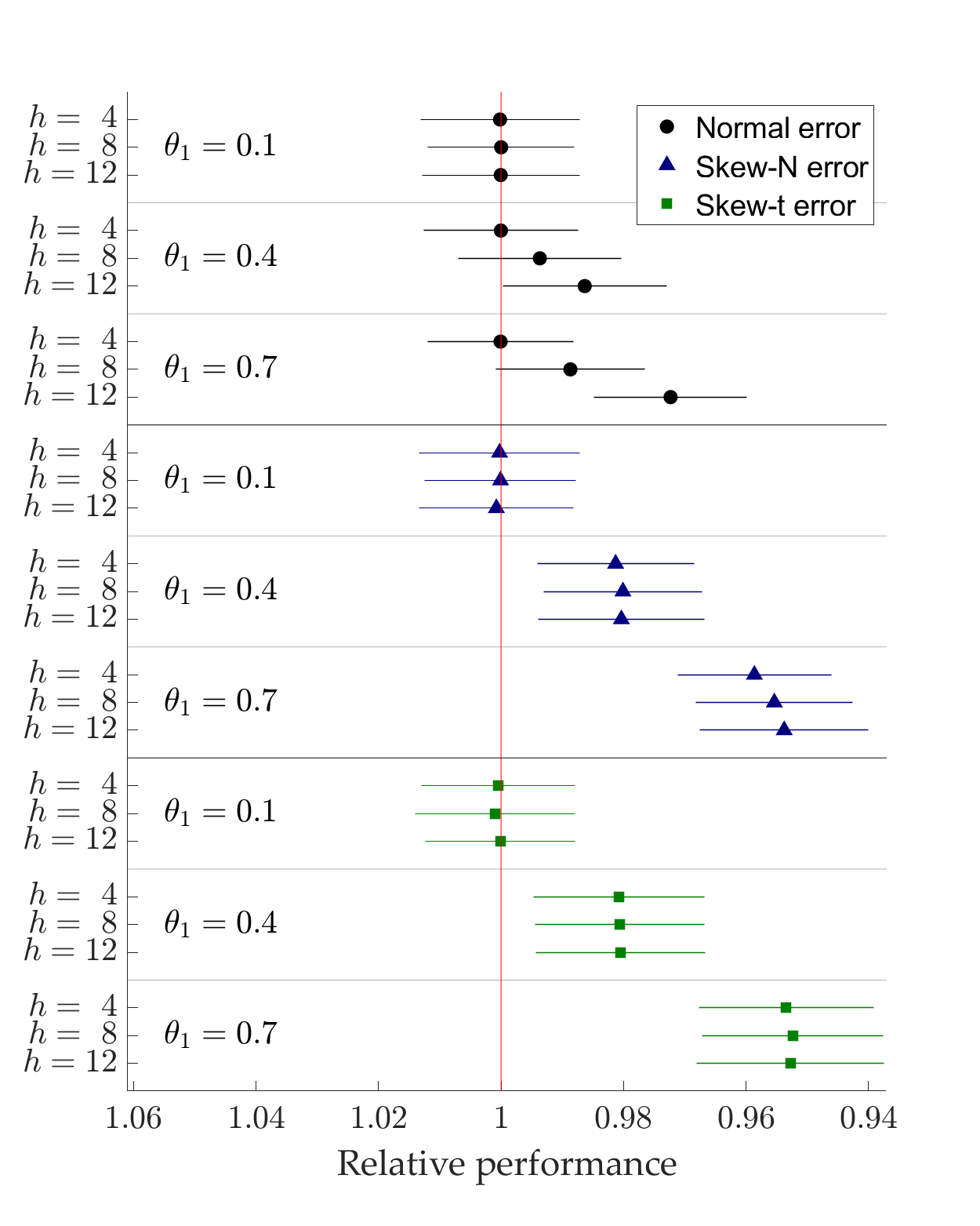}
    \vspace{-1cm}    
	\caption{Relative CRPS: copula vs benchmark}
\end{subfigure}\\
\begin{subfigure}[b]{0.49\textwidth}
	\centering
		  \includegraphics[width=\linewidth]{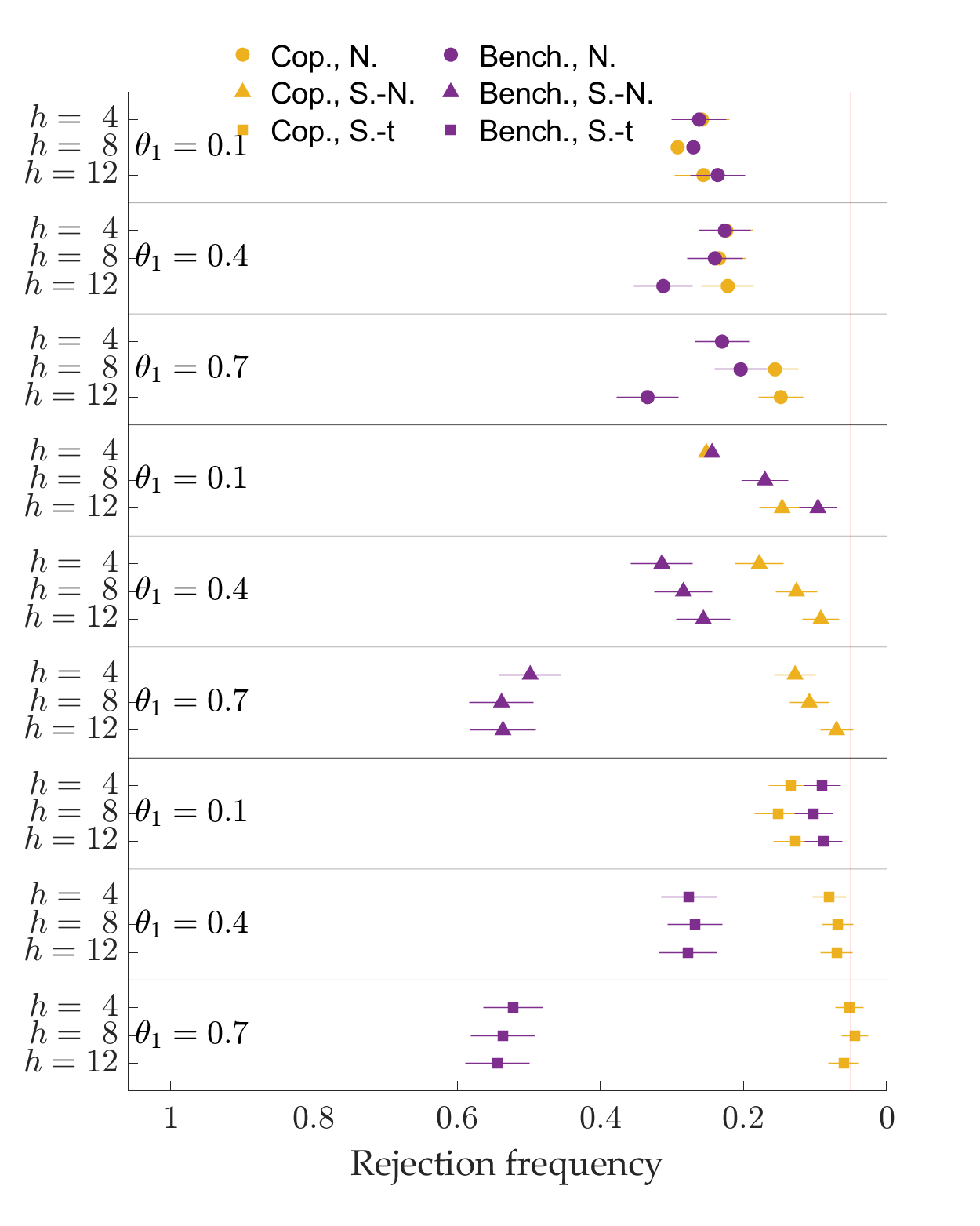}
    \vspace{-1cm}    
	\caption{EPA test: QW-CRPS}
\end{subfigure} 
\begin{subfigure}[b]{0.49\textwidth}
	\centering
		  \includegraphics[width=\linewidth]{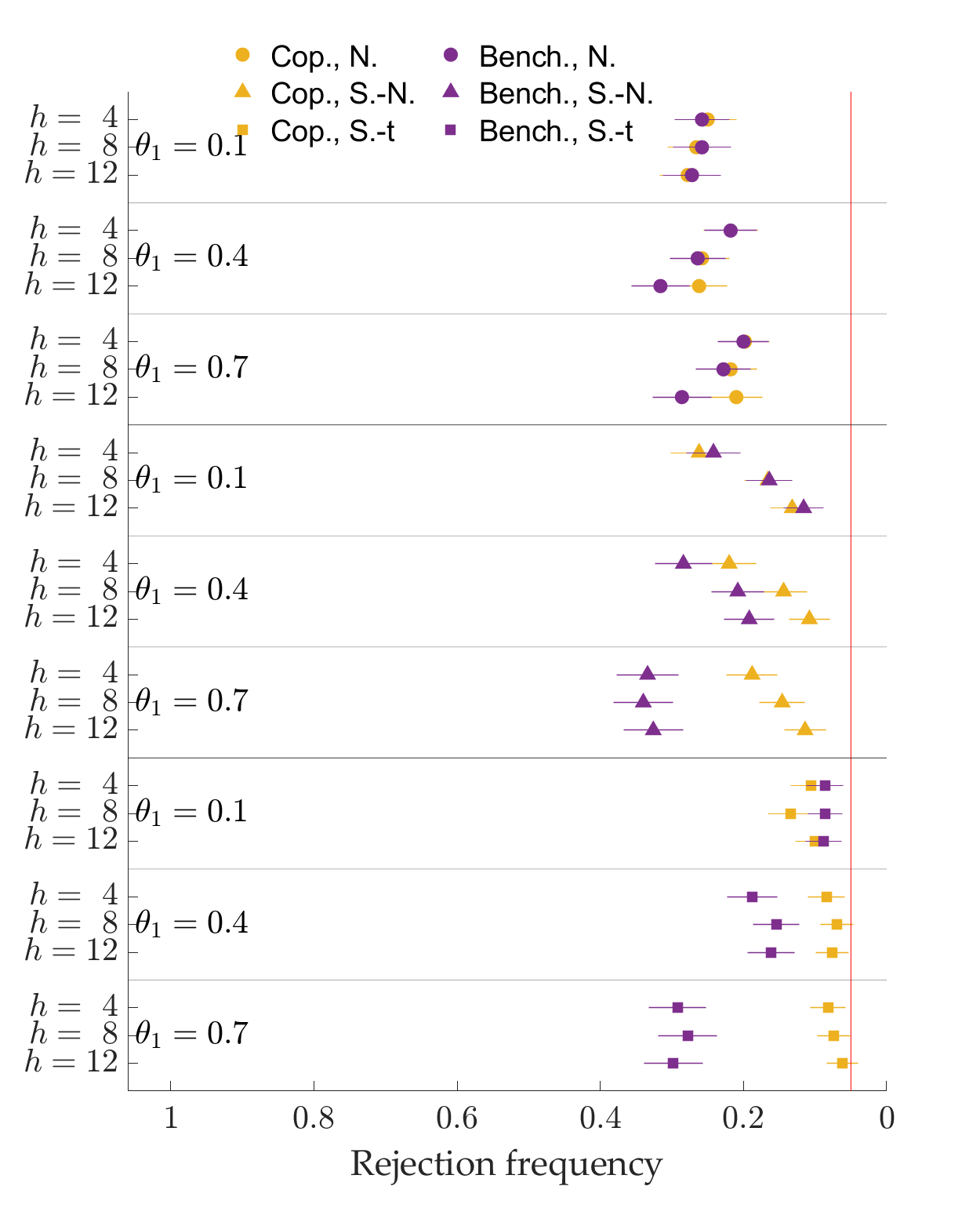}
        \vspace{-1cm}
	\caption{EPA test: CRPS}
\end{subfigure}
	\caption*{\footnotesize \textit{Note}: The $\theta_1$ indicates the autoregressive parameter of $Y_{t}$ in the DGP. The $y$-axis label $h$ denotes the year-on-year horizon, i.e., four-quarter, eight-quarter-, and 12-quarter-ahead. The $x$-axis in Panel (a) and (b) indicates the QW-CRPS (CRPS) of the copula relative to the benchmark approach, i.e., numbers smaller than one indicate a superior performance of the copula approach. The markers indicate the average score ratio across all Monte Carlo iterations. In Panel (c) and (d), the $x$-axis denotes the rejection frequency of the null hypothesis of a \cite{Giacomini2006} test of unconditional equal predictive ability against the optimal forecast. The markers indicate the average rejection frequency across all Monte Carlo iterations. The nominal size is 5\%. The horizontal lines around the markers indicate $\pm 2$ bootstrap standard errors of the rejection frequency. N., S.-N., and S.-$t$ indicate the Normal, Skew-Normal, and Skew-t distribution of the error term $\varepsilon_{1,t}$ in the DGP. Standard errors of the tests were computed using a HAC estimator with $\text{bandwidth}=h_{A}-1$.}  
\end{figure}

\begin{figure}[!t]
\caption{Monte Carlo results of tests for correct forecast density specifications}
\label{fig:MC_PIT}
\begin{subfigure}[b]{0.49\textwidth}
	\centering
		  \includegraphics[width=\linewidth]{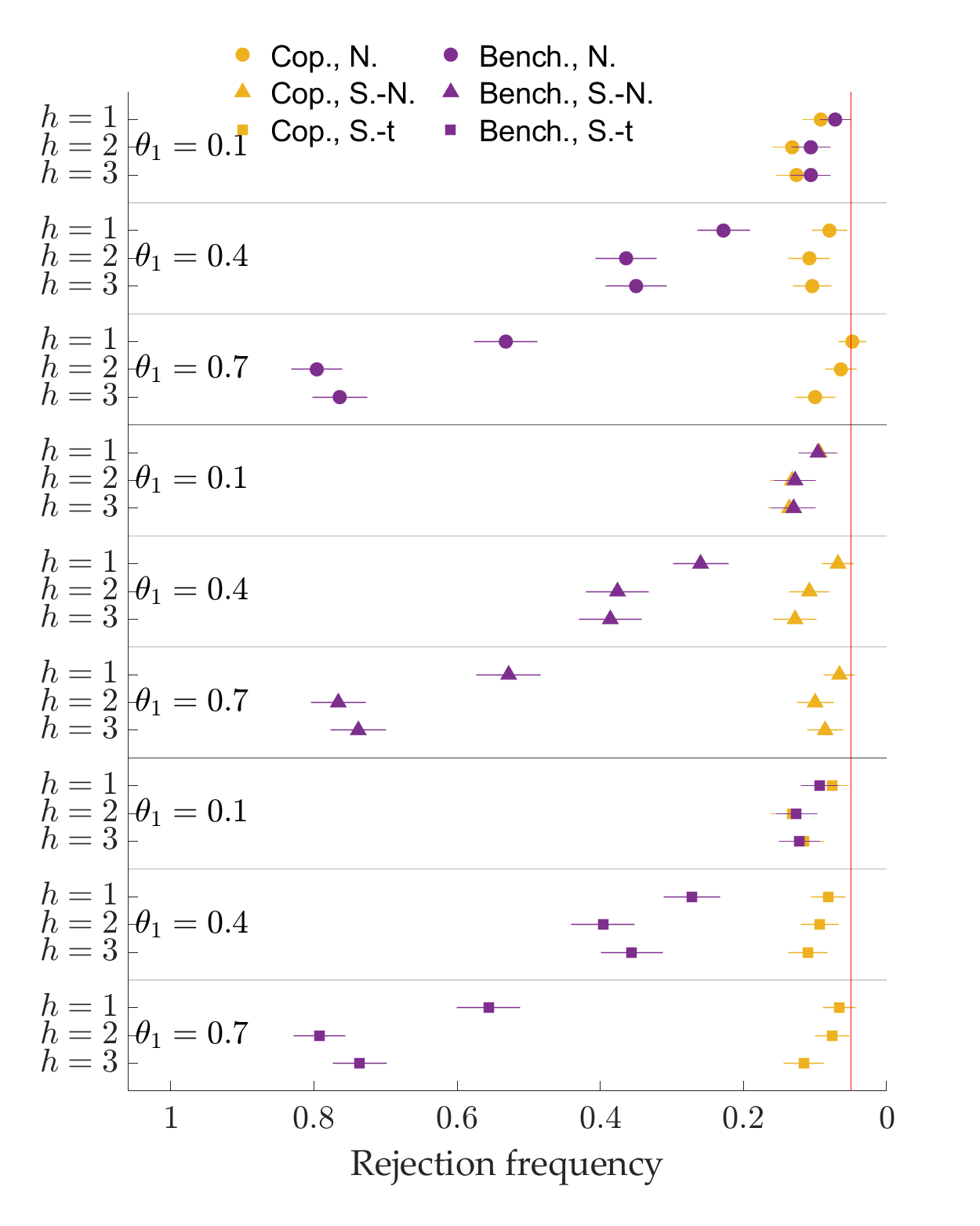}
    \vspace{-1cm}
	\caption{annual-average forecast} 
\end{subfigure} 
\begin{subfigure}[b]{0.49\textwidth}
	\centering
		  \includegraphics[width=\linewidth]{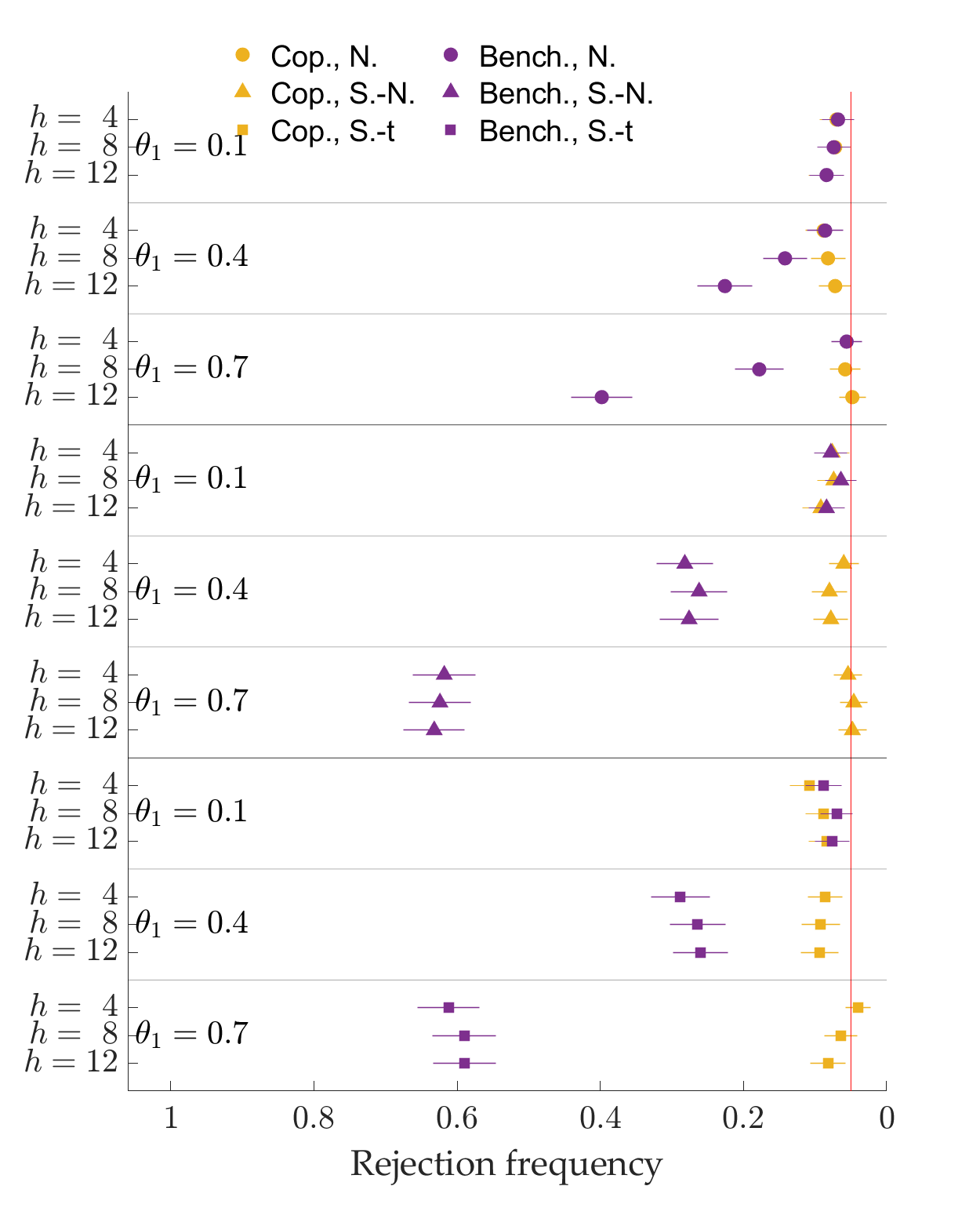}
    \vspace{-1cm}    
	\caption{Year-on-year forecasts}
\end{subfigure}
	\caption*{\footnotesize \textit{Note}: The $\theta_1$ indicates the autoregressive parameter of $Y_{t}$ in the DGP. The $y$-axis label $h$ denotes the annual-average and year-on-year horizon, respectively, in years and quarters. The $x$-axis denotes the rejection frequency of the \cite{Rossi2019} test for correct specification of the predictive density. The markers indicate the average rejection frequency across all Monte Carlo iterations. The horizontal lines around the markers indicate $\pm 2$ bootstrap standard errors of the rejection frequency. The nominal size is 5\%. N., S.-N., and S.-$t$ indicate the Normal, Skew-Normal, and Skew-t distribution of the error term $\varepsilon_{1,t}$ in the DGP.}  
\end{figure}

\cref{fig:MC_AA} shows the relative forecasting performance of the benchmark and copula approach, as well as the rejection frequencies of an equal predictive ability test when comparing either of the two approaches against the optimal forecast. The markers show the average score ratios (rejection frequencies) across all Monte Carlo iterations, while the horizontal lines denote $\pm 2$ bootstrap standard errors of the average score ratios (rejection frequencies). Panel (a) shows the ratio of the QW-CRPS of the copula approach and the benchmark approach, i.e., values $<1$ indicate superior performance of the copula approach. Panel (b) shows the analogue ratio for the CRPS. As expected, for a serial dependence close to zero there is no improvement in the forecasting performance when using the copula approach. For medium to large values of $\theta_{1}$, the copula approach outperforms the benchmark approach by about 5\% to 15\% in the case of the QW-CRPS and by about 3\% to 7\% in the case of the CRPS. Panel (c) and (d) show the average rejection frequencies of the null hypothesis of equal predictive ability using the unconditional test of \cite{Giacomini2006}. The scoring functions are the QW-CRPS (c) and the CRPS (d) and the test compares the copula (benchmark) approach against the optimal forecast. The nominal size is 5\%, i.e., rejection frequencies above 5\% indicate that the null hypothesis is rejected more often than expected given the nominal size. Panel (c) and (d) of \cref{fig:MC_AA} show that for small values of $\theta_{1}$, the copula approach performs slightly worse than the benchmark approach. We attribute this result to the additional parameter estimation uncertainty induced by the estimation of the copula parameters, which is likely to dominate the gain from taking the temporal dependence into account. In contrast, with increasing temporal dependence, i.e. increasing $\theta_{1}$, the rejection frequency associated with the copula approach tends to be close to the nominal size, whereas, for the benchmark approach, the null hypothesis of equal performance is rejected much more frequently.

It is worth noting that the rejection rate for the copula approach is close to the nominal size when compared to the true predictive density. This is an important result, since for the case with Skew-Normal and Skew-$t$ errors, the Gaussian copula is potentially misspecified. Further, for the Skew-Normal case, the marginals are potentially misspecified since we use a Skew-$t$ distribution to smooth the quantile regression predictions. Importantly, the results suggest that these misspecifications have only a small impact on the overall forecasting performance, because the copula-based predictive distributions appear often indistinguishable from true predictive distributions in terms of forecasting performance. As regards the year-on-year forecasts, the results are reported in \cref{fig:MC_YOY} and are very similar to the annual-average results.

\cref{fig:MC_PIT} shows the rejection frequencies of the null hypothesis of a correct specification of the predictive densities, evaluated using the test of \cite{Rossi2019}. Results show that a high temporal dependence implies that the benchmark approach leads to misspecified predictive distributions. In contrast, the copula approach adjusts for the temporal dependence and shows rejection frequencies only slightly above the nominal size. Hence the results in \cref{fig:MC_PIT} suggest that the consequences of the aforementioned misspecifications on the calibration of copula-based predictive densities are fairly negligible, as the rejection rates hovers around the nominal size of the test.

\FloatBarrier

\section{Robustness of the copula and alternatives}\label{sec:Discussion}
This section provides some guidance on how the training sample size and the persistence of the process affect the performance of the copula approach and discusses potential alternative approaches.
\subsection{Robustness of the copula approach}\label{sec:RobustnessSub1}
This subsection further explores to what extent the performance of the copula approach depends on (i) the training sample for the copula parameter $\boldsymbol{R}$, and (ii) the persistence of the underlying process. Our main goal here is to provide a guideline to the practitioners by setting some rules-of-thumb when choosing whether or not implementing the copula approach on specific applications. Note, however, that the conclusions drawn in this section are, to some extent, dependent on the DGP considered.

\begin{figure}[!t]
\graphicspath{{./}{Figures/}} 
\caption{Relative performance of annual-average forecasts across ranges of $T_{\text{is}}$ and $T_R$}\label{fig:Scores_ARDL_robustness}
\begin{subfigure}{\textwidth}
	\centering
		  \includegraphics[scale=0.60]{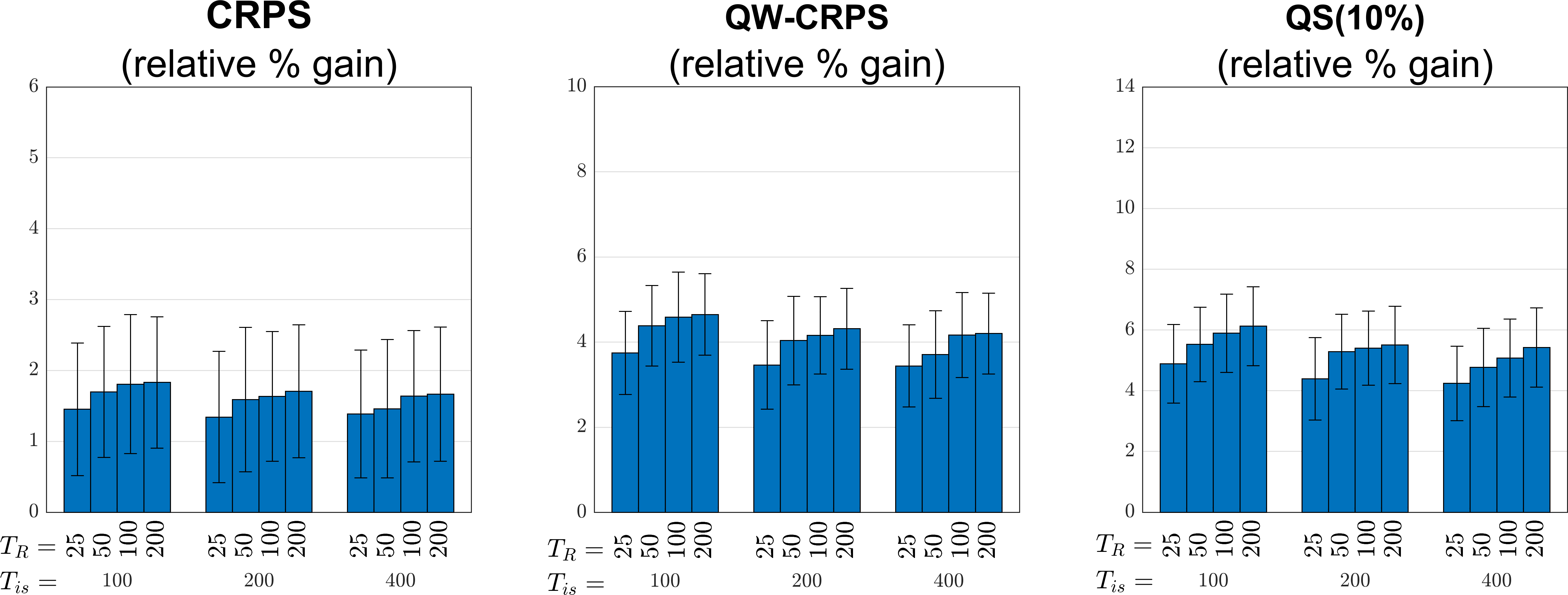}
    \vspace{0.2cm}
	\caption{One-year-ahead annual-average} 
\end{subfigure} 

\vspace{0.5cm}

\begin{subfigure}{\textwidth}
	\centering
		  \includegraphics[scale=0.60]{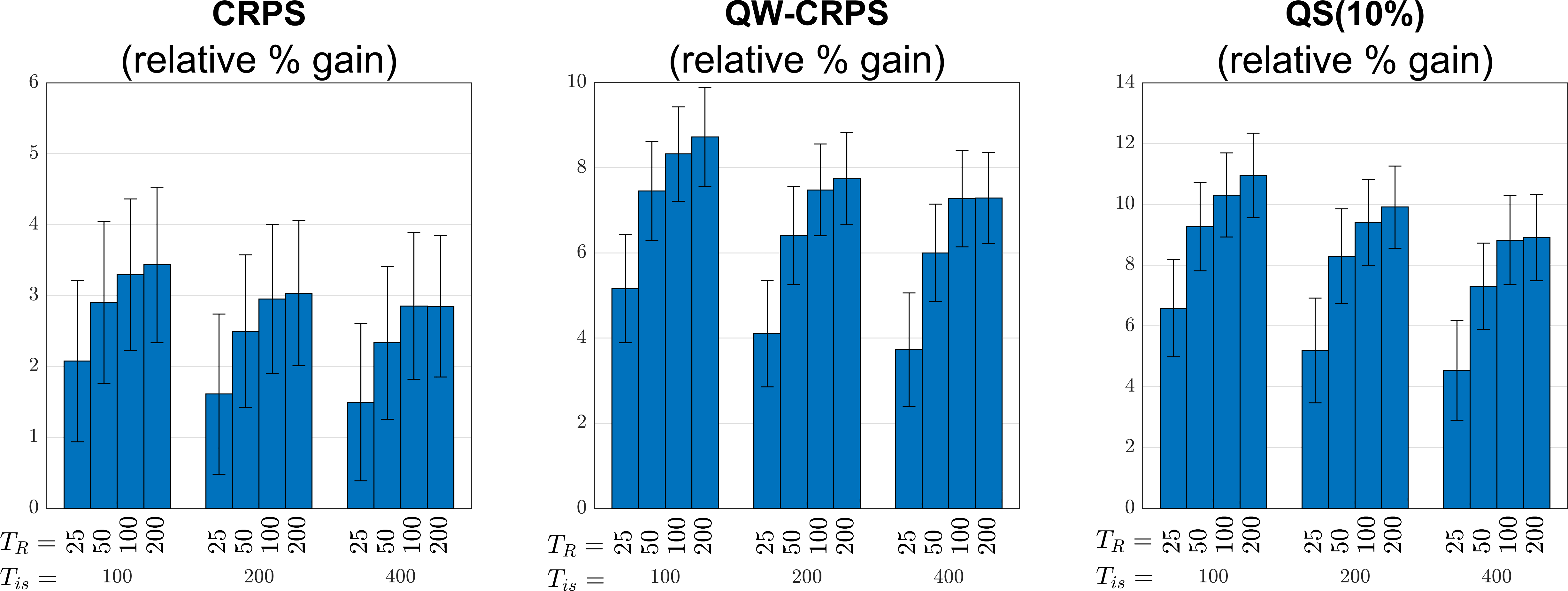}
    \vspace{0.2cm}
	\caption{Three-years-ahead annual-average} 
\end{subfigure} 
	\caption*{\footnotesize \textit{Note}: Panel (a) shows results for the one-year-ahead annual-averages. Panel (b) shows results for the three-year-ahead annual-averages. $T_\text{is}$ denotes the in-sample estimation size for the prediction model parameters. The rotated numbers on the $x$-axis denote the values of $T_R$, i.e. the estimation sample size for $\boldsymbol{R}$. The units of the $y$-axis are the average percentage point gains in the relative forecasting performance of the copula to the benchmark approach. CRPS denotes the continuous ranked probability score. QW-CRPS denotes the quantile weighted versions of the continuous ranked probability score, with emphasis on the tails. QS($10\%$) denotes the quantile score at the $10\%$ quantile. The error bars denote $\pm 2$ Monte Carlo standard errors obtained by 999 bootstrap resamples of the average scores from 1000 Monte Carlo iterations.}  
\end{figure}

For our first robustness check, we repeat the Monte Carlo experiment of \cref{sec:MonteCarlo}, keeping the design of the simulations unchanged except for the number of in-sample observations $T_\text{is}$ and the number of observations $T_{R}$ used to estimate the copula correlation matrix $R$. For this purpose, we use a grid of $T_{\text{is}}$ = 100, 200, and 400 and $T_R$ = 25, 50, 100, and 200 observations. Note that the values in the baseline Monte Carlo experiments discussed in \cref{sec:MonteCarlo} are $T_\text{is} = 200$ and $T_{R} = 50$. We focus on $\theta_{1}=0.4$, the case of intermediately strong serial correlation, we compute the same scores as in \cref{sec:MonteCarlo}, and we report the relative percentage gain of the copula approach with respect to the benchmark approach (neglecting cross-horizon dependence). 
\cref{fig:Scores_ARDL_robustness} reports the results for the one- and three-year-ahead annual-average predictions, and point to an overall stability of the relative performance of the copula approach across simulations. A slight improvement seems to arise from larger $T_\text{is}$ and $T_R$, in particular when $T_R>25$, but the results are often only marginally different and statistically indistinguishable according to (bootstrap) standard errors of the average relative scores. In other words, the Monte Carlo study suggests that \textit{i)} a moderate training out-of-sample size should be sufficient for estimating copula parameter matrix $R$ and obtaining relatively accurate forecasting results from the copula approach, and \textit{ii)} the training in-sample size used to estimate the model parameters appears less relevant.

\begin{figure}[!t]
\graphicspath{{./}{Figures/}} 
	\centering
	\caption{Distribution of the relative performance of annual-average forecasts across ranges of det$(\widehat{R})$}\label{fig:Scores_ARDL_detR}
		\includegraphics[scale=0.40]{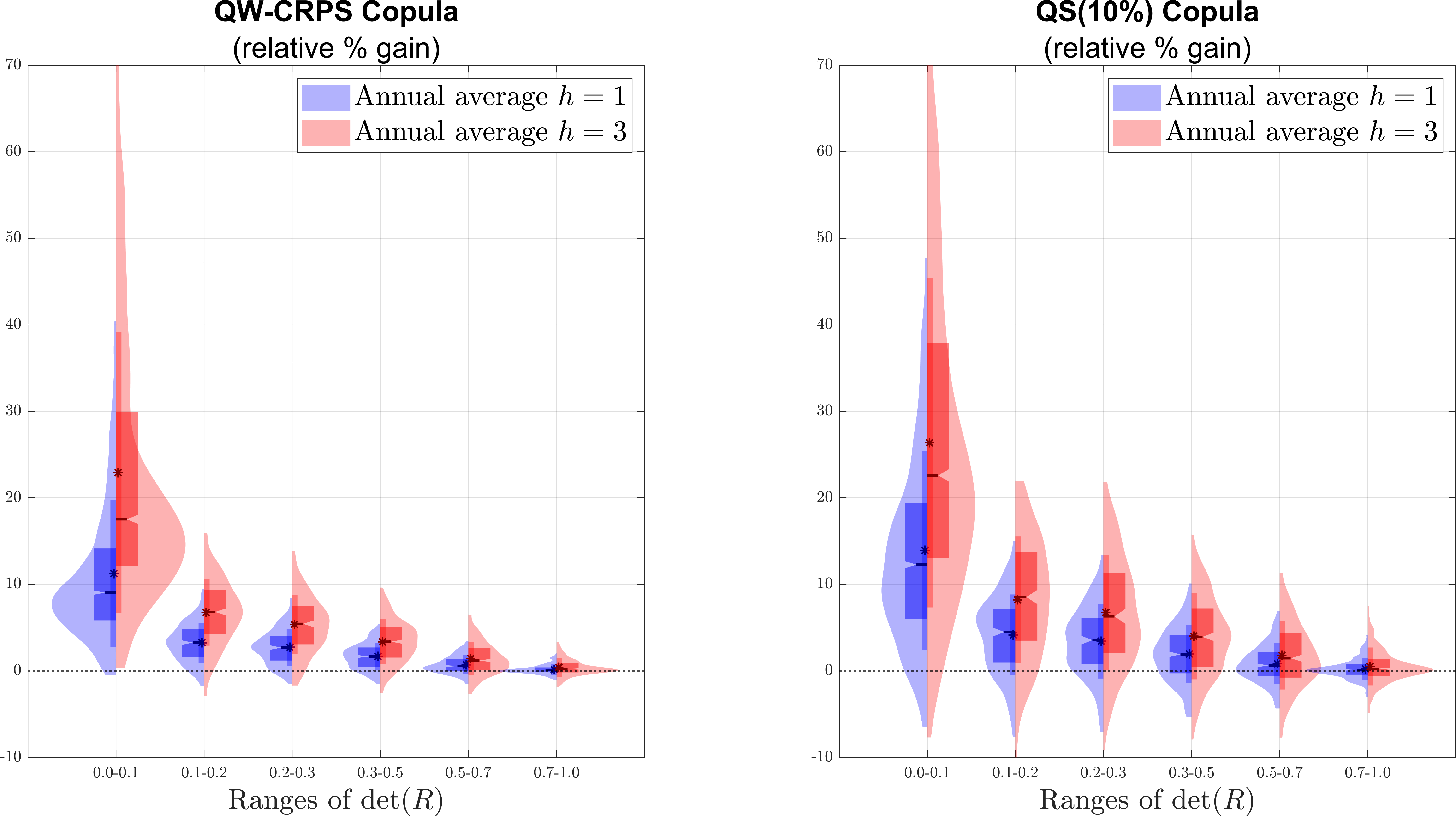}
    \vspace{0.4cm}
	\caption*{\footnotesize \textit{Note}: QW-CRPS denotes the quantile weighted versions of the CRPS. QS$(10\%)$ denotes the quantile-score at the $10\%$ quantile. The $y$-axis shows the relative percentage point gain of the copula approach with respect to the benchmark. The $x$-axis shows different levels of the determinant of $\boldsymbol{R}$; larger values imply an overall smaller persistence in the process.}
\end{figure}

For our second robustness check, we compute the relative forecasting performance of the copula approach for different levels of persistence measured by the determinant of the estimated correlation matrix $\boldsymbol{R}$, $\det(\widehat{\boldsymbol{R}})$. On the one hand, in the extreme case of completely independent marginals across all the forecast horizons, the determinant of $\boldsymbol{R}$ is equal to one. On the other hand, in the extreme case of a perfect correlation across the marginals, the determinant is equal to zero. To evaluate the effect of different degrees of path-dependence across forecast horizons, we again repeat the Monte Carlo experiment of \cref{sec:MonteCarlo}, but for each Monte Carlo iteration we now draw (with replacement) some of the VAR(1) parameters from the following uniform distributions: $\left(\theta_{1},\theta_{2},\gamma\right)\sim\mathcal{U}(-0.9,0.9)$ and $\sigma_{\epsilon_{2}}\sim\mathcal{U}(0.3,0.7)$. To attenuate the effect of finite samples on the results, we set the training sample at $T_{\text{is}}=T_R=500$ and we generate $T_{\text{oos}}=400$ observations for the out-of-sample evaluation of the transformed densities. For ease of analysis, we focus here on the Normal error case, $\epsilon_{1}\sim\mathcal{N}(0,\sigma_{\epsilon_{1}}^2)$, and on annual-average density forecasts.
Results are presented in \cref{fig:Scores_ARDL_detR}, where the $y$-axis denotes the relative performance gains and the $x$-axis shows ranges of $\det(\widehat{\boldsymbol{R}})$ categorized into bins with width 0.1: smaller values on the $x$-axis indicate a higher level of persistence. The figure shows results for the QW-CRPS and the QS(10\%) in so-called violin plots for the one- and three-year-ahead annual-average predictions. The curves in the figure depict the distribution of the gains over the 1000 Monte Carlo iterations, and the boxes show the inter-quartile range. For high levels of persistence, the gains are on average large and the distribution exhibits a large right-tail towards regions of big relative forecasting gains for the copula approach. For medium levels of $\det(\widehat{\boldsymbol{R}})$, the copula approach still shows substantial predictive gains relative to the benchmark approach. However, these gains tend to disappear with a decreasing persistence of the process. 

\FloatBarrier

\subsection{Alternative approaches}\label{sec:Alternative}	
It could be argued that the copula approach described so far may be inefficient compared to simpler alternative approaches that provide transformed forecasts in one single step. For instance, the researcher interested in the transformation in \eqref{eq:approx1} could run regressions using directly the transformed variable $Z_{t}$, instead of modelling the data at the original sample. Note that while this would violate the requirement of coherence between the moments across the different frequencies, this section nonetheless entertains the idea of a direct regression of $Z_{t}$ on its past history. However, it turns out that the forecasting regression based on the transformed sample, using the simplified example as in \cref{sec:MotivatingExample}, does not lead to a significant superior forecasting performance.

Assume again that the DGP takes the form of the autoregressive model $Y_{t+1} = \rho Y_{t} + \varepsilon_{t+1}$ and that the target forecast follows the linear transformation $Z_{t+h\vert t}=Y_{t+1\vert t} + Y_{t+2\vert t} + \dots + Y_{t+h\vert t}$. A regression of the annual-average on its past values, or in our simplified illustrative example, $Z_{t+h}$ on $Z_t$, leads to the following expression:
\begin{align}\label{eq:data_alternative}
Z_{t+h} &= Y_{t+1} + Y_{t+2} + \dots + Y_{t+h} \nonumber \\
        &= \rho^{h} \underbrace{\left(Y_{t-(h-1)}+Y_{t-(h-2)}+ \dots + Y_{t}\right)}_{Z_{t}} + \underbrace{\left[\sum_{j=1}^{h}\left(\frac{1-\rho^{j}}{1-\rho}\right)\varepsilon_{t+h+1-j} + \sum_{j=1}^{h-1}\rho^{j}\left(\frac{1-\rho^{h-j}}{1-\rho}\right)\varepsilon_{t+1-j} \right]}_{u_{t+h}} \nonumber \\
        &= \rho^{h}Z_{t} + u_{t+h} 
\end{align}

This suggests to construct a predictive density for $Z_{t+h\vert t}$ centered on the conditional mean $\rho^hZ_t$ and featuring a variance based on $u_{t+h}$. First, note that the conditional mean is not optimal because $Z_t$ is a linear combination of lags of $Y_{t}$, i.e., the forecaster partly conditions on "out-dated" information. This is akin to a situation where the forecaster only uses the lower frequency instead of taking a mixed-frequency approach. The information in the error terms, $\varepsilon_{t}$ in \eqref{eq:data_alternative} that already realized in $t$. In other words, the conditional mean forecast is not efficient. Second, note that in the regression $Z_t$ and $u_{t+h}$ in \eqref{eq:data_alternative} are correlated, i.e., the OLS estimator of $\rho$ is inconsistent.

\begin{figure}[!t]
\caption{Monte Carlo results for annual-average forecasts: relative scores and EPA tests} \label{fig:MC_AA_AlternativeReg}
\begin{subfigure}[b]{0.49\textwidth}
	\centering
		  \includegraphics[width=\linewidth]{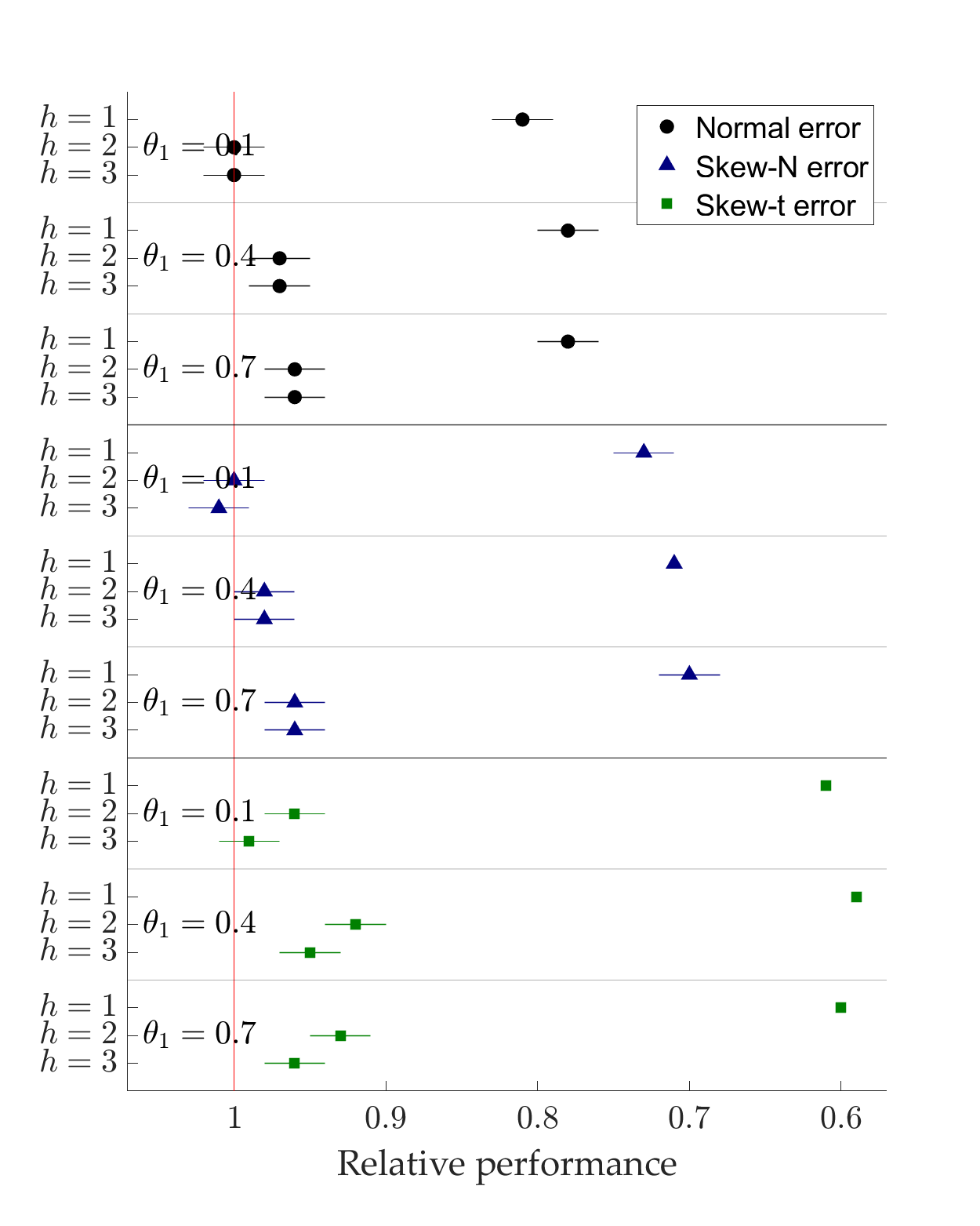}
    \vspace{-.5cm}
	\caption{QW-CRPS: copula vs annual-average regression} 
\end{subfigure} 
\begin{subfigure}[b]{0.49\textwidth}
	\centering
		  \includegraphics[width=\linewidth]{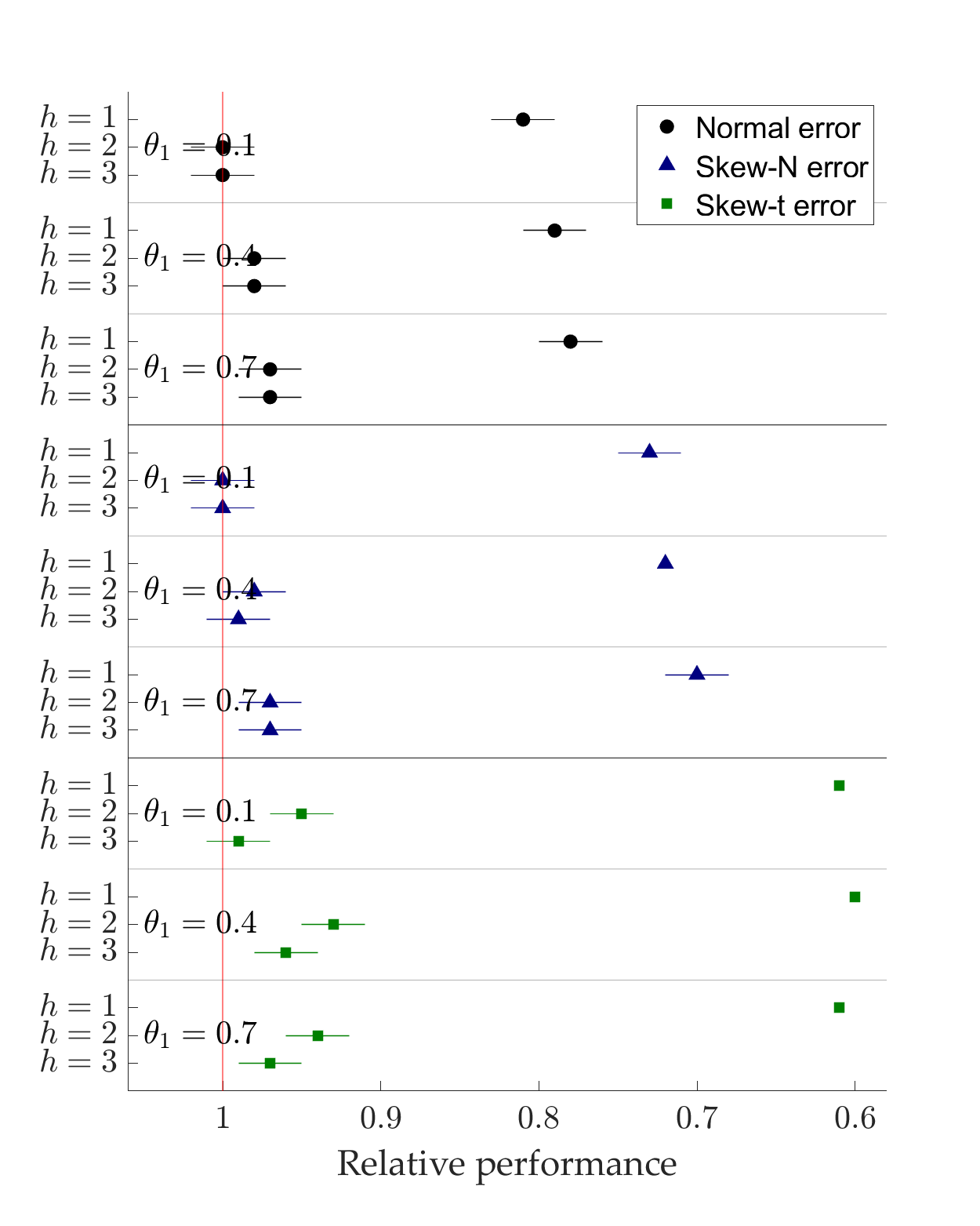}
    \vspace{-.5cm}    
	\caption{CRPS: copula vs annual-average regression} 
\end{subfigure} 
	\caption*{\footnotesize \textit{Note}: The $\theta_1$ indicates the autoregressive parameter of $Y_{t}$ in the DGP. The $y$-axis label $h$ denotes the annual-average horizon, i.e., one-year-, two-years-, and three-years-ahead. The $x$-axis in Panel (a) and (b) indicates the QW-CRPS (CRPS) of the copula relative to the annual-average regression approach, i.e., numbers smaller than one indicate a superior performance of the copula approach. The markers indicate the average score ratio across all Monte Carlo iterations. The horizontal lines around the markers indicate two standard deviations of the average score ratio frequency.}  
\end{figure}

\begin{figure}[!t]
\caption{Monte Carlo results of tests for correct forecast density specifications: direct annual-average regression} \label{fig:MC_AA_AlternativeReg_PIT}
	\centering
		  \includegraphics[width=0.6\linewidth]{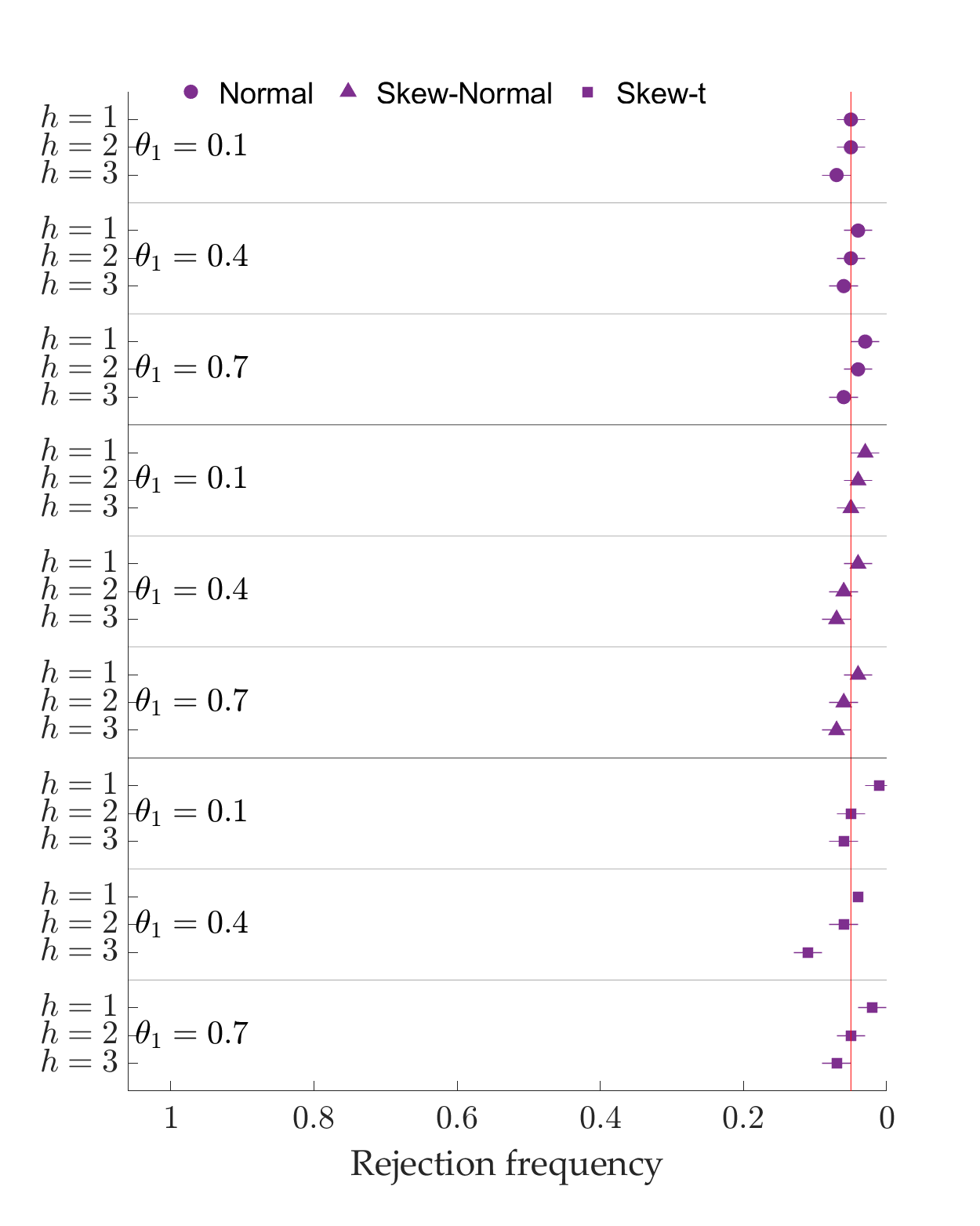}
	\caption*{\footnotesize \textit{Note}: The $\theta_1$ indicates the autoregressive parameter of $Y_{t}$ in the DGP. The $y$-axis label $h$ denotes the annual-average horizon (in years). The $x$-axis denotes the rejection frequency of the \cite{Rossi2019} test for correct specification of the predictive density. The markers indicate the average rejection frequency across all Monte Carlo iterations; the horizontal lines around the markers indicate two standard deviations of the rejection frequency. }  
\end{figure}

We repeat the Monte Carlo experiment of \cref{sec:MonteCarlo} using the same DGP and running the alternative direct annual-average regression described in (\ref{eq:data_alternative}) using $Z_{t}$ as predictor for $Z_{t+h\vert t}$. In this experiment we evaluate its forecasting performance relative to the copula approach. \cref{fig:MC_AA_AlternativeReg} reports the results for $\theta_{1}$ = 0.1, 0.4, and 0.7, three error distributions, and  one- to three-year-ahead annual-average predictions (values <1 mean that the copula approach outperforms the alternative approach). The results interestingly point to a strong predictive gain of the copula approach compared to the alternative approach for the one-year ahead prediction, while from the two-years ahead onward the two approaches provide very similar results. 
Abstracting from parameter estimation errors and the parameter consistency problem in the estimation of the direct annual-average regression, the forecasting gains by the copula approach can partly be explained by better conditional mean predictions of the higher frequency regression. Consider again the autoregressive DGP $Y_{t} = \rho Y_{t} + \varepsilon_{t}$
and assume for simplicity and without loss of generality that the calendar year is composed of two periods and that target forecast follows the weighted linear transformation: 
\begin{align*}
Z_{t+2\vert t} &= \frac{1}{2}Y_{t}  + Y_{t+1\vert t} + \frac{1}{2}Y_{t+2\vert t} \quad & \text{(for one-year-ahead predictions)} \\[1ex]
Z_{t+4\vert t} &= \frac{1}{2}Y_{t+2\vert t}  + Y_{t+3\vert t} + \frac{1}{2}Y_{t+4\vert t} \quad  & \text{(for two-year-ahead predictions)} 
\end{align*}
which is akin to an actual one-year- and two-years-ahead annual-average forecasts if the $Y_t$ were to represent semi-annual growth rates. Note that the expression for $Z_{t+2\vert t}$ embeds the observation at the forecast origin $Y_{t}$, in addition to the forecasts produced at forecast origin $t$, while $Z_{t+4\vert t}$ includes only forecast terms. 
The mean squared forecast errors ratio of the annual-average prediction provided by the copula approach ($\text{MSFE}_{C}$) and the alternative direct annual-average approach ($\text{MSFE}_{AAD}$) are:
\begin{align}
\frac{\text{MSFE}_{C}(t+2\vert t)}{\text{MSFE}_{AAD}(t+2\vert t)} &= \frac{\rho^2+4\rho+5}{6\rho^2+8\rho+6} \label{eq:MSFE1Y} \\[1ex]
\frac{\text{MSFE}_{C}(t+4\vert t)}{\text{MSFE}_{AAD}(t+4\vert t)} &= \frac{\rho^6+4\rho^5+7\rho^4+8\left(\rho^3+\rho^2+\rho\right)+6}{6\rho^6+8(\rho^5+\rho^4+\rho^3+\rho^2+\rho)+6} \label{eq:MSFE2Y}
\end{align}
For $h=1$, it can be shown that the obtained expression in \eqref{eq:MSFE1Y} (a ratio of two second-order polynomials) is always lower than 1 for $\vert\rho\vert<1$, and it ranges between 0.8 and 0.6 for $0.1\le\rho\le 0.7$. Conversely, for $h=2$, the ratio of mean square forecast errors simplifies to a ratio of two sixth-order polynomials, which is very close to 1 for $-0.6\le\rho\le 0.5$ and still above 0.9 up to $\rho=0.8$. These analytical results are consistent with the simulation results reported in \cref{fig:MC_AA_AlternativeReg}, where a slightly different DGP and additional metrics are considered. 
In summary, the expected forecasting performance of the direct annual-average approach is substantially worse than that of the copula approach, in particular for $h=1$ and in presence of strong persistence. The intuition is that, compared with the copula approach, the direct annual-average approach likely makes an inefficient use of the available short-term information. However, for $h>1$ this inefficiency loses importance, and the predictive performance converges across the two approaches, because the underlying DGP is mean reverting and for longer horizons the initial information advantage in the copula approach becomes irrelevant. Finally, \cref{fig:MC_AA_AlternativeReg_PIT} reports the results for the PIT test for the direct annual-average approach only. The results point overall to a correct specification of density forecasts, with rejection frequencies even closer to the nominal size than those of the copula approach reported in \cref{fig:MC_PIT}. The findings hence suggest that the alternative direct annual-average approach may be inefficient but correctly specified (see \citealp{Gneiting2007}, for other examples of such predictive densities).

\FloatBarrier

\section{Empirical applications}\label{sec:Empirical}		

This section provides three distinct empirical exercises where the copula approach is evaluated in a forecasting environment with actual macroeconomic data. Note that we do not use real-time vintages in the forecasting exercises, so all ``out-of-sample'' forecasts are strictly speaking pseudo out-of-sample forecasts.

\subsection{Large bivariate exercise}\label{sec:EmpiricalBivariate}

In this section, we provide the results of a large-scale forecasting exercise based on monthly data from FRED-MD \citep{McCracken2016}. In the construction of the forecasting environment, we closely follow \cite{McCracken2019} and consider pairs, $\left(y_{1,t},y_{2,t}\right)$, of two monthly series randomly drawn from the dataset. We first compute up to 12-months-ahead density forecasts of cumulative growth at a monthly frequency $y_{i,t+h\vert t}$, with $h=1,\dots,12$ months and $i=1,2$. Density forecasts are obtained through horizon-specific monthly-frequency autoregressive distributed lag (ARDL) direct-multistep regressions, estimated via OLS at each forecast origin. Then, we apply our proposed copula approach to the monthly density forecasts to compute up to four-quarter-ahead density forecasts of cumulative growth at a quarterly frequency $z_{i,t+h\vert t}$ up to four quarters ahead. In particular, for each for $i=\{1,2\}$, we run the following regression to obtain the monthly density forecasts:

\begin{equation}\label{eq:DMS_empapp1}
y_{i,t+h}= \alpha + \sum_{j=0}^{p-1}\beta_{j}y_{1,t-j} + \sum_{j=0}^{p-1}\gamma_{j}y_{2,t-j} + \varepsilon_{t+h}   
\end{equation}
with
\begin{equation}\label{eq:Definition_GrowthRates}
y_{i,t+h}=\left\{
\begin{array}{cc}
    Y_{i,t+h}-Y_{i,t} & \text{if~} Y_{i,t}\sim I(1) \\
    Y_{i,t+h}-Y_{i,t}-h\Delta Y_{i,t} & \text{if~} Y_{i,t}\sim I(2)
\end{array}
\right.
\end{equation}
and $Y_{i,t+h}$ denotes the levels or log-levels. Forecasts $y_{i,t+h\vert t}$ are then used to get $Y_{i,t+h\vert t}$ and the quarterly average $Z_{i,t+h\vert t}=\frac{1}{3}\sum_{j=1}^{3}Y_{i,t+h-j+1\vert t}$. Finally, the quarterly frequency transformation is obtained as:
\begin{displaymath}
z_{i,t+h\vert t}=\left\{
\begin{array}{cc}
    Z_{i,t+h\vert t}-Z_{i,t} & \text{if~} Z_{i,t}\sim I(1) \\
    Z_{i,t+h\vert t}-Z_{i,t}-h\Delta Z_{i,t} & \text{if~} Z_{i,t}\sim I(2)
\end{array}
\right.
\end{displaymath}

For the sake of simplicity, we only consider the first month of each quarter as a forecast origin. The number of lags in \eqref{eq:DMS_empapp1} is either fixed at $p=4$ or selected through the Bayesian Information Criterion (BIC) among $p\in \left\{1,\dots,12\right\}$.

We consider two samples: a full sample starting in 1974:M1 and a reduced sample starting in 1984:M1, covering only the Great Moderation period. Different samples may help to assess the impact of potential breaks in the persistence of the series over time. The sample ends in both cases in 2019:M12. The sample is partitioned into (i) an in-sample part, of size $T_\text{is}$, for the estimation of the parameters $\alpha,\beta_j$ and $\gamma_j$, (ii) a calibration part, of size $T_R = 60-h$ and spanning from 1995:M$h$ to 2004:M12, for the estimation of the copula parameters, and (iii) a forecasting part, of size $T_\text{oos} = 60-h_q$ and spanning from 2005:M$h$ to 2019:M12, for the out-of-sample evaluation in the target frequency. Note that as in \cref{sec:MonteCarlo}, $T_\text{oos}$ denotes the number of periods in the target frequency, i.e., in quarterly observations whereas $T_\text{is}$ and $T_R$ are denoted in the monthly frequency. The estimation is carried out using a rolling-window approach, updating at each forecast origin the model parameters, the PITs, and the copula parameters.

As a datasource we use the February 2023 vintage of FRED-MD. After dropping series not meeting some minimal conditions, the number of variables used in the bivariate exercise is 101, organized into 5 different groups as in \citet{Marcellino2006} and \citet{McCracken2019}: (1) income, output, sales, and capacity utilization; (2) employment and unemployment; (3) construction, inventories, and orders; (4) interest rates and asset prices; (5) nominal prices, wages, and money.\footnote{We exclude from the dataset three series starting after the 1970 (new orders for consumers goods, new orders for non-defense capital goods, trade weighted US dollar index), one series presenting missing values (consumer sentiment index), and one series switching to negative over the sample (non-borrowed reserves of depository institutions). In addition, we exclude 21 series that should be used in levels or log-levels, as we focus on series that can be expressed in first/second difference (or log-difference) in our application.}

\begin{figure}[!t]
\caption{Forecast comparison of copula vs benchmark approach} \label{fig:Emp1_RelativeScore}
\begin{subfigure}{\textwidth}
	\centering
		  \includegraphics[scale=0.45]{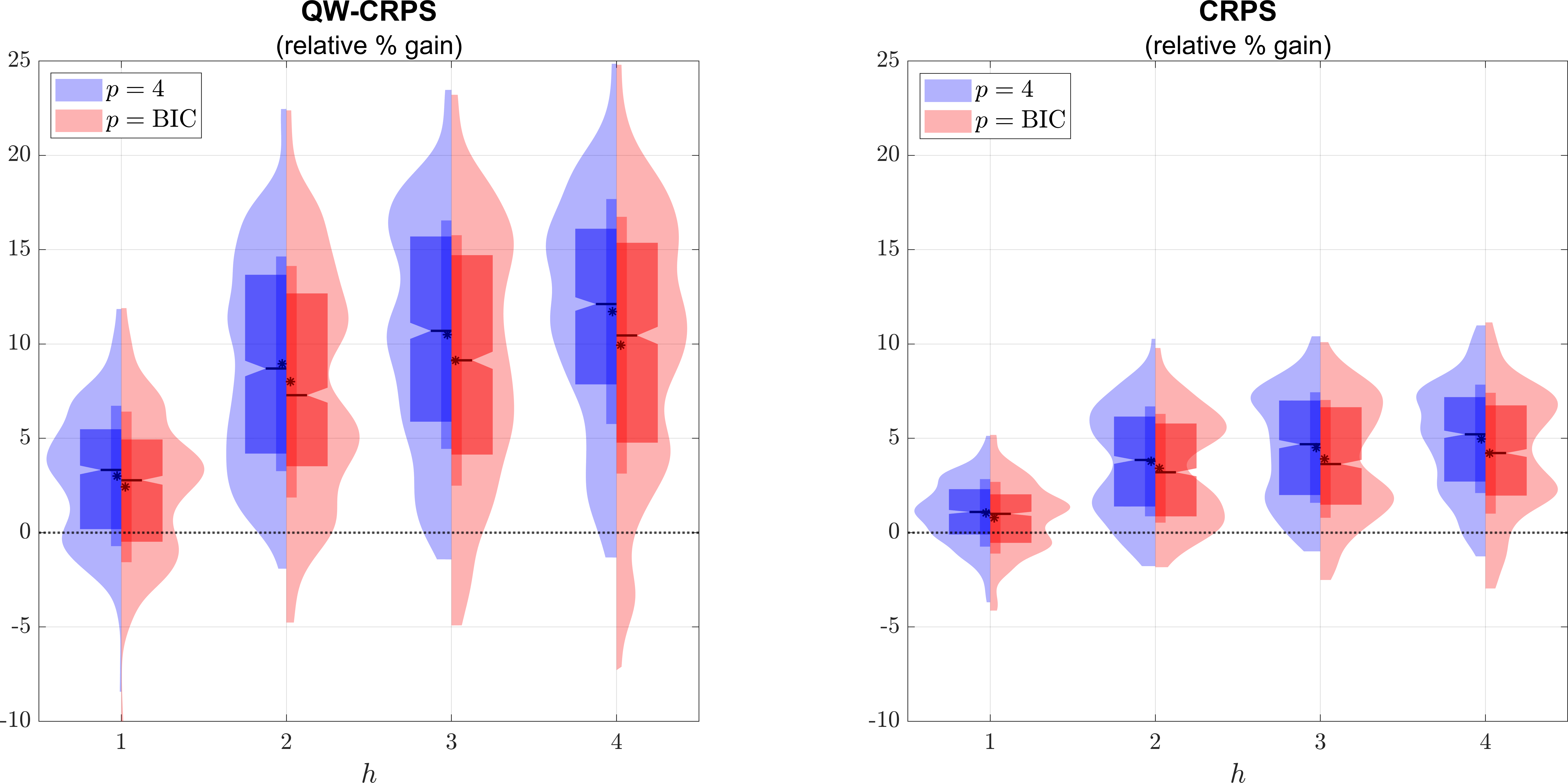}
    \vspace{0.2cm}
	\caption{Full sample (1974:M1-2019:M12)} 
\end{subfigure} 

\vspace{0.5cm}

\begin{subfigure}{\textwidth}
	\centering
		  \includegraphics[scale=0.45]{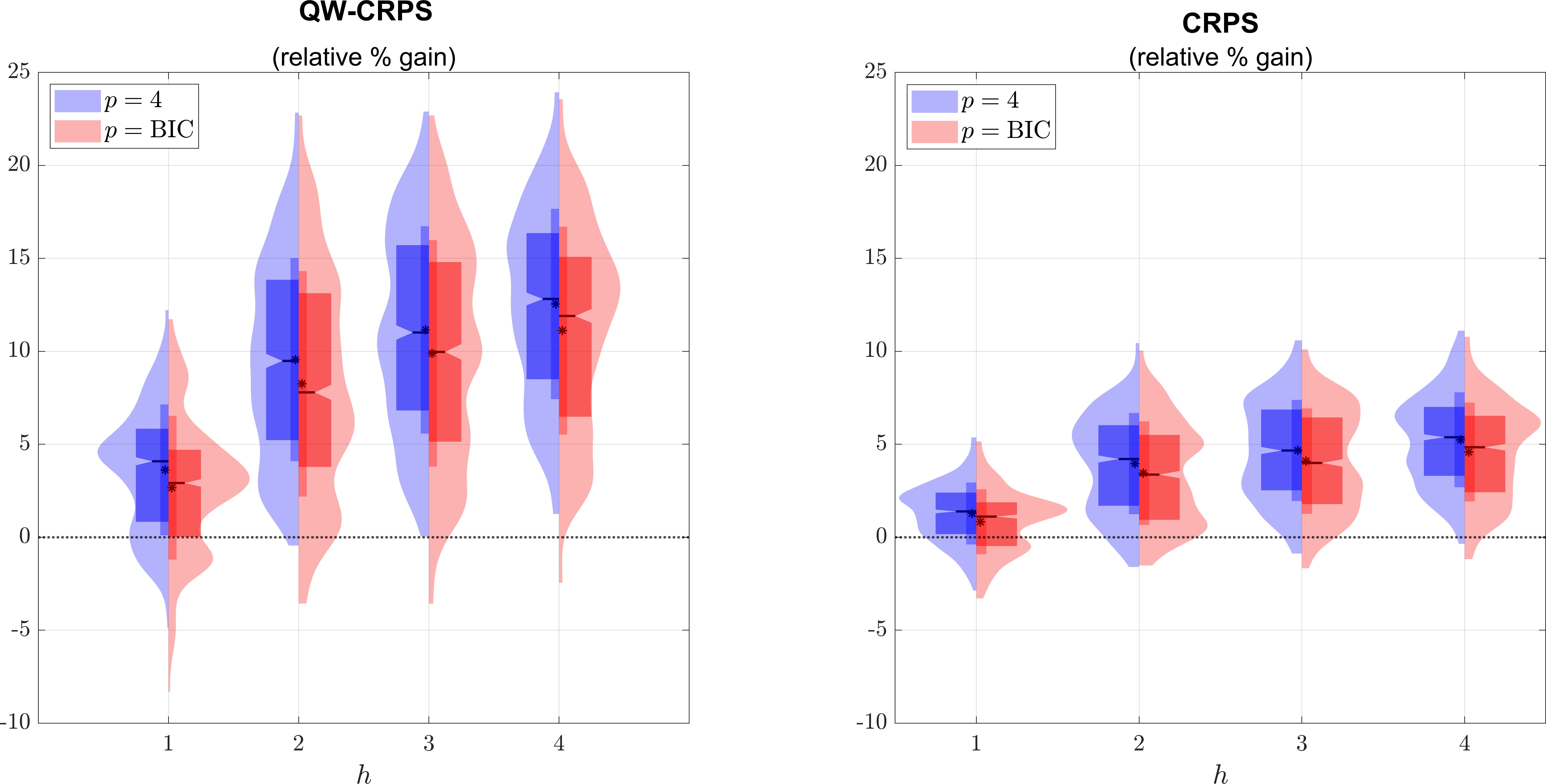}
    \vspace{0.2cm}    
	\caption{Great Moderation sample (1984:M1-2019:M12)} 
\end{subfigure}
	\caption*{\footnotesize \textit{Note}: The figure shows the distribution of the relative \% predictive gain in terms of the respective loss function of the copula approach vs the benchmark approach over the 1,300 forecasting exercises. Values larger than zero indicate positive gains for the copula approach. The $x$-axis denotes the forecast horizon in quarters.}  
\end{figure}

Then, 650 random pairs of variables are selected from the database such that $y_{1}$ and $y_{2}$ come from distinct groups and an equal number of series pairs $(y_{1}, y_{2})$ comes from each of the 10 possible group pairings. For each permutation of the series, we compute density forecasts at the quarterly target frequency, based on the monthly density forecasts, using either the copula approach or the benchmark approach, and we compare their forecasting performance using their average QW-CRPS and CRPS over the out-of-sample. For each lag selection method and forecast horizon, the 650 random pairs forming the systems $(y_{1}, y_{2})$ and $(y_{2}, y_{1})$ provide a distribution of 1,300 average score ratios.

\begin{figure}[!t]
\caption{Rejection frequencies EPA test: copula vs benchmark approach} \label{fig:Emp1_RejectFreq}
	\centering
		  \includegraphics[scale=0.55]{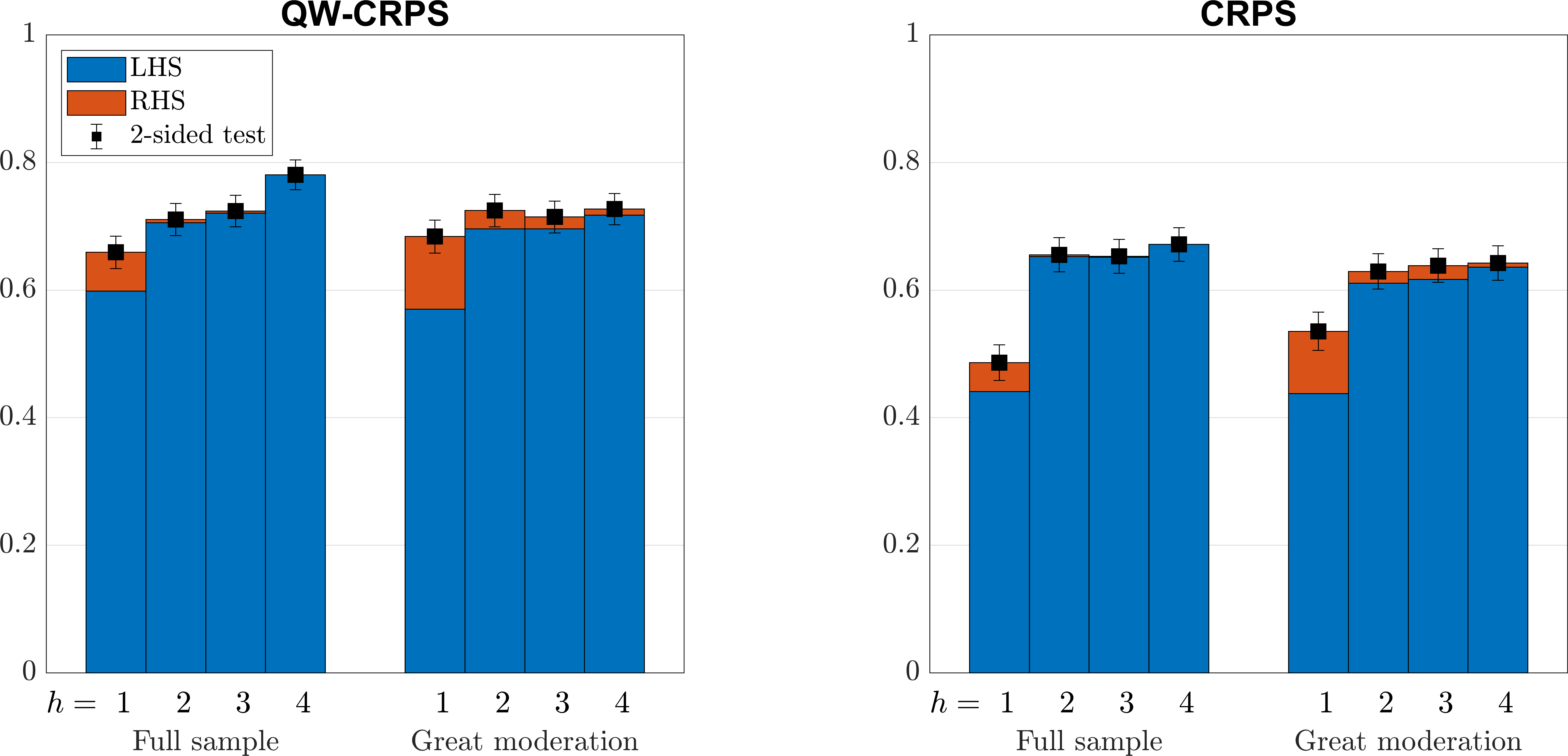}
	\vspace{0.1cm} 
    \caption*{\footnotesize \textit{Note}: The figure show the rejection frequency of the EPA tests \citep{Giacomini2006} over the 1,300 forecasting exercises (the black square) and its decomposition into the rejection frequency when the average loss of the copula is smaller than the average loss of the benchmark (the blue bar) and when the average loss of the copula is larger than the average loss of the benchmark (the red bar). $h$ denotes the forecast horizon in quarters. The labels ``Full sample'' and ``Great Moderation'' indicate the sample period used for the forecasting exercise. The nominal size is 5\%. The error bars denote $\pm 2$ standard errors computed through 1000 bootstrap replications.}  
\end{figure}

\begin{figure}[!t]
\caption{Empirical distribution of det($\widehat{R}$)} \label{fig:Distribution_detR}

\begin{subfigure}[b]{0.49\textwidth}
	\centering
		  \includegraphics[scale=0.50]{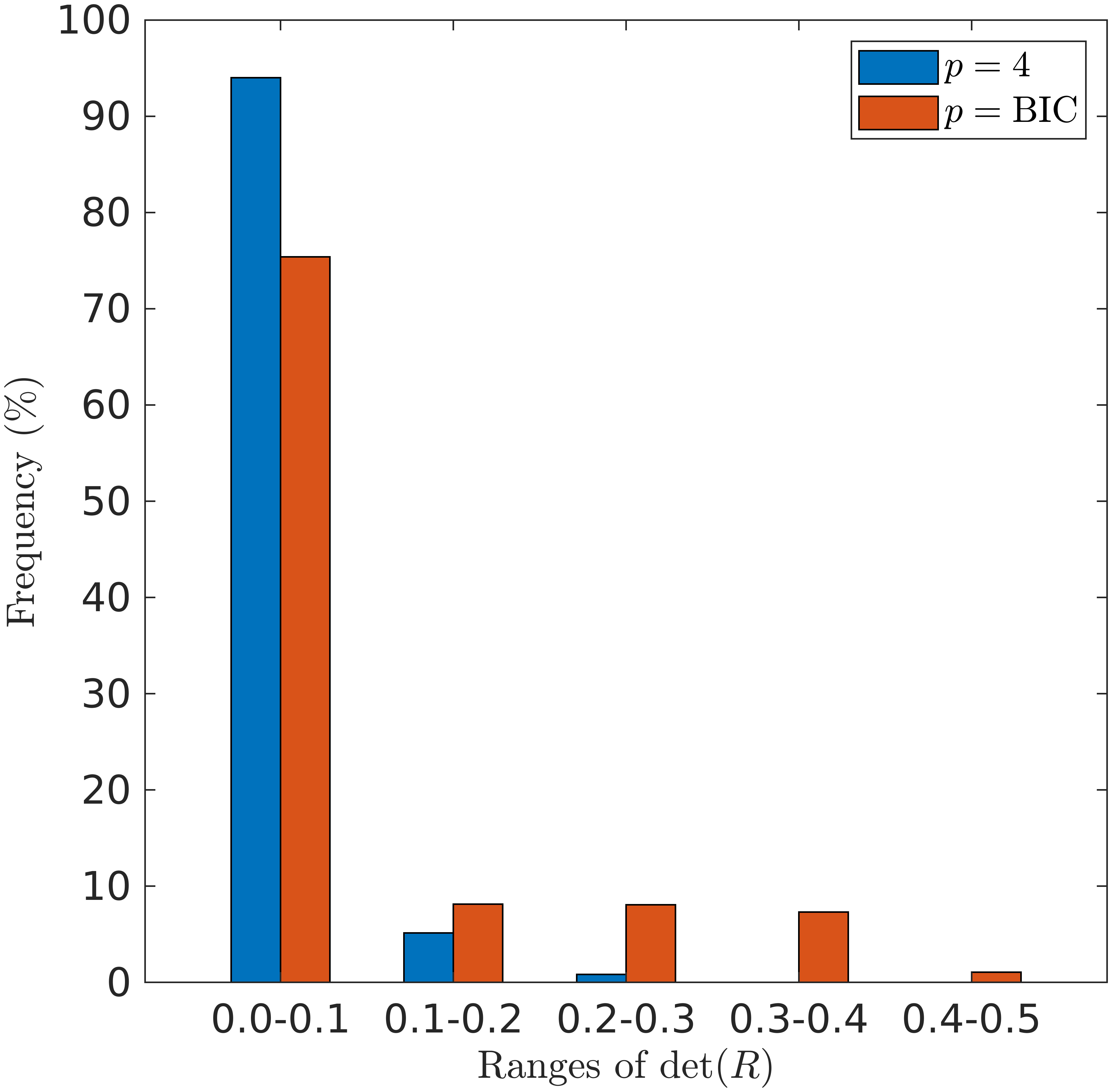}
    \vspace{0.5cm}
	\caption{Full sample (1974:M1-2019:M12)} 
\end{subfigure} 
\begin{subfigure}[b]{0.49\textwidth}
	\centering
		  \includegraphics[scale=0.50]{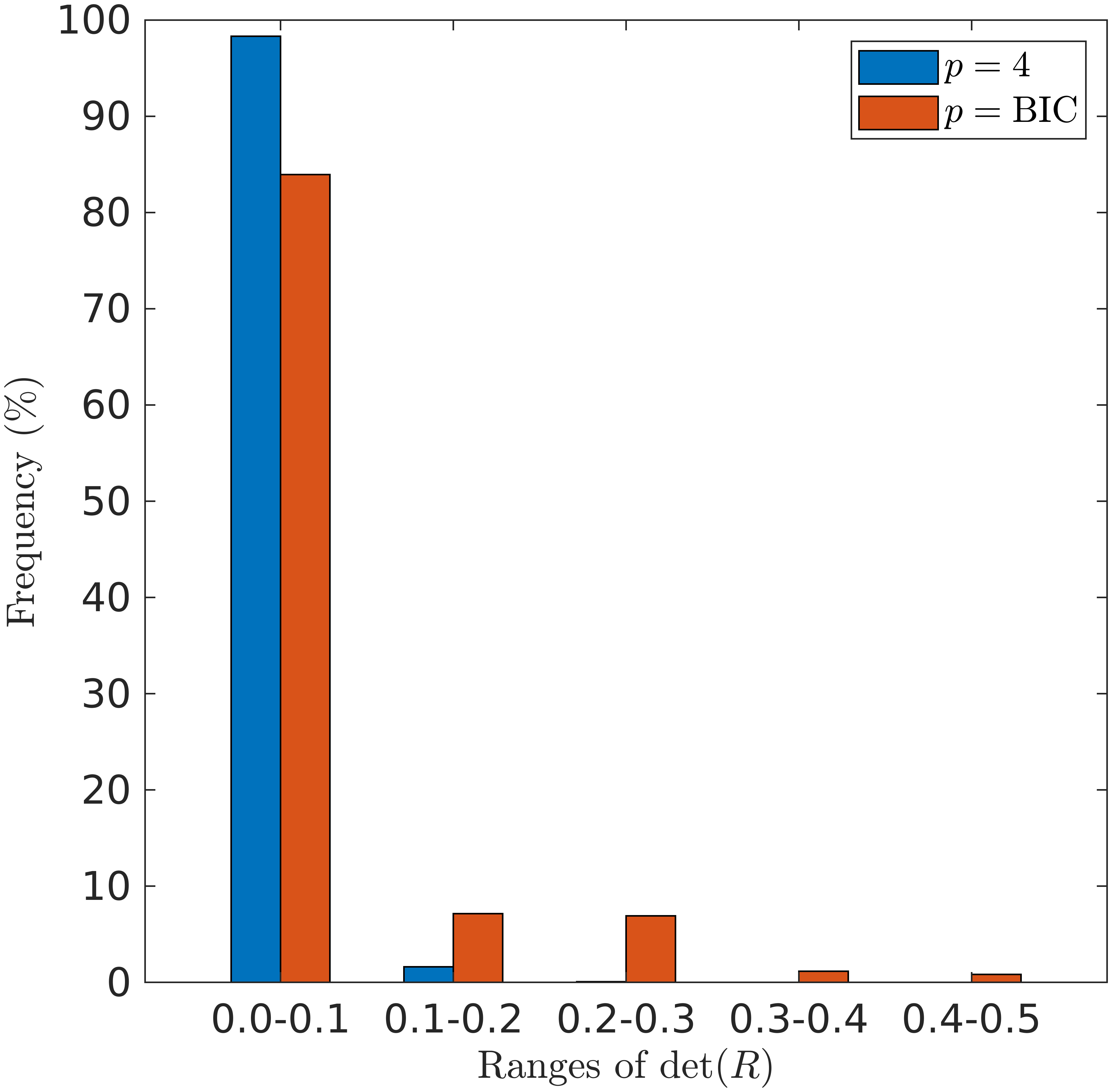}
    \vspace{0.5cm}    
	\caption{Great Moderation sample (1984:M1-2019:M12)}
\end{subfigure}
	\centering
	\caption*{\footnotesize \textit{Note}: The figure shows distribution of det($\widehat{\boldsymbol{R}}$) over the 1,300 forecasting exercises. Smaller numbers of the determinant imply a larger degree of cross-horizon persistence in the forecasts.}  
\end{figure}

\cref{fig:Emp1_RelativeScore} shows these distributions after computing the relative \% gain of the copula approach with respect to the benchmark. The labels ``Full sample'' and ``Great Moderation'' indicate the samples used for the forecasting exercise. The copula approach provides, on average, a better forecasting performance for all horizons, with the predictive gain increasing with the horizon up to around 10\% according to the QW-CRPS and around 5\% according to the CRPS, irrespective of the sample used for the estimation and the lag length $p$. It is worth noting that these empirical predictive gains appear overall statistically positive, as the 95\% confidence interval for the median and the inter-quartile range tend to exclude the zero value. To grasp the statistical significance of these results in a more formal way, we also compute for each pairing a test of unconditional equal predictive accuracy \citep{Giacomini2006}. The results are summarized in \cref{fig:Emp1_RejectFreq}, where we provide the overall rejection frequency of the null hypothesis of equal predictive accuracy and its decomposition into the rejection frequency when the alternative is that the average loss of the copula is inferior to the average loss of the benchmark (the blue bar) and the respective opposite alternative (the red bar). The rejection frequencies appear relatively high across all samples and horizons, hovering around 0.6-0.8\% for the QW-CRPS and 0.5-0.6\% for the CRPS, and it can be largely attributed to the copula approach.

\cref{fig:Distribution_detR} shows the empirical distribution of the determinant of the estimated copula parameter matrices $\widehat{\boldsymbol{R}}$. The determinant is a summary measure of the persistence of the forecasts $y_{i,t+h}$ across the forecast horizons $h$ (see \cref{sec:RobustnessSub1}). The results suggest that for a majority of the randomly selected bivariate systems, the persistence in $y_{i,t+h}$ is considerably large. This should explain the superior average forecasting performance of the copula approach relative to the benchmark approach depicted in  \cref{fig:Emp1_RelativeScore}. The large average cross-horizon persistence is, in part, driven by defining the monthly density forecasts as cumulative growth rates such that shocks are accumulated across the horizons $h$. However, predicting cumulative growth rates is not at all unusual and, in fact, the definition of the predictands in \eqref{eq:Definition_GrowthRates} is taken from \cite{McCracken2019}.

\FloatBarrier

\subsection{Inflation-at-Risk}\label{sec:EmpiricalInflation}		

In this section, we provide estimates of inflation at risk for year-on-year and annual-average inflation based on the U.S. Consumer Price Index. We use quantile regressions in combination with a Lasso, to select among many potential predictors, and produce monthly forecasts of the year-on-year inflation rate. Then, we construct predictive distributions of the (calendar year) annual-average inflation rates out of the year-on-year predictions. This forecasting environment aims at replicating the situation where a professional forecaster dispose of predictive densities for year-on-year inflation, but she  is required to transform them into lower-frequency annual-average densities. This is a pretty common operational framework across institutions such as central banks, where the communication around the expected inflation environment is usually performed around annual-average rates, even though the monthly year-on-year rate represents the operational target.\footnote{Note that, in fact, both the ECB and the Federal Reserve publish annual-average forecasts as part of their institutional projection exercises.} 
Importantly, the year-on-year and annual-average predictive densities need to be coherent, i.e., they should be based on the same predictors, model type, and the moments of the baseline frequency should be reflected in the moments of the new target frequency.

The underlying price index is the monthly and seasonally adjusted Consumer Price Index for all Urban Consumers\footnote{Results are robust to using the non-seasonally adjusted CPI for all items, which has the mnemonic USACPIALLMINMEI on FRED.} (henceforth CPI) from 1960 to 2022. The baseline model predicts monthly year-on-year inflation, computed via the log-difference of $t$ and $t-12$, via a quantile regression model that uses a Lasso to select among a number of potential predictors \citep{Belloni2011}. The quantile regression coefficient vector $\boldsymbol{\beta}_\tau$, for the $\tau$th quantile, is estimated by minimizing the following objective function: 

\begin{equation} \label{eq:BQRreg} \widehat{\boldsymbol{\beta}}_\tau =  \argmin_{\boldsymbol{\tilde{\beta}}_\tau \in \mathbb{R}^p} \ \sum_{t=t_{0}}^{T_\text{is}+t_{0}-1} \rho_\tau(y_{t+h}-\boldsymbol{x}_{t}'\boldsymbol{\tilde{\beta}}_\tau) + \frac{\lambda\sqrt{\tau(1-\tau)}}{T_\text{is}} \sum_{j=1}^p \widehat{\sigma}_j^2|\tilde{\beta}_{j,\tau}|, \end{equation}
where $y_{t+h}$ denotes the monthly year-on-year inflation rate, $\boldsymbol{x}_t$ denotes a $p \times 1 $ vector of predictors (including a constant), $\rho_\tau(z) = (\tau - \mathbb{1}\{z \leq 0\})z$, $\lambda$ is a hyperparameter that determines the degree of penalization (set here as recommended by \citealp{Belloni2011}), $p$ is the number of predictors, $h=1,...,12$ is the forecast horizon, $T_\text{is}$ is the in-sample estimation, $t_{0}=1,\dots,565$, denotes the initial observation which changes due to a rolling window estimation scheme, and $\widehat{\sigma}_j^2 = \sum_{t=t_{0}}^{T_\text{is}+t_{0}-1} x_{j,t}^2$.

Note that because not all predictors are available over the entire range of the sample, not all predictors enter the model in all forecast origins (see \cref{app:InflationatRisk} for details on the data series). Therefore, the predictor vector $x_t$ contains at most 23 exogeneous predictors, similar to the predictors used in \cite{Korobilis2017}, as well as two lags of the endogenous variable and the constant.\footnote{Results are robust to increasing the number of lags to six.}

We use the model in \eqref{eq:BQRreg} to produce out-of-sample forecasts for horizons $h=1,...,12$ months ahead, with the first forecast origin being 1974:M12 and the last forecast origin being 2021:M12. At each forecast origin, the quantile regression is re-estimated over a rolling window of 15 years of monthly data, that is, $T_\text{is} = 180$. For example, for the forecast origin 1974:M12, the first observation used in the model estimation is 1960:M1.
We let the model produce quantile predictions for $\tau \in [0.01,0.02,...,0.98,0.99]$ and we calculate the predictive CDF based on linearly interpolating between adjacent quantiles (see \citealp{Mitchell2024}, for a similar approach; details on the interpolation are provided in \cref{app:InflationatRisk}).\footnote{Quantile crossing is dealt with by sorting the predictive quantiles.}
Finally, the generated year-on-year predictive densities are then transformed into annual-average inflation densities by drawing from the joint distribution of the year-on-year forecasts using either the copula or the benchmark approach. In order to approximate the annual-average, we first compute month-on-month rates from the year-on-year forecasts, and then we transform the month-on-month predictions into annual-averages using the approximation of \cite{Mariano2003}.

For the estimation of the copula parameter matrix $\boldsymbol{R}$, we use about 10 years of data over a rolling window starting with the observations forecast origins 1974:M12 to 1984:M12, that is, $T_R=121$. More precisely, we start by evaluating the empirical PITs of the predictive distributions of monthly year-on-year inflation using observations 1975:M$h$ to 1985:M$h$ for forecast horizon $h=1,...,12$. We then estimate the copula parameter based on the empirical PITs and we use the estimated copula parameters in combination with the forecasts for 1985:M$h$, $h=1,...,12$, to construct an annual-average inflation predictive distribution for 1986. Next, we use the predictive distributions for year-on-year inflation from 1976:M$h$ to 1986:M$h$ for horizon $h=1,...,12$, and re-do the steps to construct the annual-average inflation predictive distribution for 1987. We then repeat this until we have an annual-average forecast for each year from 1986 to 2022. In addition, we construct annual-average forecasts as described above but using the benchmark approach, which is identical to setting all off-diagonal elements of $\boldsymbol{R}$ equal to zero. Therefore, the out-of-sample evaluation period for annual-average inflation forecasts spans from 1986 to 2022, leading to 37 out-of-sample (calendar year) annual-average predictive distributions, that is, $T_{\text{oos}} = 37$.

The results can be summarized as follows. Regarding the performance at the tails of the predictive distributions, the QW-CRPS ratio of the copula approach relative to the benchmark is 0.79, i.e., the copula approach provides a substantial predictive gain of about 21\%. Similarly, the ratio of the QS(10\%) and QS(90\%) is respectively 0.72 and 0.85, i.e., the copula approach provides a relative gain of 28\% and 15\% for predicting the risks of low and high inflation, respectively. When evaluating the entire density forecasts, the CRPS ratio of the copula approach relative to the benchmark is 0.91, i.e., the copula approach provides a relative predictive gain of 9\%. All these results are statistically significant at the 1\% level according to the test for unconditional equal predictive ability of \cite{Giacomini2006}.

\begin{figure}[htbp]
	\caption{annual-average predictive densities for inflation}\label{fig:IatRisk}
\graphicspath{{Figures/}}
\begin{subfigure}[b]{0.49\textwidth}
	\centering
		\includegraphics[width=\linewidth,keepaspectratio]{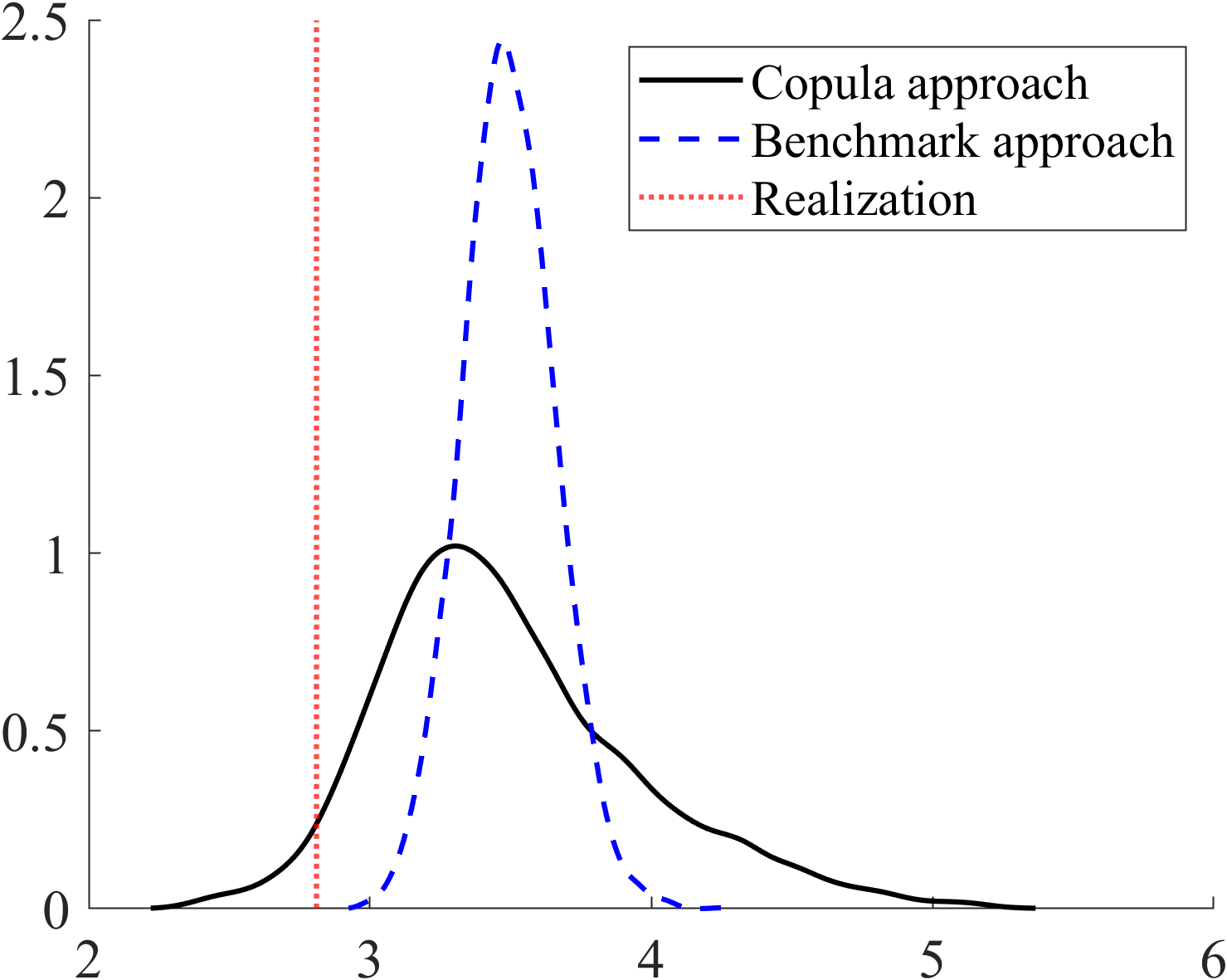}
	\caption{For 2001}\label{fig:IatRisk2000}
\end{subfigure}
\begin{subfigure}[b]{0.49\textwidth}
	\centering
		\includegraphics[width=\linewidth,keepaspectratio]{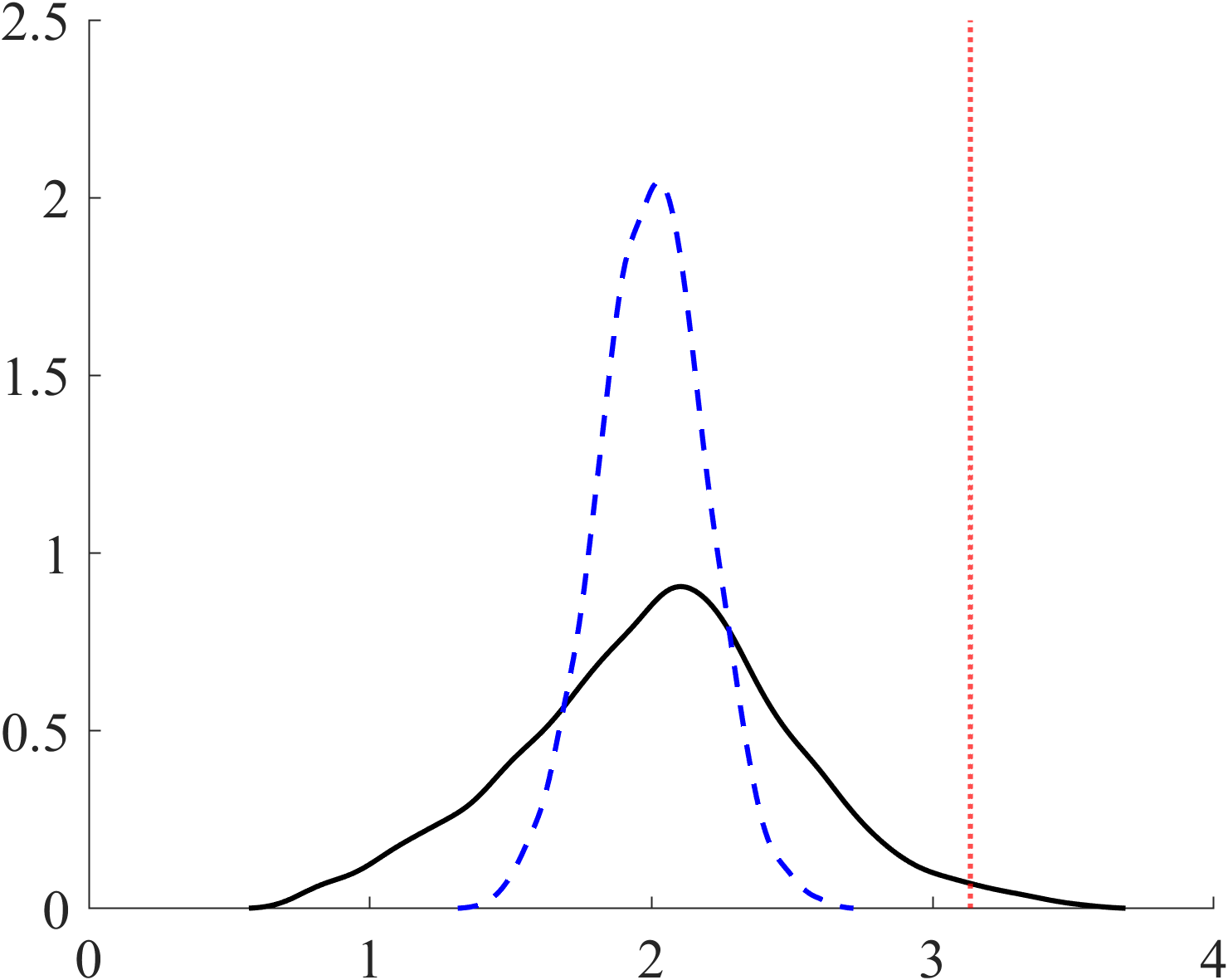}
	\caption{For 2011}\label{fig:IatRisk2010}
\end{subfigure}

	\caption*{\footnotesize \textit{Note}: The figure shows the results for the annual-average inflation predictive density for 2001, panel (a), and for 2011, panel (b), for both the copula approach (solid line) as well as the benchmark approach (dashed line), alongside the realization (dotted line).}
\end{figure}

\cref{fig:IatRisk} illustrates the superior predictive ability of the copula approach. Panel (a) shows the predictive density of the annual-average inflation for 2001 and panel (b) shows the predictive density of the annual-average inflation for 2011. In both cases, the benchmark approach assigns ex-ante zero-probability to the realizations, whereas the copula-based density forecasts can better predict tail-risks. In particular, the latter displays fatter tails as a result of the high correlation between adjacent months' year-on-year growth rates. It is also worth noting that the predictive distributions of the copula-based approach exhibit notably stronger asymmetry.

\FloatBarrier

\subsection{Growth-at-Risk}\label{sec:EmpiricalGrowth}		

In a recent influential contribution, \cite{Adrian2019} use quantile regressions to produce \textit{direct} density forecasts of quarterly quarter-on-quarter U.S. real GDP growth and show that financial conditions, captured by the National Financial Conditions Index (NFCI), are an important predictor for downside risks to GDP growth. We apply our methodology to transform their quarter-on-quarter density forecasts into annual-average densities. 

The total data sample of \cite{Adrian2019} ranges from 1973:Q1 to 2015:Q4 and, starting with forecast origin 1993:Q1, the authors produce (pseudo) out-of-sample forecasts using quantile regressions for up to four quarters ahead.\footnote{The predictors of their preferred specification are one lag of both the endogenous variable (the real GDP growth) and the NFCI. For details on the exact specification and variable definitions, see \cite{Adrian2019}.} Thus, using their exact specification leaves us with a series of out-of-sample forecasts from 1993:Q$h$ to 2015:Q4, for forecast horizons $h=1,\dots,4$. 

To construct annual-average predictive distributions, we proceed as follows. We first evaluate the predictive distributions for the forecast targets 1993:Q$h$ to 2001:Q$h$, with $h=1,\dots,4$, and we obtain a series of 32 empirical PITs for each forecast horizon $h=1,\dots,4$. From the PITs, we calculate the copula parameter as described in \cref{sec:Method}. We then compute the annual-average predictive distribution for 2002 from the quarter-on-quarter densities for horizon $h=1,\dots,4$, with origin 2001:Q4 (and forecasting targets 2002:Q1 to 2002:Q4). Then, we move four quarters ahead and repeat the exercise to construct the annual-average predictive distribution for 2003. In total, we repeat this algorithm until we have a series of 14 annual-average predictive distributions spanning from 2002 to 2015. The obtained annual-average predictive distributions are hence based on the original quarterly predictive distributions of \cite{Adrian2019} and takes into account the serial correlation across the quarterly growth rates. The benchmark annual-average predictive distributions are constructed similarly. 

\begin{figure}[htbp]
\graphicspath{{./}{Figures/}} 
	\centering
	\caption{annual-average predictive densities for 2008}\label{fig:Adrian2008}
		\includegraphics[width=0.6\linewidth,keepaspectratio]{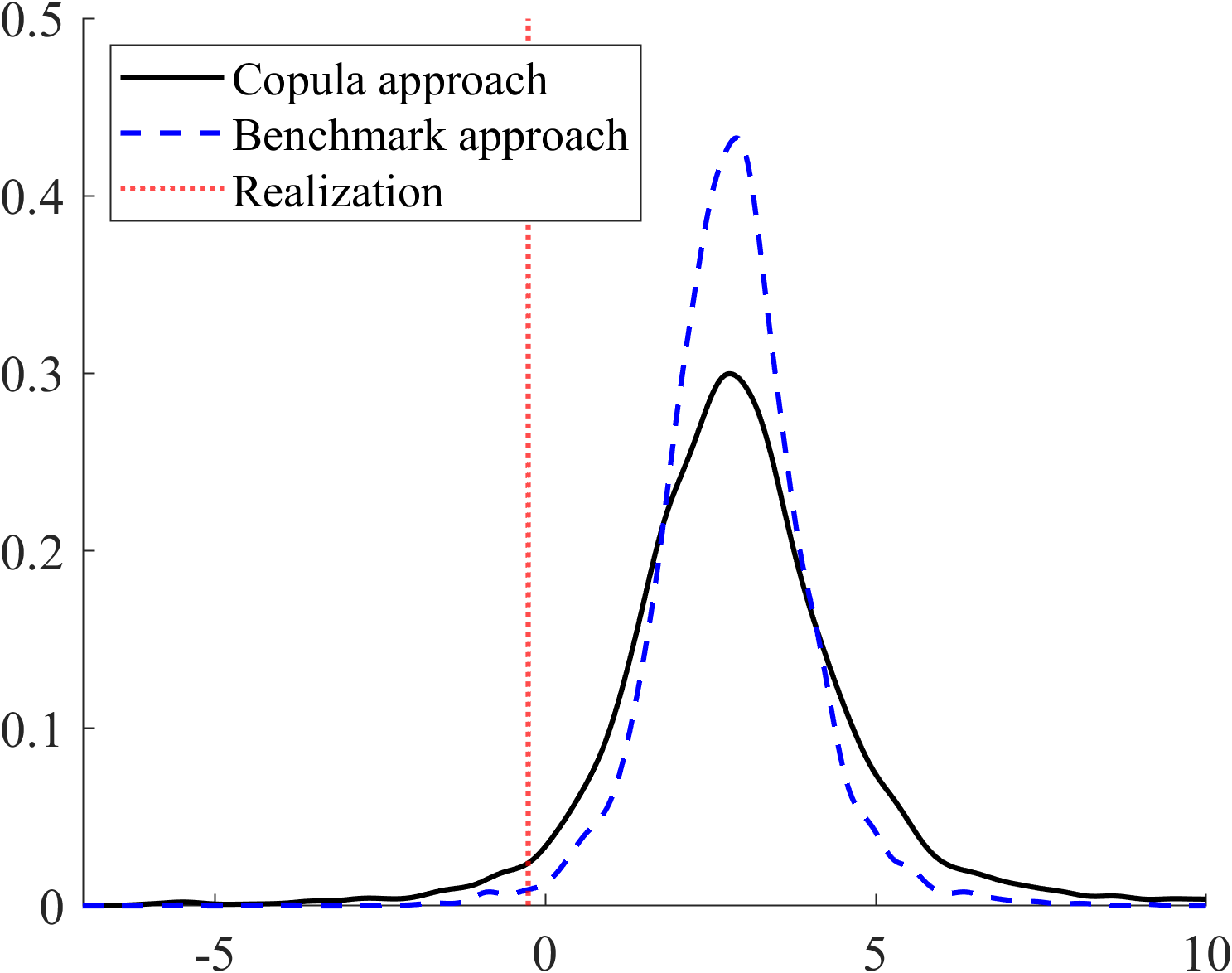}
	\caption*{\footnotesize \textit{Note}: The figure shows the results for the annual-average predictive density of 2008 for both the copula approach (solid line) as well as the benchmark approach (dashed line), alongside the realization (dotted line).}
\end{figure}

\cref{fig:Adrian2008} shows the results for the annual-average predictive density of 2008 for both the copula approach (solid line) and the benchmark approach (dashed line), alongside the realized annual-average growth rate (dotted line). The copula approach leads to annual-average forecasts with larger tails due to the positive correlation between the quarterly growth rates. Indeed, the rank correlation, i.e., the elements in the Gaussian copula correlation matrix, is around 50\% to 60\% for adjacent quarters, depending on the forecast horizon. Importantly, the larger tails of the copula approach help to assess the downside risk to the annual-average real GDP growth for 2008, i.e., during the onset of the financial crisis. This underlines the usefulness of our approach and provides anecdotal evidence of incorrect risk assessment if the serial correlation between the original predictive distributions is not taken into account when constructing multi-horizon objects.

Turning to the full evaluation sample, the QW-CRPS and CRPS ratios point to a predictive for the copula approach with respect to the benchmark of about 2.3\% and 2.5\%, respectively. However, the QS(10\%) ratio points to a substantial predictive gain of about 25\% for the copula approach. In other words, even though the sample is too short for formal testing procedures, the empirical analysis suggests that the copula approach may provides a considerable improvement for predicting lower tail risks on activity, which are carefully and closely monitored by policy-makers.

\FloatBarrier

\section{Conclusion}\label{sec:Conclusion}		

In this work, we propose a method to combine \textit{direct} forecasts to obtain new predictive objects that are function of several horizons. The approach is useful in a situation where the forecaster has a set of \textit{direct} forecasts available and has to use the same set of \textit{direct} predictive densities to construct a new predictive object; for instance, if the forecasters already has a set of \textit{direct} quarter-on-quarter growth rates but also needs annual-average predictions. These type of situations typically arise, but are not limited to, in institutions where the forecasting process is rigid.

In a Monte Carlo exercise, we show that our methodology outperforms the benchmark approach in terms of forecasting performance whenever the serial correlation across different forecasting horizons is close to zero. In terms of the absolute forecasting performance the copula approach provides density forecasts that largely pass a correct specification test based on evaluating the uniformity of the PIT, whereas the benchmark approach fails to pass this test whenever the serial correlation is not close to zero.

In a first empirical application, we investigate in a large-scale forecasting exercise, based on monthly data from FRED-MD \citep{McCracken2016}, the performance of our methodology for a large number of outcome variable and predictor combinations. In this exercise, we transform month-on-month predictive densities to quarter-on-quarter density forecasts through our proposed copula approach and results show that copula approach outperforms the benchmark approach for the majority of outcome variable and predictor combinations.

In the second empirical application, we show the usefulness of the approach by transforming year-on-year predictive densities for inflation into annual-average predictive densities. The copula approach significantly outperforms the benchmark approach both in terms of the CRPS and quantile tick loss evaluation. 

In the third empirical application, we transform the quarter-on-quarter \textit{direct} forecasts of U.S. real GDP growth of \cite{Adrian2019} into annual-average forecasts, and provide anecdotal evidence that the copula approach provides a better forecasts of the growth at risk during the Great Recession period.

\clearpage
\bibliographystyle{apalike}
\bibliography{library}

\begin{thebibliography}{}

\bibitem[Adrian et~al., 2019]{Adrian2019}
Adrian, T., Boyarchenko, N., and Giannone, D. (2019).
\newblock Vulnerable growth.
\newblock {\em American Economic Review}, 109(4):1263--89.

\bibitem[Azzalini and Capitanio, 2003]{Azzalini2003}
Azzalini, A. and Capitanio, A. (2003).
\newblock Distributions {{Generated}} by {{Perturbation}} of {{Symmetry}} with
  {{Emphasis}} on a {{Multivariate Skew}} t-{{Distribution}}.
\newblock {\em Journal of the Royal Statistical Society. Series B (Statistical
  Methodology)}, 65(2):367--389.

\bibitem[Belloni and Chernozhukov, 2011]{Belloni2011}
Belloni, A. and Chernozhukov, V. (2011).
\newblock L1-penalized quantile regression in high-dimensional sparse models.
\newblock {\em The Annals of Statistics}, 39(1):82--130.

\bibitem[Clark et~al., 2020]{Clark2020}
Clark, T.~E., McCracken, M.~W., and Mertens, E. (2020).
\newblock {Modeling Time-Varying Uncertainty of Multiple-Horizon Forecast
  Errors}.
\newblock {\em The Review of Economics and Statistics}, 102(1):17--33.

\bibitem[Fan and Patton, 2014]{Patton2014}
Fan, Y. and Patton, A.~J. (2014).
\newblock Copulas in econometrics.
\newblock {\em Annual Review of Economics}, 6:179--200.

\bibitem[Ferrara et~al., 2022]{Ferrara2022}
Ferrara, L., Mogliani, M., and Sahuc, J.-G. (2022).
\newblock High-frequency monitoring of growth at risk.
\newblock {\em International Journal of Forecasting}, 38(2):582--595.

\bibitem[Ganics et~al., 2024]{Ganics2024}
Ganics, G., Rossi, B., and Sekhposyan, T. (2024).
\newblock From fixed-event to fixed-horizon density forecasts: Obtaining
  measures of multihorizon uncertainty from survey density forecasts.
\newblock {\em Journal of Money, Credit and Banking}, 56(7):1675--1704.

\bibitem[Giacomini and Komunjer, 2005]{Giacomini2005}
Giacomini, R. and Komunjer, I. (2005).
\newblock Evaluation and {{Combination of Conditional Quantile Forecasts}}.
\newblock {\em Journal of Business \& Economic Statistics}, 23(4):416--431.

\bibitem[Giacomini and White, 2006]{Giacomini2006}
Giacomini, R. and White, H. (2006).
\newblock Tests of {{Conditional Predictive Ability}}.
\newblock {\em Econometrica}, 74(6):1545--1578.

\bibitem[Gneiting et~al., 2007]{Gneiting2007}
Gneiting, T., Balabdaoui, F., and Raftery, A.~E. (2007).
\newblock Probabilistic forecasts, calibration and sharpness.
\newblock {\em Journal of the Royal Statistical Society: Series B (Statistical
  Methodology)}, 69(2):243--268.

\bibitem[Gneiting and Ranjan, 2011]{Gneiting2011}
Gneiting, T. and Ranjan, R. (2011).
\newblock Comparing density forecasts using threshold- and quantile-weighted
  scoring rules.
\newblock {\em Journal of Business \& Economic Statistics}, 29(3):411--422.

\bibitem[Grothe et~al., 2023]{Grothe2023}
Grothe, O., K\"{a}chele, F., and Kr\"{u}ger, F. (2023).
\newblock From point forecasts to multivariate probabilistic forecasts: The
  {Schaake} shuffle for day-ahead electricity price forecasting.
\newblock {\em Energy Economics}, 120.

\bibitem[Hafner and Manner, 2012]{Hafner2012}
Hafner, C.~M. and Manner, H. (2012).
\newblock Dynamic stochastic copula models: Estimation, inference and
  applications.
\newblock {\em Journal of Applied Econometrics}, 27:269--295.

\bibitem[Korobilis, 2017]{Korobilis2017}
Korobilis, D. (2017).
\newblock Quantile regression forecasts of inflation under model uncertainty.
\newblock {\em International Journal of Forecasting}, 33(1):11--20.

\bibitem[Marcellino et~al., 2006]{Marcellino2006}
Marcellino, M., Stock, J.~H., and Watson, M.~W. (2006).
\newblock A comparison of direct and iterated multistep {{AR}} methods for
  forecasting macroeconomic time series.
\newblock {\em Journal of Econometrics}, 135(1-2):499--526.

\bibitem[Mariano and Murasawa, 2003]{Mariano2003}
Mariano, R.~S. and Murasawa, Y. (2003).
\newblock A new coincident index of business cycles based on monthly and
  quarterly series.
\newblock {\em Journal of Applied Econometrics}, 18(4):427--443.

\bibitem[McCracken and McGillicuddy, 2019]{McCracken2019}
McCracken, M.~W. and McGillicuddy, J.~T. (2019).
\newblock An empirical investigation of direct and iterated multistep
  conditional forecasts.
\newblock {\em Journal of Applied Econometrics}, 34(2):181--204.

\bibitem[McCracken and Ng, 2016]{McCracken2016}
McCracken, M.~W. and Ng, S. (2016).
\newblock {{FRED}}-{{MD}}: {{A Monthly Database}} for {{Macroeconomic
  Research}}.
\newblock {\em Journal of Business \& Economic Statistics}, 34(4):574--589.

\bibitem[Mitchell et~al., 2024]{Mitchell2024}
Mitchell, J., Poon, A., and Zhu, D. (2024).
\newblock Constructing density forecasts from quantile regressions:
  Multimodality in macrofinancial dynamics.
\newblock {\em Journal of Applied Econometrics}, 39(5):790--812.

\bibitem[Nelsen, 2006]{Nelsen2006}
Nelsen, R.~B. (2006).
\newblock An introduction to copulas.
\newblock {\em Springer Series in Statistics}.

\bibitem[Patton, 2006]{Patton2006}
Patton, A.~J. (2006).
\newblock Modelling asymmetric exchange rate dependence.
\newblock {\em International Economic Review}, 47:527--556.

\bibitem[Rossi and Sekhposyan, 2019]{Rossi2019}
Rossi, B. and Sekhposyan, T. (2019).
\newblock Alternative tests for correct specification of conditional predictive
  densities.
\newblock {\em Journal of Econometrics}, 208(2):638--657.

\bibitem[Sklar, 1959]{Sklar1959}
Sklar, A. (1959).
\newblock Fonctions de repartition a n dimensions et leurs marges.
\newblock {\em Publ. Inst. Statist. Univ. Paris (in French)}, 8:229--231.

\bibitem[Smith and Vahey, 2016]{Smith2016}
Smith, M.~S. and Vahey, S.~P. (2016).
\newblock Asymmetric forecast densities for u.s. macroeconomic variables from a
  gaussian copula model of cross-sectional and serial dependence.
\newblock {\em Journal of Business \& Economic Statistics}, 34(3):416--434.

\end{thebibliography}

\FloatBarrier
\begin{appendices}

\section*{Appendix}\label{sec:AppMain}
\crefalias{section}{appsec}
\section{Additional results and data}

\counterwithin{figure}{section}
\counterwithin{table}{section}
\counterwithin{equation}{section}
\crefalias{section}{appsec}

\setcounter{figure}{0}
\setcounter{table}{0}
\setcounter{equation}{0}

\subsection{Additional simulation results}

\begin{figure}[H]
\graphicspath{{./}{Figures/}} 
	\centering
	\caption{Scores for density forecasts with negative autoregressive coefficients: "dependence-attentive" \textit{vs} "dependence-inattentive" approach}\label{fig:Scores_AR_neg}
		\includegraphics[scale=0.45]{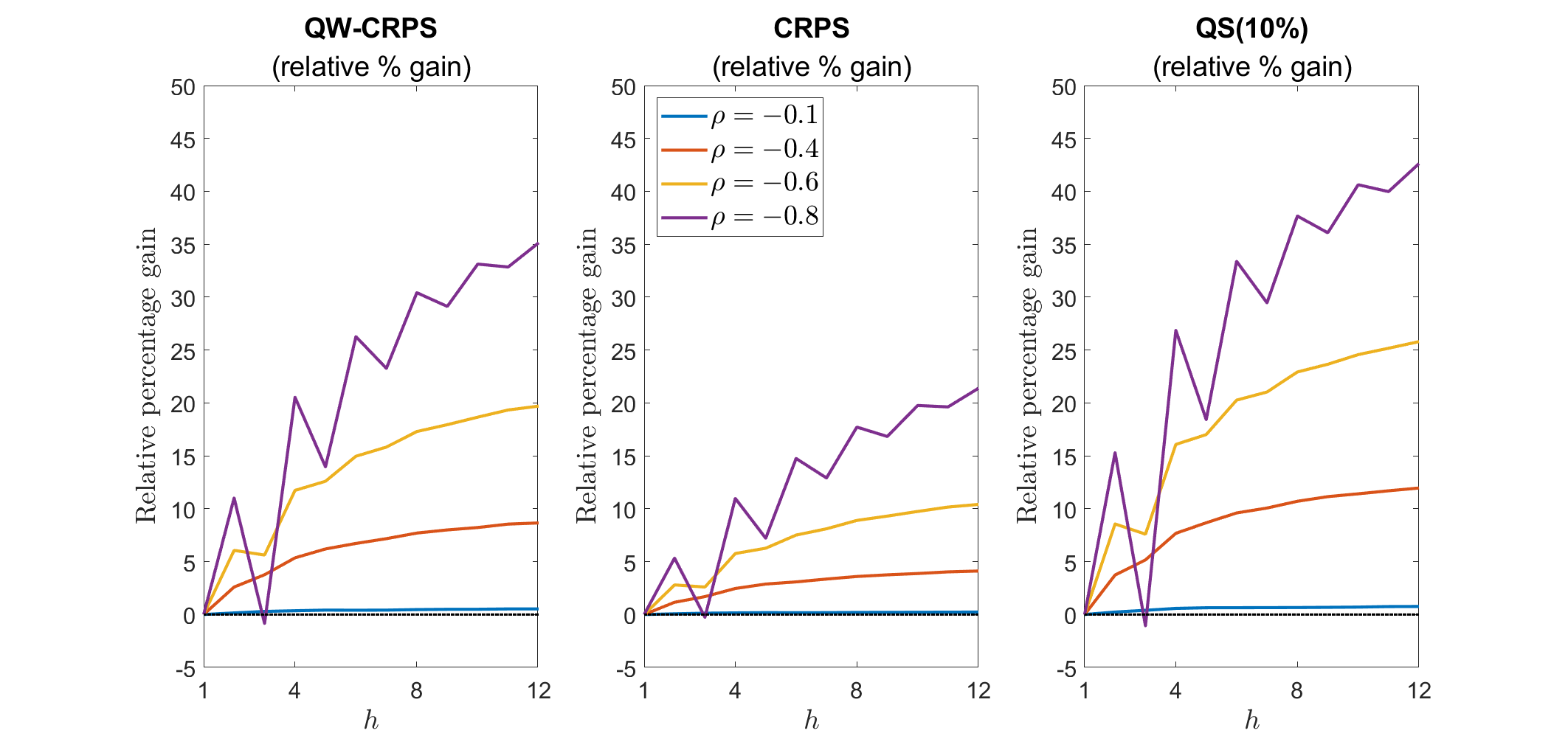}
    \vspace{0.4cm}
	\caption*{\footnotesize \textit{Note}: 
   QW-CRPS denotes the quantile weighted versions of the continuous ranked probability score, with emphasis on the tails. CRPS denotes the continuous ranked probability score. QS(10\%) denotes the quantile score at the 10\% quantile. QW-CRPS, CRPS, and QS are expressed in relative \% gain of the "dependence-attentive" forecaster with respect to the "dependence-inattentive" forecaster.}
\end{figure}

\begin{figure}[H]
\caption{Monte Carlo results for QS(10\%): relative scores and EPA tests} \label{fig:MC_TickLoss}
\begin{subfigure}[b]{0.49\textwidth}
	\centering
		  \includegraphics[width=\linewidth]{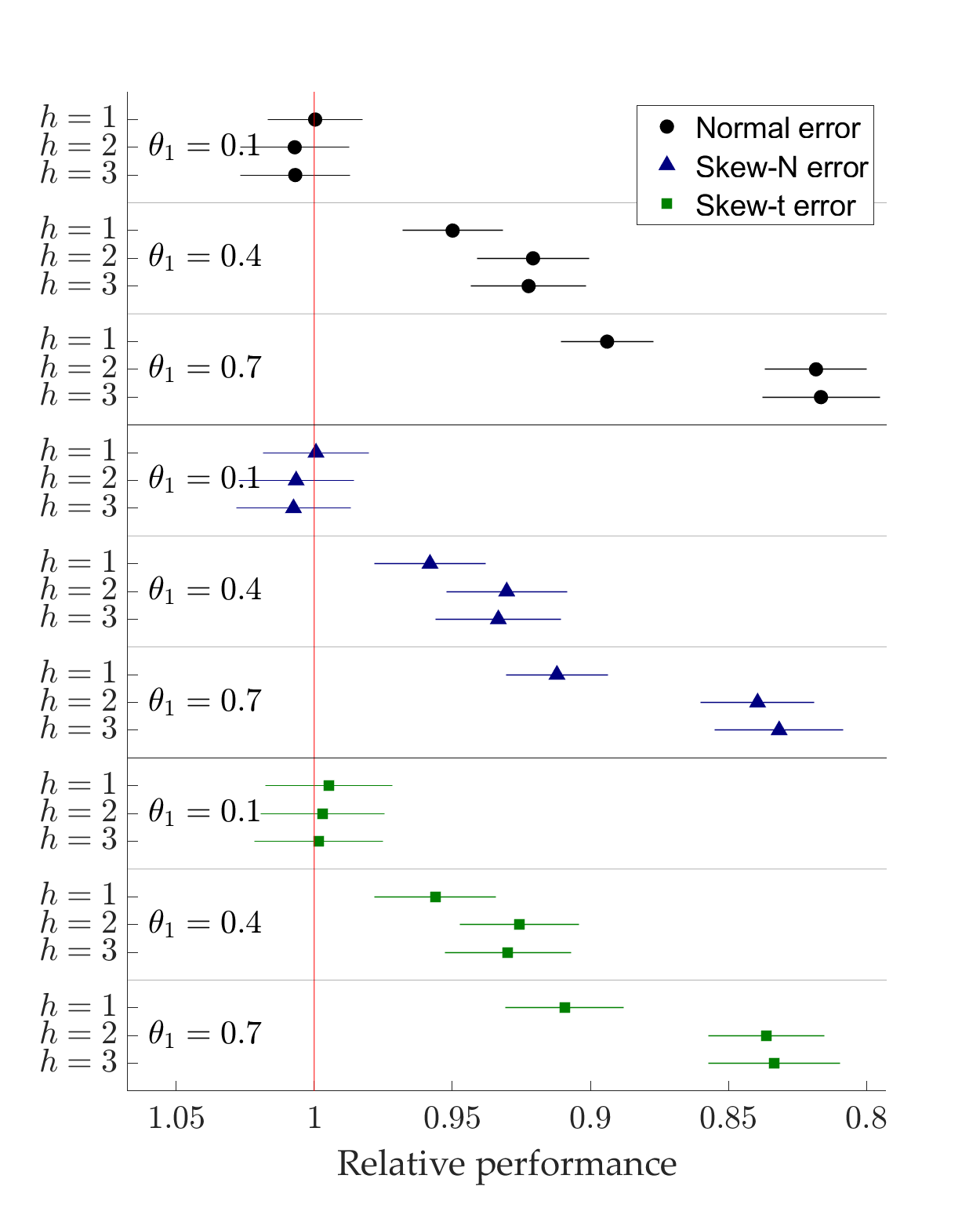}
    \vspace{-1cm}
	\caption{Relative score: annual-average} 
\end{subfigure} 
\begin{subfigure}[b]{0.49\textwidth}
	\centering
		  \includegraphics[width=\linewidth]{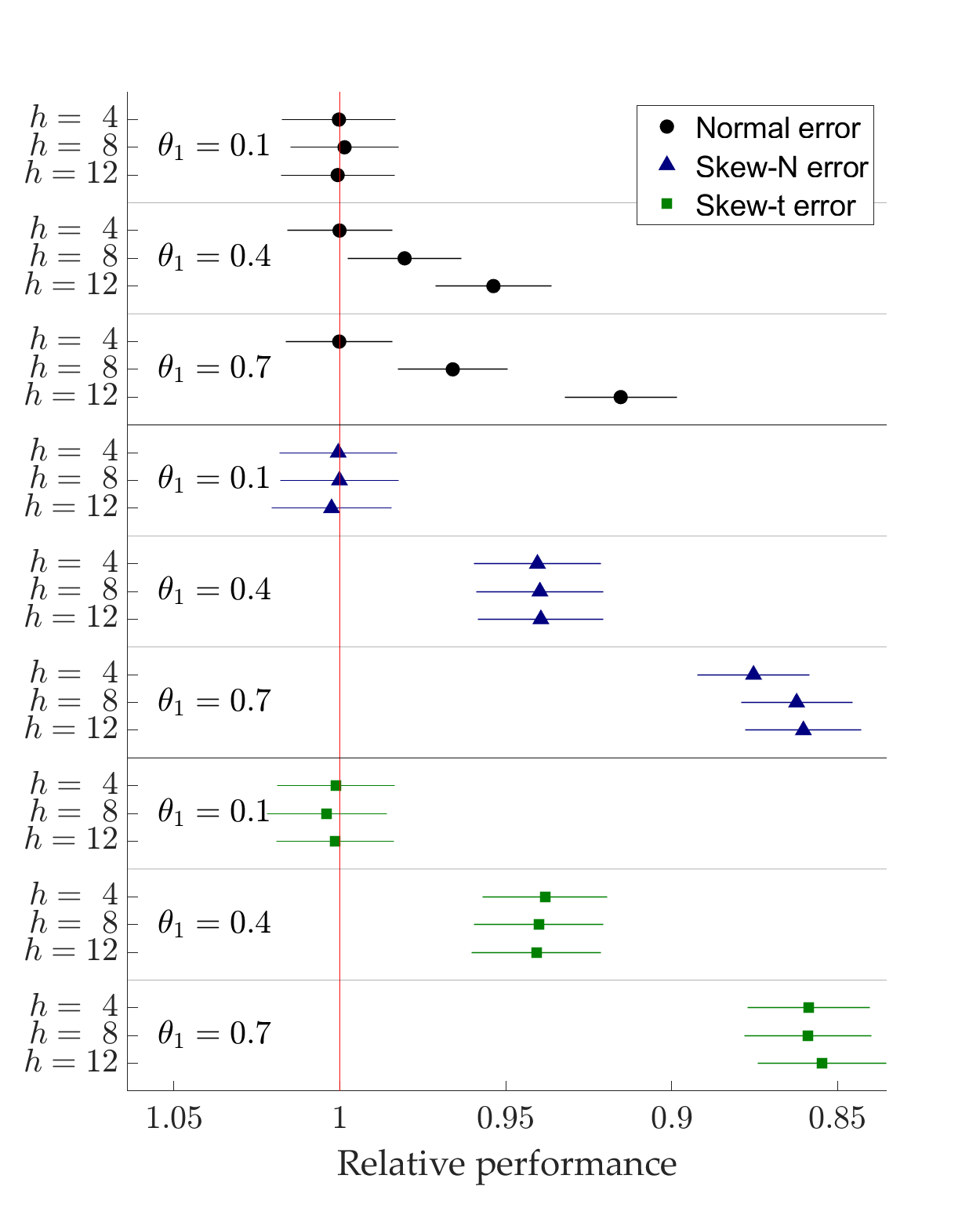}
    \vspace{-1cm}    
	\caption{Relative score: y-o-y} 
\end{subfigure}
\begin{subfigure}[b]{0.49\textwidth}
	\centering
		  \includegraphics[width=\linewidth]{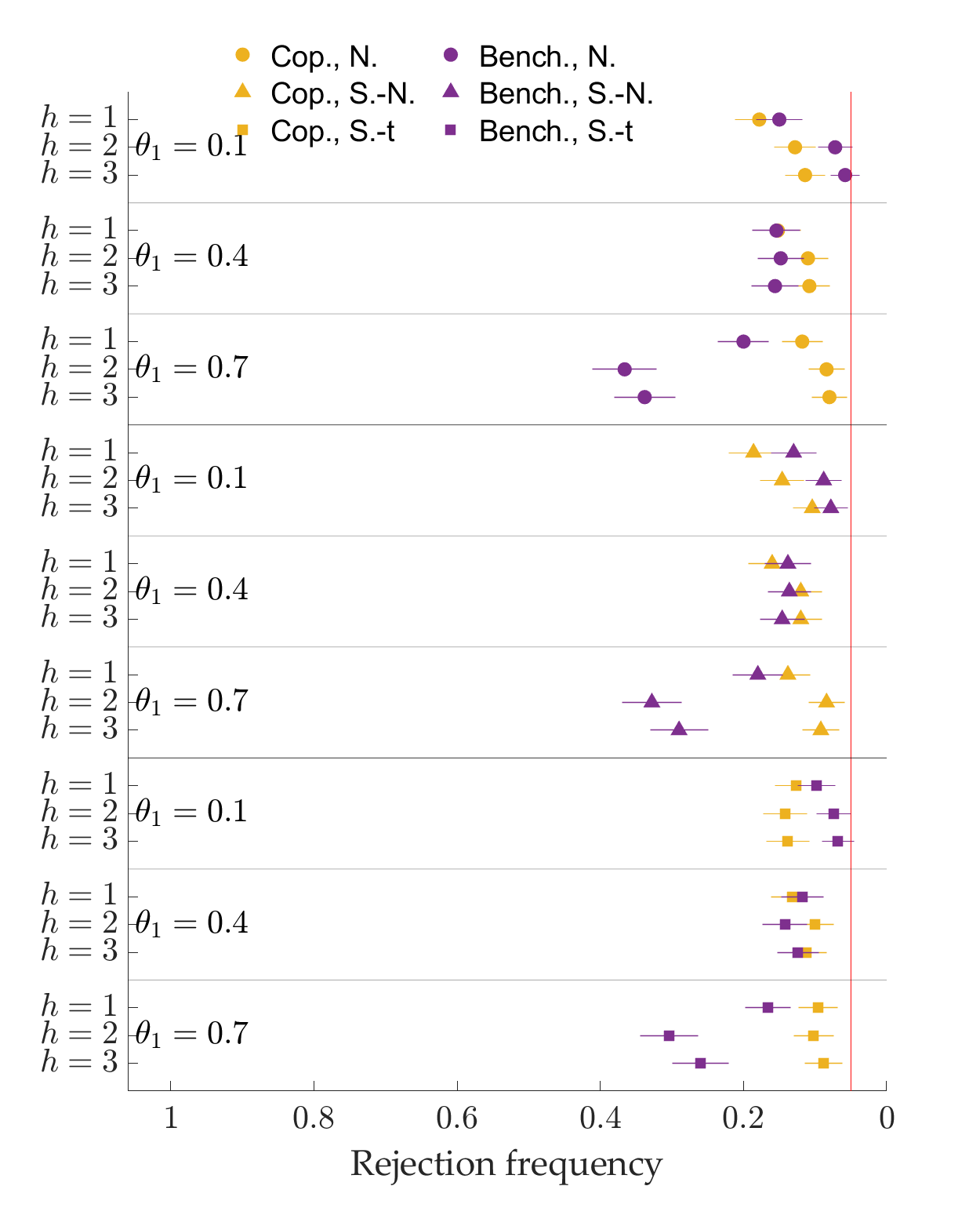}
    \vspace{-1cm}    
	\caption{Rejection frequencies of EPA test: annual-average}
\end{subfigure} 
\begin{subfigure}[b]{0.49\textwidth}
	\centering
		  \includegraphics[width=\linewidth]{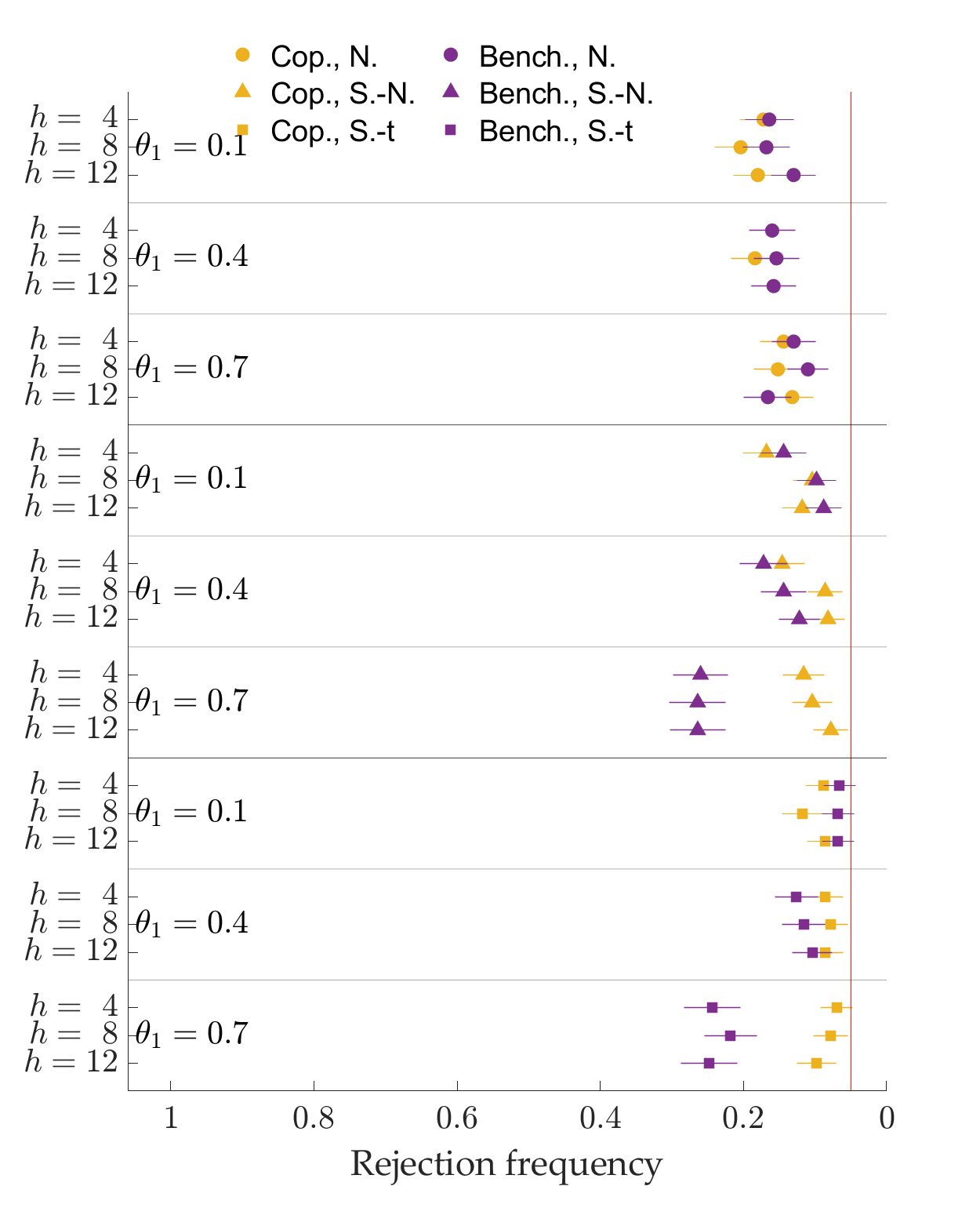}
        \vspace{-1cm}
	\caption{Rejection frequencies of EPA test: y-o-y}
\end{subfigure}
	\caption*{\footnotesize \textit{Note}: The QS(10\%) denotes the quantile score at the 10\% quantile. The $\theta$ indicates the autoregressive parameter of $Y_{t}$ in the DGP. The $y$-axis label $h$ denotes the annual-average horizon, i.e., one-year-, two-years-, and three-years-ahead. The $x$-axis in Panel (a) and (b) indicates the QS(10\%) of the copula relative to the benchmark approach, i.e., numbers smaller than one indicate a superior performance of the copula approach. In Panel (c) and (d), the $x$-axis denotes the rejection frequency of the null hypothesis of a \cite{Giacomini2006} test of unconditional equal predictive ability using the QS (10\% ). The nominal size is 5\%. N., S.-N., and S.-$t$ indicate the Normal, Skew-Normal, and Skew-t distribution of the error terms in the DGP. Standard errors of the tests were computed using a HAC estimator with $\text{bandwidth}=h_{A}-1$.}  
\end{figure}
\FloatBarrier

\subsection{Inflation at Risk data}\label{app:InflationatRisk}

\cref{tab:PredictorsInflationatRisk} shows the predictors used in \cref{sec:EmpiricalInflation}. All data is seasonally adjusted were applicable. The transformation codes imply the following: 1 --- no transformation; 4 --- log($x_t$); 5 --- 100[log($x_t$) - log($x_{t-12}$)]		

\begin{table}[htb!]
	\centering
	\renewcommand\arraystretch{1.5}
	\setlength{\tabcolsep}{1pt}
	\begin{threeparttable}
		\caption{Predictors for inflation at risk forecasts}\label{tab:PredictorsInflationatRisk}
		\small
		\begin{tabular}{lcc}
			\toprule
			Variable&Transformation & Mnemonic \\
			\cmidrule{1-3}
    Aggregate weekly hours & 5& AWHI\\
    Commercial + industrial loans & 5& BUSLOANS \\
     Labor force participation rate &4&CIVPART \\
   Consumer loans & 5&CONSUMER  \\
   CPI All Urban Consumers   & 5&CPIAUCSL\\
Canadian dollar to U.S. exchange rate   & 5&EXCAUS \\
Japanese Yen to U.S. exchange rate    & 5&EXJPUS \\
 British Pound to U.S. exchange rate  & 5& EXUSUK \\
  Federal Funds Target  & 1& FEDFUNDS \\
  Private Housing starts  & 4&HOUST \\
   New Family houses sold & 4&HSN1F \\
  Industrial production  & 5&INDPRO \\
  Fixed-rate 30-year mortgage rate  & 1&MORTG \\
   Bank prime loan rate & 1& MPRIME \\
    Motor Vehicle assemblies& 1&MVATOTASSS \\
   Total non-farm employees & 5&PAYEMS \\
  Real estate loans  & 5& REALLN \\
  Capacity utilization  & 4&TCU \\
   Number unemployed for 15 weeks \& over & 4&UEMP15OV \\
  Number unemployed for less than 5 weeks  & 4&UEMPLT5 \\
 University of Michigan: consumer sentiment   & 1&UMCSENT \\
  Unemployment rate  & 1&UNRATE \\
   WTI spot price & 5&WTISPLC \\
			\\
			\bottomrule
		\end{tabular}
		\begin{tablenotes}\setlength\labelsep{0pt}
			\item {\footnotesize\textit{Note:} All data was downloaded from the database FRED of the St. Louis Federal Reserve Bank.}
		\end{tablenotes}
	\end{threeparttable}
\end{table}

\FloatBarrier
\newpage

\subsection{Inflation at Risk quantile interpolation}
Let $F_{t+h|t}(y_{t+h})$ denotes the predictive cumulative distribution function in $t$ and $h$ horizons ahead, evaluated at $y_{t+h}$. Then, given the set $\{Q_{t+h|t}(\tau_{i})\}_{i=1}^{99}$ of predictive quantiles, we compute $F_{t+h|t}(y_{t+h})$ as follows:
\begin{equation} \label{eq:QuantileInterpolation} F_{t+h|t}(y_{t+h}) = \tau_i + \frac{\tau_{i+1}-\tau_{i}}{Q_{t+h|t}(\tau_{i+1})-Q_{t+h|t}(\tau_{i})} (y_{t+h}-Q_{t+h|t}(\tau_{i})) , \end{equation}
where $Q_{t+h|t}$ denotes predictive value of quantile $\tau_i$, and $\tau_i$ and $\tau_{i+1}$ are such that $y_{t+h} \in [Q_{t+h|t}(\tau_{i}),Q_{t+h|t}(\tau_{i+1})]$. For values of $y_{t+h} < Q_{t+h|t}(\tau_1)$ and $y_{t+h} > Q_{t+h|t}(\tau_{99})$, we approximate the slope as $\frac{\tau_{2}-\tau_{1}}{Q_{t+h|t}(\tau_{2})-Q_{t+h|t}(\tau_{1})}$ and  $\frac{\tau_{99}-\tau_{98}}{Q_{t+h|t}(\tau_{99})-Q_{t+h|t}(\tau_{98})}$ and the distance as $(Q_{t+h|t}(\tau_{1})-y_{t+h})$ and $(y_{t+h}-Q_{t+h|t}(\tau_{99}))$.

Similarly, we sample from the distribution given by the conditional quantiles $Q_{t+h|t}(\tau_i)$ as follows. Let $u_j$ denote the $j$-th draw from the uniform distribution and $u_j \in [\tau_i^j,\tau_{i+1}^j]$. Then, 
\begin{equation} y_{t+h|t}^j = Q_{t+h|t}(\tau_{i}^j) + \frac{Q_{t+h|t}(\tau_{i+1}^j)-Q_{t+h|t,}(\tau_{i}^j)}{\tau_{i+1}^j-\tau_{i}^j} (u_{j}-\tau_{i}^j) , \end{equation}
where $y_{t+h|t}^j$ denotes draw $j$ of the predictive distribution for $y_{t+h}$, conditional on information in $t$. The two endpoints are treated analogously to the procedure described for equation \eqref{eq:QuantileInterpolation}.

\end{appendices}
\end{document}